\documentclass[twocolumn,showpacs,amsmath,amssymb,prc,superscriptaddress,nofootinbib]{revtex4-1}
\usepackage{graphicx}

\usepackage{dcolumn}% Align table columns on decimal point
\usepackage{bm}% bold math
\usepackage{times}
\usepackage{siunitx}
\usepackage{booktabs}
\usepackage{etoolbox}
\usepackage{braket}
\usepackage{threeparttable}
\usepackage{rotating}
\usepackage{multirow}
\usepackage{subcaption}
\usepackage{aas_macros}
\usepackage{dirtytalk}
\usepackage{xcolor}
\robustify\itshape
\allowdisplaybreaks

\AtBeginDocument{
\heavyrulewidth=.08em
\lightrulewidth=.05em
\cmidrulewidth=.03em
\belowrulesep=.65ex
\belowbottomsep=0pt
\aboverulesep=.4ex
\abovetopsep=0pt
\cmidrulesep=\doublerulesep
\cmidrulekern=.5em
\defaultaddspace=.5em
}

\newcolumntype{.}{D{.}{.}{1}}
\newcolumntype{X}{D{X}{X}{1}}

\newcommand{\printthis}[2][true]{%
\ifbool{#1}{%
#2}{}}% End of \printthis

\begin{document}
\def\sun{\odot}

%%%%%%%%%%%%%%%% TITLE PAGE %%%%%%%%%%%%%%%%%%%%%%%%
\title{New Constraints on Sodium Production in Globular Clusters From the $^{23}$Na$(^3$He$, \textbf{d})^{24}$Mg Reaction}
 
\author{C.~Marshall}
\altaffiliation{Present Address: Institute of Nuclear \& Particle Physics, Department of Physics \& Astronomy, Ohio University, Athens, Ohio 45701, USA}
 \affiliation{Department of Physics, North
  Carolina State University, Raleigh, NC 27695, USA}
\affiliation{Triangle Universities Nuclear Laboratory, Durham, NC
  27708, USA}

\author{K.~Setoodehnia}
\altaffiliation{Present Address: Facility For Rare Isotope Beams, Michigan State University, East Lansing, Michigan 48824, USA}
\affiliation{Department of Physics, North
  Carolina State University, Raleigh, NC 27695, USA}
\affiliation{Triangle Universities Nuclear Laboratory, Durham, NC
  27708, USA}

\author{G.~C.~Cinquegrana} \affiliation{School of Physics \& Astronomy, Monash University
, Clayton VIC 3800, Australia} \affiliation{ARC Centre of Excellence for All Sky Astrophysics in 3 Dimensions (ASTRO 3D)}

\author{J. H.~Kelly} \affiliation{Department of Physics, North
  Carolina State University, Raleigh, NC 27695, USA}
\affiliation{Triangle Universities Nuclear Laboratory, Durham, NC
  27708, USA}

\author{F.~Portillo Chaves} \affiliation{Department of Physics, North
  Carolina State University, Raleigh, NC 27695, USA}
\affiliation{Triangle Universities Nuclear Laboratory, Durham, NC
  27708, USA}

\author{A.~Karakas} \affiliation{School of Physics \& Astronomy, Monash University
, Clayton VIC 3800, Australia} \affiliation{ARC Centre of Excellence for All Sky Astrophysics in 3 Dimensions (ASTRO 3D)}

\author{R.~Longland} \affiliation{Department of Physics, North
  Carolina State University, Raleigh, NC 27695, USA}
\affiliation{Triangle Universities Nuclear Laboratory, Durham, NC
  27708, USA}

\begin{abstract}

  The star to star anticorrelation of sodium and oxygen {is} a defining feature of globular clusters, but, to date, the astrophysical site responsible for this unique chemical signature remains unknown. Sodium enrichment within these clusters depends sensitively on reaction rate of the sodium destroying reactions $^{23}$Na$(p, \gamma)$ and $^{23}$Na$(p, \alpha)$.   
  In this paper, we report the results of a $^{23}$Na$(^3\text{He}, d)^{24}$Mg transfer reaction carried out at Triangle Universities Nuclear Laboratory using a $21$ MeV $^3$He beam. Astrophysically relevant states {in $^{24}$Mg} between $11 < E_x < 12$ MeV were studied using high resolution magnetic spectroscopy, thereby allowing the extraction of excitation energies and spectroscopic factors. Bayesian methods are combined with the distorted wave Born approximation to assign statistically meaningful uncertainties to the extracted spectroscopic factors. For the first time, these uncertainties are propagated through to the estimation of proton partial widths. Our experimental data are used to calculate the reaction rate. The impact of the new rates are investigated using asymptotic giant branch star models. It is found that while the astrophysical conditions still dominate the total uncertainty, intra-model variations on sodium production from the $^{23}$Na$(p, \gamma)$ and $^{23}$Na$(p, \alpha)$ reaction channels are a lingering source of uncertainty.

\end{abstract}

% \keywords{Nuclear reactions, nucleosynthesis, abundances --- Stars:
% AGB and post-AGB --- Stars: massive}%Use showkeys class option if keyword
% display desired

\maketitle

% \tableofcontents

\section{Introduction}

Globular clusters are among the oldest objects in the Milky Way. Comprised of tens to hundreds of thousands of stars that are gravitationally bound, they offer a unique probe of galactic and stellar evolution \cite{gratton_2004, gratton_2012}. Despite decades of intense study, we have an incomplete understanding of their unique chemical evolution \cite{gratton_2019}. In particular, high resolution photometry has unambiguously determined the presence of multiple stellar populations within these clusters \cite{piotto_2007}, with the youngest of these populations displaying a star-to-star variation in light elements. The anti-correlation between sodium and oxygen is the most ubiquitous chemical signature, and as such can be considered a defining feature of globular clusters \cite{gratton_2019}. The Na-O anti-correlation indicates that some amount of cluster material has undergone hydrogen burning at elevated temperatures  \cite{kudryashov_1988, d_and_d_1989, prantzos_2017}. However, at this time the source of this enriched material is still unknown, with {models of massive asymptotic branch stars, fast rotating massive stars, interacting massive binaries, and very massive stars all} failing to reproduce the observed chemical signatures \cite{problems_with_hbb}.

Sodium is synthesized from a series of proton capture reactions that occur during hydrogen burning at $50\text{-}100$ MK. Known as the NeNa cycle, this group of proton induced reactions and $\beta$-decays around $A = 20 \text{-} 24$ are of critical importance to understanding the creation of sodium in globular clusters. Within the NeNa cycle, sodium may be destroyed via the $^{23}$Na$(p, \gamma)$ or $^{23}$Na$(p, \alpha)$ reactions, both of which proceed through the compound nucleus $^{24}$Mg. For decades direct measurements have aimed to constrain these astrophysical reaction rates for the $(p, \gamma)$ and $(p, \alpha)$ channels \cite{Zyskind_1981, goerres_1989}. The study of {G\"orres} \textit{et al.} (Ref.~\cite{goerres_1989}) is of particular note, as it was one of the first to directly search for a resonance around $138$-keV. Corresponding to the $E_x \approx 11830$-keV state in $^{24}$Mg, this state was first observed in the indirect measurements of Refs.~\cite{moss_1976, vermeer_1988} and is thought to dominate the $^{23}$Na$(p, \gamma)$ rate at the temperatures important to globular cluster nucleosynthesis. Since the study of {G\"orres} \textit{et al.}, several direct searches have been performed, all with the intent of measuring the $138$-keV resonance {strength}. {The authors of Ref.~\cite{Rowland_2004} reported} an upper limit of $\omega \gamma_{(p, \gamma)} \leq 1.5 \times 10^{-7}$ eV. Subsequently, {the authors of} Ref.~\cite{Cesaratto_2013} used a high intensity proton beam of $\approx 1$ mA to give a further reduced upper limit of $\omega \gamma_{(p, \gamma)} \leq 5.17 \times 10^{-9}$ eV, in the process ruling out its importance for the $(p, \alpha)$ channel. Recently, nearly thirty years after the first direct search was carried out, detection of the $138$-keV resonance with a statistical significance above $2 \sigma$ came in Ref.~\cite{BOELTZIG_2019} {reporting} $\omega \gamma_{(p, \gamma)} = 1.46^{+0.58}_{-0.53} \times 10^{-9}$ eV. These efforts have solidified the important role of the $138$-keV resonance in globular cluster nucleosynthesis.

At the present time, {direct measurements of the $138$-keV resonance strength have greatly reduced the uncertainty of the $^{23}$Na$(p, \gamma)$ reaction rate at the temperatures of relevance to globular clusters to approximately $30 \%$}. However, much of the rate is still dependent on the results and evaluation presented in Ref.~\cite{hale_2004}. In that study, a $(^{3}$He$,d)$ transfer reaction was performed, and a state at $E_x = 11831.7(18)$ keV was observed. The $^{23}$Na$(^{3}$He$,d)$ measurement we present in this paper was carried out {to further reduce the reaction rate uncertainty.} Earlier results from our experiment have been published in Ref.~\cite{Marshall_2021}, and provided evidence that the $138$-keV resonance lies at a lower energy of $133$ keV, resulting in a factor of $2$ increase in the {$^{23}$Na}$(p, \gamma)$ reaction rate. {This paper uses the same data set as as Ref.~\cite{Marshall_2021} but expands upon the analysis of that paper by providing} a more complete set of updated excitation energies, and reports spectroscopic factors for levels of astrophysical interest. Bayesian analysis methods are applied to extract excitation energies, spectroscopic factors, and $\ell$ values. Our analysis is the first of its kind, where every quantity extracted from the transfer measurement is assigned uncertainties based on Bayesian statistical arguments, allowing these quantities and their uncertainties to be incorporated into thermonuclear reaction rate libraries. 

% The present day constraints on the energy of the $138$-keV resonance have a strong dependence on this precise energy measurement. On the other hand, the angular distribution for the $11831$-keV state was inconclusive and could not provide a constraint on either its spin or parity. Perhaps most importantly, the authors of that work presented a detailed evaluation of the literature to formulate the basis for the current $(p, \gamma)$ and $(p, \alpha)$ rates.  

Our paper is organized as follows: Sec.~\ref{sec:experiment-details} provides an overview of the experimental techniques, Sec.~\ref{sec:energy_level_update} gives an in depth discussion of the necessary corrections to the current nuclear data in order to extract accurate excitation energies for the current experiment, Sec.~\ref{sec:energy_cal_na} reports our energy values and gives updated recommended values, Sec.~\ref{sec:bayes-dwba-analys} presents the analysis of the transfer angular distributions using a Bayesian method for the distorted-wave Born approximation (DWBA), and Sec.~\ref{sec:prot-part-widths} reports our values for the proton partial widths derived from this experiment.
Sec.~\ref{sec:23nap-gamma-23nap} presents our updated astrophysical reaction rate and its incorporation into an \textbf{a}symptotic \textbf{g}iant \textbf{b}ranch (AGB) model, one of the possible sites for the Na-O abundance anomaly in globular clusters. Sec.~\ref{sec:conclusions-outlook} provides additional outlook and discussion.

\section{Experiment Details}
\label{sec:experiment-details}
The $^{23}$Na$(^{3}\textnormal{He}, d)^{24}$Mg experiment was carried out at Triangle Universities Nuclear Laboratory (TUNL) using the Split-pole spectrograph (SPS) { \cite{setoodehnia_2016}}. A beam of $^3$He$^{+2}$ was accelerated to $21$ MeV using the 10 MV TUNL FN tandem accelerator, and the beam energy was set using a set of high resolution $90 \text{-} 90$ dipole magnets. While the amount of beam that made it to the target varied throughout the experiment, typical beam currents were $100-200$ enA of $^3$He$^{+2}$.

For the experiment reported here, NaBr was selected as the target material based on the observations of Ref.~\cite{hale_2004}. The authors of that study noted that NaBr targets were stable to beam bombardment, reasonably resistant to oxygen contamination, and found no evidence of contaminant states arising from reactions on $^{79, 81}$Br in the region of interest. Our targets were fabricated by using thermal evaporation to deposit a layer of NaBr on $22$ $\mu$g$/$cm$^2$ thick $^{\textnormal{nat}}$C foils. The carbon foils were purchased from Arizona Carbon Foil Co., Inc. \cite{acfmetals}, and floated onto target frames to create the backing for the NaBr layer. A quartz crystal thickness monitor measured the rate of deposition and total thickness of the targets. A total of six targets were placed into the evaporator, and evaporation was halted once they reached a thickness of $70$ $\mu$g/cm$^2$. After the evaporation was complete,  the targets were brought up to atmosphere and then immediately placed into a container for transfer to the target chamber of the SPS. This container was brought down to rough vacuum to reduce exposure to air during transport. Three of the targets were mounted onto the SPS target ladder. In addition to the NaBr targets, the ladder was also mounted with a $1$ mm diameter collimator for beam tuning, a $^{\textnormal{nat}}$C target identical to the backing of the NaBr targets for background runs, and thermally evaporated $^{27}$Al on a $^{\textnormal{nat}}$C backing to use for an external energy calibration. {All three NaBr targets were used during the 120 hour beam time. No degradation for any of the targets was observed in the elastic scattering spectra (discussed below), nor was there any sign of significant oxidation.}

The $^{23}$Na$(^{3}\textnormal{He}, d)^{24}$Mg reaction was measured at angles between $3^{\circ} \text{-} 21^{\circ}$ in steps of $2^{\circ}$ with a field of $1.14 \text{-} 1.15$ T. Additionally, the elastic scattering reaction, $^{23}$Na$(^{3}\textnormal{He}, ^{3}\textnormal{He})$, was measured at angles between $15^{\circ} \text{-} 55^{\circ}$ in $5^{\circ}$ steps and $59^{\circ}$ using fields of $0.75 \text{-} 0.80$ T. The solid angle of the SPS was fixed throughout the experiment at $\Omega_{\textnormal{SPS}} = 1.00(4)$ msr. After the reaction products were momentum to charge analyzed by the spectrograph, they were detected at the focal plane of the SPS. The focal plane detector consists of two position sensitive avalanche counters, a $\Delta E$ proportionality counter, and a residual $E$ scintillator. Additional detail about this detector can be found in Ref.~\cite{marshall_2018}.

Due to potential for uncontrolled systematic effects from the charge integration of the SPS beamstop and target degradation, it was decided to determine the absolute scaling of the data relative to $^{23}$Na$(^{3}\textnormal{He}, ^{3}\textnormal{He})$. Elastic scattering was measured continuously during the course of the experiment by a silicon $\Delta E/E$ telescope positioned at $\theta_{lab} = 45^{\circ}$. The telescope was double-collimated using a set of brass apertures to define the solid angle. A geometric solid angle of $\Omega_{\textnormal{Si}} = 4.23(4)$ msr was measured.

\section{Updates to Energy Levels Above $11$ M\lowercase{e}V}
\label{sec:energy_level_update}

Spectrograph measurements like the current experiment are dependent on previously reported excitation energies for energy calibration of the focal plane. In the astrophysical region of interest ($ 11 \lessapprox E_x \lessapprox 12$ MeV) the current ENSDF evaluation, Ref.~\cite{firestone_2007}, was found to be inadequate for an accurate calibration of our spectra. Discussion of the issues with the evaluation are available in Ref.~\cite{Marshall_2021}, which first reported the astrophysically relevant results of the energy measurements of this work. 
In addition to the issues mentioned in the prior work, the current ENSDF evaluation recommends energies that include calibration points from the spectrograph measurements of Ref.~\cite{moss_1976} and Ref.~\cite{zwieglinski_1978}. The inclusion of these calibration points is an error because calibration points are not independent measurements and increase the weight of the values they are based on in the resulting average. Every compilation and evaluation since 1978 \cite{ENDT_1978} includes this error. The measurements of Hale \textit{et al}.~\cite{hale_2004} have been excluded from our compiled values. Discussion of this decision can be found in Appendix A. Our compiled values are based on the most precise available literature, but are limited to a narrow excitation region selected for the purpose of accurately energy calibrating the current experiment and subsequently for calculating the astrophysical reaction rate. We made no attempt to update values outside of the region of interest.         

Our compiled energies are presented in Table.~\ref{tab:energy_comp}. Note that in the case of Ref.~\cite{endt_1990}, resonant capture was used to excite $^{24}$Mg, but the excitation energies were deduced from gamma ray energies making these values independent of the reaction $Q$-value. For the measurements that report the laboratory frame resonance energies, the excitation energies are deduced from:
\begin{equation}
  \label{eq:lab_to_ex}
  E_x = Q + E_{P} \frac{M_T}{M_T + M_{P}}, 
\end{equation}
where $E_P$ is the projectile energy measured in the laboratory frame, and $M_{P}$ and $M_T$ are the \textit{nuclear} masses for the projectile and target nuclei, respectively. We have used the \textit{atomic} masses from Ref.~\cite{Wang_2017} assuming the difference is negligible compared to the statistical uncertainty in $E_{P}$. $Q$ is the $Q$-value for either the $(p,\gamma)$ or $(\alpha, \gamma)$ reaction. The $2020$ mass evaluation \cite{AME_2020_1, AME_2020_2} was released after our compilation, but leads to a difference of $1$ eV in the central $Q$-value, well below its associated uncertainty. The column in Table \ref{tab:energy_comp} from Ref.~\cite{endt_eval_1990} shows energies deduced from a weighted average of several $(p, \gamma)$ measurements, and that paper should be referred to for additional details. For the present work, the suggested value of these weighted averages is treated as a single measurement that is updated according to Eq.~(\ref{eq:lab_to_ex}). The weighted averages of all measurements are presented in the last column. In order to reduce the effects of potential outliers, when there are three or less measurements, the lowest measured uncertainty was used instead of the weighted average uncertainty. 
%
% \begin{equation}
%     \label{eq:weighted_mean}
%     \bar{x} = \frac{\sum_i^N w_i x_i}{\sum_j^N w_j},
% \end{equation}
% %
% with uncertainty given by:
% \begin{equation}
%   \label{eq:weighted_unc}
%     \bar{\sigma} = \frac{1}{\sqrt{\sum_j^N w_j}},
% \end{equation}
%
%where the weight is $w_i = 1/\sigma_i^2$, $\sigma_i$ is the uncertainty of measurement $i$, $x_i$ is the reported value of measurement $i$, and $N$ is the total number of measurements. 

\begin{turnpage}
  \begin{table*}
    \centering
    \setlength\tabcolsep{8pt}
  \def\arraystretch{1.2}
  \caption{ \label{tab:energy_comp} Previously measured energies {in units of keV}. An * indicates that the listed energy was used as a calibration point in the listed experiment. These values, therefore, have been excluded from the weighted average. Excitation energies derived from resonance energies have been updated based on the 2016 mass evaluation \cite{Wang_2017}. Note the $\dagger$ on the $12259.6$-keV state. This value was taken from Ref.~\cite{endt_eval_1990}, and is actually the unweighted average of a pair of states with updated energies of $12259.4(4)$ keV and $12259.8(4)$ keV, respectively.}

    \begin{tabular}{lllllllllll}
    \toprule
    \toprule
         $(p, p^{\prime})$ & $(p, p^{\prime})$ & $(^{16} \textnormal{O}, \alpha)$ & $(p,\gamma)$ & $(p,\gamma)$ & $(\alpha,\gamma)$ & $(\alpha,\gamma)$ & $(\alpha,\gamma)$ & $(\alpha,\gamma)$ & Weighted Average \\
    \cite{moss_1976}      & \cite{zwieglinski_1978} & \cite{vermeer_1988}  & \cite{endt_1990}       & \cite{endt_eval_1990} & \cite{fiffield_1978} & \cite{schmalbrock_1983} & \cite{goerres_1989}  & \cite{smulders_1965}   \\  \hline
 11389(3)*                             & 11391(7)                                     & 11390(4)                                 &                                       &              &                                            & 11394(3)                                     &                                                               & 11393(5)                                  & 11392.5(21)      \\
11456(3)                              & 11452(7)                                     & 11455(4)                                 & 11452.8(4)                            &              &                                            & 11456(3)                                     &                                                               &                                           & 11452.9(4)       \\ 
11521(3)*                             & 11520(7)                                     & 11519(4)                                 &                                       &              &                                            & 11522(2)                                     &                                                               & 11523(5)                                  & 11521.5(16)      \\
11694(3)*                             & 11694(7)*                                    & 11694(4)                                 &                                       &              &                                            & 11699(2)                                     &                                                               & 11694(5)                                  & 11698(2)         \\
11727(3)*                             & 11727(7)                                     & 11727(4)                                 &                                       &              &                                            & 11731(2)                                     &                                                               & 11728(5)                                  & 11729.8(16)      \\
11828(3)                              &                                              & 11827(4)                                 &                                       &              &                                            &                                              &                                                               &                                           & 11828(3)         \\
11862(3)*                             & 11860(7)                                     & 11860(4)                                 &                                       &              & 11861(5)                                   & 11868(3)                                     & 11859.4(20)                                                   & 11862(5)                                  & 11861.6(15)      \\
11935(3)*                             &                                              & 11930(4)                                 &                                       & 11933.05(19) &                                            &                                              & 11933.2(10)                                                   &                                           & 11933.05(19)      \\
11967(3)*                             & 11965(7)                                     & 11963(4)                                 &                                       & 11966.6(5)   & 11967(5)                                   & 11974(3)                                     & 11966.7(10)                                                   & 11968(5)                                  & 11966.7(5)       \\
11989(3)*                             & 11990(7)                                     & 11985(4)                                 & 11988.0(3)                            & 11988.47(6)  &                                            &                                              & 11988.7(10)                                                   &                                           & 11988.45(6)      \\
12015(3)*                             & 12016(7)*                                    &                                          &                                       & 12017.1(6)   & 12016(5)                                   &                                              & 12016.5(10)                                                   &                                           & 12016.9(5)       \\
12050(3)*                             & 12050(7)                                     &                                          &                                       & 12051.3(4)   & 12050(5)                                   &                                              &                                                               &                                           & 12051.3(4)       \\
12121(3)*                             & 12124(7)                                     &                                          &                                       & 12119(1)     & 12121(5)                                   &                                              &                                                               &                                           & 12119(1)         \\
12181(3)*                             &                                              &                                          &                                       & 12183.3(1)   &                                            &                                              &                                                               &                                           & 12183.3(1)       \\
12258(3)*                             & 12261(7)                                     &                                          &                                       & 12259.6(4)$^{\dagger}$   & 12258(5)                                   &                                              &                                                               &                                           & 12259.6(4)       \\
12342(3)*                             &                                              &                                          &                                       & 12341.0(4)   &                                            &                                              &                                                               &                                           & 12341.0(4)       \\
12402(3)                              & 12402(7)                                     &                                          &                                       & 12405.3(3)   & 12405(5)                                   &                                              &                                                               &                                           & 12405.3(3)       \\
12528(3)*                             &                                              &                                          &                                       & 12528.4(6)   &                                            &                                              &                                                               &                                           & 12528.4(6)       \\
12577(3)*                             & 12578(7)                                     &                                          &                                       &              & 12578(5)                                   &                                              &                                                               &                                           & 12578(5)         \\
12669(3)                              &                                              &                                          & 12669.9(2)                            & 12670.0(4)   &                                            &                                              &                                                               &                                           & 12669.9(4)       \\
12736(3)                              & 12739(7)                                     &                                          &                                       & 12739.0(7)   & 12740(5)                                   &                                              &                                                               &                                           & 12738.9(7)       \\
                                      &                                              &                                          &                                       & 12817.77(19) &                                            &                                              &                                                               &                                           & 12817.77(19)     \\
12849(3)                              & 12850(7)                                     &                                          &                                       & 12852.2(5)   &                                            &                                              &                                                               &                                           & 12852.1(5)       \\
12921(3)*                             &                                              &                                          &                                       & 12921.6(4)   & 12923(5)                                   &                                              &                                                               &                                           & 12921.6(4)       \\
12963(3)                              &                                              &                                          &                                       & 12963.9(5)   &                                            &                                              &                                                               &                                           & 12963.9(5)      \\
   
    \bottomrule
    \bottomrule
\end{tabular}
\end{table*}

\end{turnpage}
% \end{landscape}
%\end{lscapenum}
%\restoregeometry

\section{Energy Calibration}
\label{sec:energy_cal_na}

Our energy calibration is the same one as reported in Ref.~\cite{Marshall_2021}, but is reiterated and expanded here for clarity and completeness. Excitation energies were extracted from the focal plane position spectrum using a third-order polynomial fit to parameterize the bending radius of the SPS in terms of the ADC channels, $x$:

\begin{equation}
  \label{eq:energy_fit}
  \rho = Ax^3 + Bx^2 + Cx + D.
\end{equation}

An updated version of the Bayesian method presented in Ref.~\cite{marshall_2018} was used to fit the polynomial. Briefly, the method accounts for uncertainties in both $x$ and $\rho$ while also estimating an additional uncertainty based on the quality of the fit. The update uses the python package \texttt{emcee} to more efficiently sample the posterior \cite{emcee}. As a result, excitation energies can be directly calculated from posterior samples, ensuring the correlations between the fit parameters are correctly accounted for in our reported energies.   
 
\begin{figure*}
    \centering
    \includegraphics[width=\textwidth]{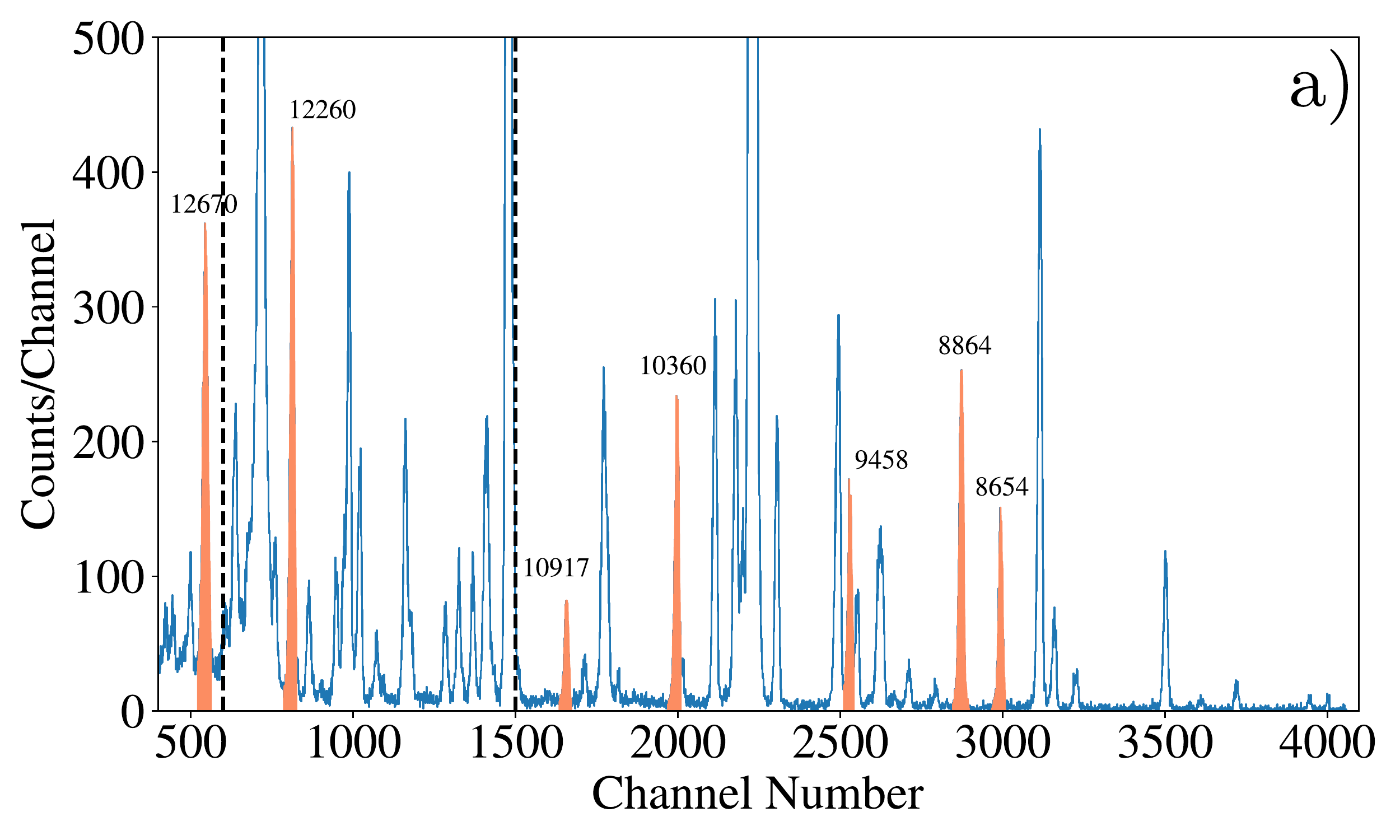}
    \includegraphics[width=\textwidth]{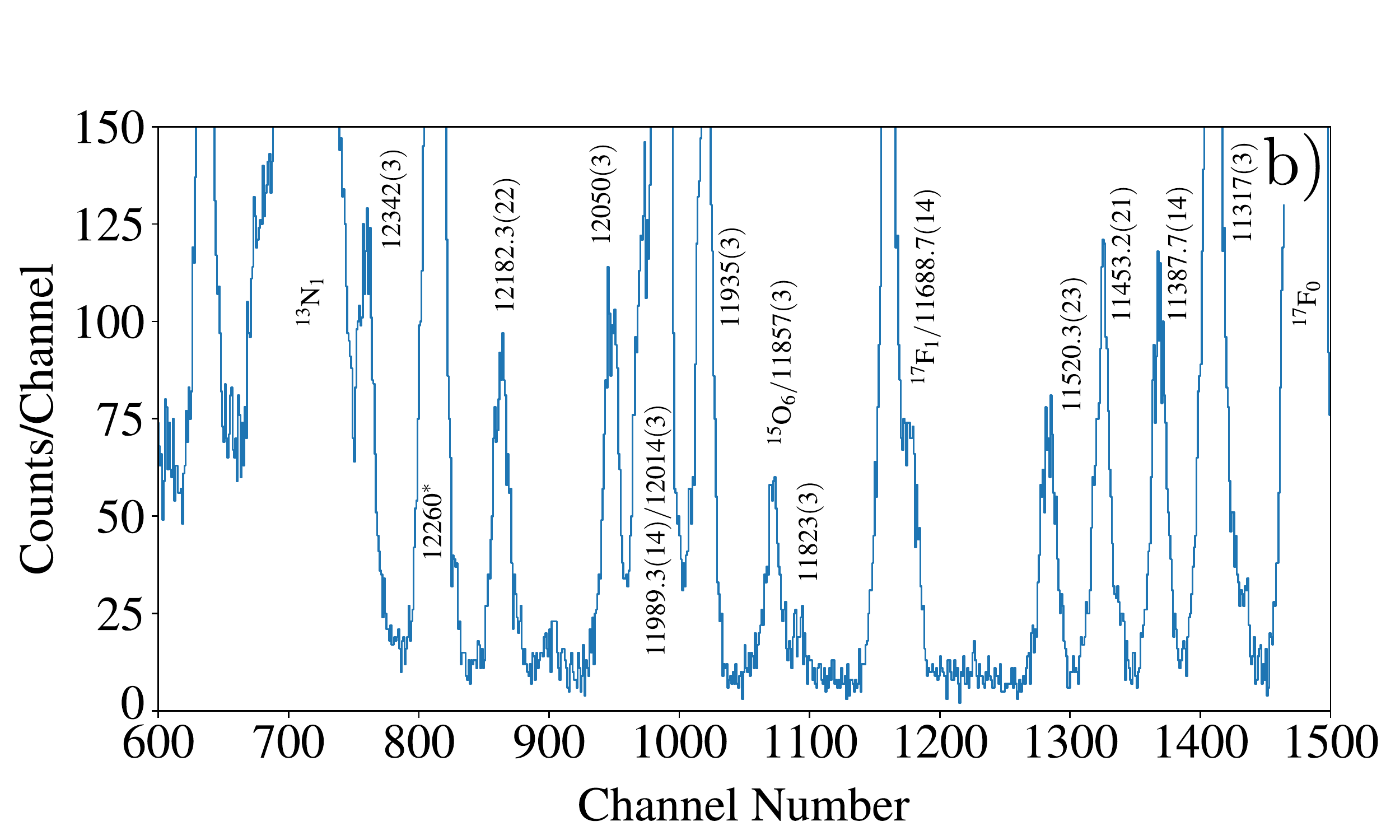}
    \caption{Full and partial focal plane position spectrum at $\theta_{lab} = 11^{\circ}$ {after 6 hours ($10^{15}$ particles) of data accumulation.} The top panel (\say{a)}) shows the entire focal plane spectrum with the calibration states ({given in keV}) highlighted in orange and {the astrophysical region of interest is between the dashed black lines}. The energy values for the states below the proton threshold {($11692.69$ keV)} are taken from Ref.~\cite{firestone_2007}, while the rest are from Table~\ref{tab:energy_comp}. {All values in the top panel are rounded to the nearest integer.} The bottom panel (\say{a)}) is zoomed in on the astrophysical region of interest. Peaks from $^{24}$Mg have been identified with their final weighted average energy value in keV.}
    \label{fig:na_cal_spec}

\end{figure*}

Calibration states were methodically selected to span the majority of the focal plane. Care was taken to avoid introducing additional systematic errors that would come with misidentifying a state used for calibration. As such, some intensely populated peaks were excluded due to the possibility of misidentifying them with nearby levels that differed in energy by more than a few keV. The chosen calibration states at $\theta_{lab} = 11^{\circ}$ are shown in the top panel of Fig.~\ref{fig:na_cal_spec}. The validity of this internal calibration in the astrophysical region of interest between $11$ and $12$ MeV was checked at $\theta_{lab} = 11^{\circ}$ against a separate external calibration using the $^{27}$Al$(^3$He$, d)^{28}$Si reaction. The aluminum states were selected based on the spectrum shown in Ref.~\cite{champagne_1986}. When applying the external aluminum calibration to the sodium states an energy offset of $\approx 7$ keV compared to the internal calibration was observed. Using the stopping powers of SRIM \cite{Ziegler_2010}, it was found that the energy offset could be ascribed to the difference in energy loss between the Al and NaBr targets. Taking the above as confirmation of its validity, the internal calibration was adopted for all angles.

The energies of this work are presented in Table~\ref{tab:comp_and_recommended}. They are the weighted average of the energies deduced at each angle. The bottom panel of Fig.~\ref{fig:na_cal_spec} shows the location of the peaks in the astrophysical region of interest at $11^{\circ}$. Only states that were seen at three or more angles are reported. Calibration states are given without uncertainties, italicized, and marked with an asterisk for clarity in Table~\ref{tab:comp_and_recommended}.

The additional uncertainty estimated by our Bayesian energy calibration also introduces a further complication into the weighted averaging between angles. Since this uncertainty is estimated directly from the data, it will be influenced by systematic effects. These systematic effects introduce correlations between the deduced energies and uncertainties at each angle, which can become significant because of the high number of angles measured in this experiment. A clear indication of a correlation was seen in the deduced energies of our calibration points. The energy of the calibration points predicted by the fit tend to agree with their input values at each angle, but {exhibit little statistical scatter from angle to angle producing some disagreement larger than $2\sigma$ for final values if a simple weighted average is adopted}. To account for possible correlations, the uncertainties on the weighted averages were estimated using the methods of Ref.~\cite{Schmelling_1995}. This correction is done by calculating the $\chi^2$ value of the data with respect to the weighted average, $\bar{x}$, which is given by:
\begin{equation}
  \label{eq:chi_sq}
  \chi^2 = \sum_i^N \frac{(x_i - \bar{x})^2}{\sigma_i^2}.
\end{equation}
Since the expected value of $\chi^2$ is $N - 1$, the uncertainties from the weighted average, $\bar{\sigma}$, are adjusted based on the deviation from $N-1$. For the case of positive correlations, $\chi^2 < 1$, and, therefore, $\bar{\sigma}$ will need to be adjusted by:
\begin{equation}
  \label{eq:positive_corr_chi}
  \sigma_{adj} = \sqrt{(N-\chi^2) \bar{\sigma}^2}.
\end{equation}
A separate estimate can also be made if the scatter in the data is not well described by the weighted average. In this case, $\chi^2 > 1$, which gives the adjustment: 
\begin{equation}
  \label{eq:negative_corr_chi}
  \sigma_{adj} = \sqrt{\frac{\chi^2}{N-1} \bar{\sigma}^2}.
\end{equation}
To be conservative, we adopt the larger of these two values. It can be seen from Table \ref{tab:comp_and_recommended} that our energies are in good agreement with previous measurements. The sole exceptions are the pair of states at $E_x = 9838(7)$ {keV} and $9977(6)$ {keV}, which lie $10$ keV above the values reported in Ref.~\cite{firestone_2007}. However, both of these states show clear bimodal behavior as a function of spectrograph angle, undergoing shifts of over $10$ keV, and as a result skewing the average towards higher energies. Behavior of this nature is inconsistent with the kinematic shift seen from contaminants, but did not appear to impact the corresponding angular distributions, which were in agreement with the known $1^+$ spin-parities. These states have no bearing on the astrophysical measurement, so while their unique behavior is puzzling, we have opted to report the average of all angles using the \textit{expected value method} of Ref.~\cite{BIRCH_2014} to give a more conservative estimate of the uncertainties.

\begin{table*}[]
  \caption{\label{tab:comp_and_recommended} $^{24}$Mg excitation energies from this work compared {to the values of Table~\ref{tab:energy_comp} for states within the region of interest and those of Ref.~\cite{firestone_2007} for all others.} In some cases, due to the presence of a high number of states in certain regions, a unique identification of the observed state could not be made so all nearby states are listed. For states in the region of interest we compare to the values of Table.~\ref{tab:energy_comp}. The recommended excitation and resonance energies are also given for these states.  
States used for this {work's} energy calibration are reported in italics, marked with $*$, and listed without uncertainties, but they do represent the mean value obtained after calibration. {All energies are given in units of keV.}}
  \begin{threeparttable}
    \setlength\tabcolsep{8pt}
    \def\arraystretch{1.2}
    \centering
  \begin{tabular}{ll|llll}
    \toprule
    \toprule
ENSDF (\cite{firestone_2007}) & This Work                  & Compilation (Table~\ref{tab:energy_comp}) & This Work               & Recommended $E_x$ & {Recommended} $E_r$ \\ \hline
    $7349.00(3)$              & $7364(14)$                 & $11392.5(21)$                             & $11387.7(14)$           & $11389.2(12)$     & $-303.5(12)$                       \\
    $7555.04(15) $            & $7555(13)$                 & $11452.9(4)   $                           & $ 11453.2(21)$          & $11452.9(4) $     & $-239.8(4)$                        \\
 $7747.51(9)  $               & $7752(10)$                 & $11521.5(16)  $                           & $ 11520.3(23)$          & $11521.1(13)$     & $-171.6(13)$                       \\
 $8357.98(13) $               & $8362(4) $                 & $11698(2)     $                           & $ 11688.7(14)$          & $11694(4)   $     & $1(4)$                             \\
 $8437.31(15) $               & $8441(4) $                 & $11729.8(16)  $                           & $            $          & $11729.8(16)$     & $37.1(16)$                         \\
 $8439.36(4)  $               &                            & $11828(3)     $                           & $ 11823(3)   $          & $11826(3)   $     & $133(3)$                           \\
 $8654.53(15) $               & \textit{8654}\tnote{*}     & $11861.6(15)  $                           & $ 11857(3)   $          & $11860.8(14)$     & $168.1(14)$                        \\
 $8864.29(9)  $               & \textit{8864}\tnote{*}     & $11933.05(19) $                           & $ 11935(3)   $          & $11933.06(19)$    & $240.37(19)$                       \\
 $9003.34(9)  $               & $ 9002.9(24) $             & $11966.7(5)   $                           & $            $          & $11966.7(5) $     & $274.0(5)$                         \\
 $9145.99(15) $               & $ 9145.0(16) $             & $11988.45(6)  $                           & $ 11989.3(14)$          & $11988.45(6)$     & $295.76(6)$                        \\
 $9284.22(14) $               & $ 9292.6(12) $             & $12016.9(5)   $                           & $  12014(3)  $          & $12016.8(5) $     & $324.1(5)$                         \\
 $9299.77(24) $               &                            & $12051.3(4)   $                           & $ 12050(3)   $          & $12051.3(4) $     & $358.6(4)$                         \\
 $9457.81(4)  $               & \textit{9460}\tnote{*}     & $12119(1)$                                & $12121.5(17)$           & $12119.7(8) $     & $427.0(8)$                         \\
 $9516.28(4)  $               & $9520(3)$                  & $12183.3(1)$                              & $12182.3(22)$           & $12183.3(1) $     & $490.6(1)$                         \\
 $9527.8(21)  $               &                            & $12259.6(4)$                              & \textit{12260}\tnote{*} &                   &                                    \\
 $9828.11(11) $               & $9838(7)$\tnote{$\dagger$} & $12341.0(4)$                              & $12342(3)$              & $12341.0(4) $     & $648.3(4)$                         \\
 $9967.19(22) $               & $9977(6)$\tnote{$\dagger$} & $12405.3(3)$                              & $12406.0(22)$           & $12405.3(3) $     & $712.6(3)$                         \\
 $10027.97(9) $               & $10021(3)$                 & $12528.4(6)$                              & $12530.5(24)$           & $12528.5(6) $     & $835.8(6)$                         \\
 $10058.54(16)$               & $10055(3)$                 & $12578(5)  $                              & $12576(3)   $           & $12577(3)   $     & $884(3) $                          \\
 $10161(3)    $               & $10163.2(19)$              & $12669.9(4)$                              & \textit{12670}\tnote{*} & $12669.9(4)  $    & $977.2(4)$                         \\
 $10333.29(13)$               & $10328.1(18)$              & $12738.9(7)$                              & $12738(3)$              & $12738.8(7)  $    & $1046.1(7)$                        \\
 $10360.51(13)$               & \textit{10358}\tnote{*}    & $12817.77(19)$                            & $12819(4)$              & $12817.77(19)$    & $1125.08(19)$                      \\
 $10576.02(7) $               & $10572.7(21)$              & $12852.1(5)$                              & $12854(3)$              & $12852.2(5)  $    & $1159.5(5) $                       \\
 $10659.58(13)$               & $10660.1(21)$              & $12921.6(4)$                              & $12924(4)$              & $12921.6(4)  $    & $1228.9(4) $                       \\
 $10660.03(4) $               &                            & $12963.9(5)$                              & $12965(4)$              & $12963.9(5)  $    & $ 1271.2(5) $                      \\
 $10711.74(17)$               & $10713.9(12)$              &                                           &                         &                   &                                    \\
 $10730.79(11)$               & $10732.5(16)$              &                                           &                         &                   &                                    \\
 $10820.7(4)  $               & $10824.3(13)$              &                                           &                         &                   &                                    \\
 $10916.96(17)$               & \textit{10918}\tnote{*}    &                                           &                         &                   &                                    \\
 $11010.5(14) $               & $11011(3)$                 &                                           &                         &                   &                                    \\
 $11015.8(7)  $               &                            &                                           &                         &                   &                                    \\
 $11208.4(16) $               & $11201(5)$                 &                                           &                         &                   &                                    \\
 $11314.4(15) $               & $11317(3)$                 &                                           &                         &                   &                                    \\
    \bottomrule
    \bottomrule
  \end{tabular}
  \begin{tablenotes}
    \item[*] State used for calibration.
    \item[$\dagger$] These two states show bimodal behavior as a function of angle. The \textit{expected value method} of Ref. \cite{BIRCH_2014} was adopted to give a more conservative uncertainty since only one state is expected around each energy. See text for additional details.     
  \end{tablenotes}
  \end{threeparttable}
\end{table*}

\subsection{Suggested Energies for Astrophysically Relevant States }
\label{sec:recommended-energies}

Our angle averaged excitation energies have been combined with our compilation of literature values (Sec.~\ref{sec:energy_level_update}), to produce the recommended resonance energies given in the second half of Table \ref{tab:comp_and_recommended}. 
The energies of Ref.~\cite{hale_2004} have been excluded from the averaging (see Appendix.A). Note that some states not directly measured in the present work are included since they play a role in the reaction rate. All values come from a weighted average of the separate measurements, except for the $11694$-keV state. For this state an extreme tension {of 10 keV} exists between the two most precise measurements, which are this work and the value of Ref.~\cite{schmalbrock_1983}. In order for our recommended value to reflect this disagreement, we again adopt the \textit{expected value method} of Ref.~\cite{BIRCH_2014} to combine our measurement and the measurement of Ref.~\cite{schmalbrock_1983} leading to a more realistic uncertainty given the discrepant data.

\section{Bayesian DWBA Analysis}
\label{sec:bayes-dwba-analys}
Proton partial widths necessary for the calculation of the reaction rate can be estimated from the spectroscopic factors extracted from single-particle transfer reactions. Uncertainties arising from the optical potential and bound state wave function will typically dominate the total uncertainties of the spectroscopic factors. Analysis of our data would be incomplete if we ignored these sources of uncertainty; therefore, we adopt the Bayesian {distorted wave Born approximation (DWBA)} methods of Ref.~\cite{Marshall_2020} to quantify these uncertainties for the present measurement. All DWBA calculations were carried out using \texttt{FRESCO} \cite{fresco}. Ref.~\cite{Marshall_2020} should be consulted for a more complete discussion of the Bayesian DWBA method, but a brief overview is given here in the context of the present study.

\subsection{Overview of Bayesian DWBA}
\label{sec:overv-bayes-dwba}
Elastic scattering data are used to constrain the parameters of a Woods-Saxon potential given by:
\begin{equation}
  \label{eq:woods_saxon}
  V(r) = -\frac{V}{1+\exp{(\frac{r-r_0A_t^{1/3}}{a_0}})},
\end{equation}
where $V$ is the depth of the well in MeV, $r_0$ is the radius in fm, and $a_0$ is the diffuseness in fm.
The optical model uses a linear combination of both real and imaginary Woods-Saxon potentials, and by adjusting the parameters of these potentials the observed elastic scattering data can be reproduced. Bayesian statistics treats parameters as probability distributions. By assigning each parameter a \textit{prior} probability distribution before considering the data, Bayesian statistics allows the data, $\mathbf{D}$, and Bayes' theorem to update the prior distributions in light of our observations. Bayes' theorem is given by: 
\begin{equation}
  \label{eq:bayes_theorem}
  P(\boldsymbol{\theta}|\mathbf{D}) = \frac{P(\mathbf{D}|\boldsymbol{\theta}) P(\boldsymbol{\theta})}
  {P(\mathbf{D})},
\end{equation}
where $P(\boldsymbol{\theta})$ are the prior probability distributions of the model parameters, 
$P(\mathbf{D}|\boldsymbol{\theta})$ is the likelihood function, $P(\mathbf{D})$ is the evidence, and $P(\boldsymbol{\theta}|\mathbf{D})$ is the posterior \cite{bayes}. Informally we can state: the priors are what we believe about the model parameters considering the new data, the likelihood is the probability of measuring the observed data given a set of model parameters, the evidence is the probability of the observed data, and the posterior is what we know about the model parameters after analyzing the new data.

The goal of the present experiment is to extract spectroscopic factors and assign $\ell$ values to states in $^{24}$Mg in the astrophysical region of interest. Spectroscopic factors are extracted from experimental angular distributions, $\frac{d \sigma}{d \Omega}_{\textnormal{Exp}}$,  according to:
\begin{equation}
  \label{eq:spec_factor}
  \frac{d \sigma}{d \Omega}_{\textnormal{Exp}} = C^2S_{\textnormal{proj}} C^2S_{\textnormal{targ}} \frac{d \sigma}{d \Omega}_{\textnormal{DWBA}},
\end{equation}
where $C$ and $S$ denotes the isospin Clebsch-Gordan coefficient and spectroscopic factor, while the subscripts $\textnormal{proj}$ and $\textnormal{targ}$ refer to the projectile and target systems, respectively. For this work, we approximate $C^2S$ for the projectile system as $\frac{3}{2}$ according to Ref.~\cite{satchler}. Any further mention of $C^2S$ should be understood to be in reference to $C^2S_{\textnormal{targ}}$. 

It is essential to recognize that Eq.~(\ref{eq:spec_factor}) establishes $C^2S$ as a parameter in the framework of DWBA. The only meaningful way to estimate its uncertainty in the presence of both the measured uncertainties of $\frac{d \sigma}{d \Omega}_{\textnormal{Exp}}$ and the optical model uncertainties that affect $\frac{d \sigma}{d \Omega}_{\textnormal{DWBA}}$ is to treat it as a parameter in the statistical analysis. Using Bayesian statistics this entails assigning a prior distribution. The excited states of interest to this work lie above $11$ MeV, where it can be safely assumed that the majority of the single particle strength of the proton shells has been exhausted. Thus, $C^2S \ll 1$ and we assign an informative prior:
\begin{equation}
  \label{eq:spec_factor_prior}
  C^2S \sim \textnormal{HalfNorm}(1.0^2),
\end{equation}
where $\sim$ means ``distributed according to''. $\textnormal{HalfNorm}$ stands for the half normal distribution, which is strictly positive and has one free parameter the standard deviation, $\sigma$. In the case of Eq.~(\ref{eq:spec_factor_prior}), $\sigma = 1.0$ is chosen to reflect our assumption that $C^2S$ is more than likely to be less than one in the astrophysical region of interest. 

Assigning probabilities to $\ell$ values requires a subcategory of Bayesian inference called model selection. In this context, the model is $\ell_l$, which is shorthand for $\ell = l$ (for example $\ell = 0$ is written $\ell_0$). Posterior distributions for $\ell_l$ can be determined through a modified version of Bayes' theorem:
\begin{equation}
  \label{eq:m_theorem}
  P(\ell_l|\mathbf{D}) = \frac{P(\mathbf{D}|\ell_l) P(\ell_l)}
  {\sum_k P(\mathbf{D}|M_k)P(M_k)}.
\end{equation}
Each $\ell_l$ is implicitly dependent on a set of model parameters $\boldsymbol{\theta}_l$ which have been marginalized. Expanding $P(\mathbf{D}|\ell_l)$ to show the explicit dependence gives:
\begin{equation}
  \label{eq:marg}
  P(\mathbf{D}|\ell_l) = \int P(\mathbf{D}|\ell_l, \boldsymbol{\theta}_l) P(\boldsymbol{\theta}_l|\ell_l) d\boldsymbol{\theta}_l.
\end{equation}
This equation shows that $P(\mathbf{D}|\ell_l)$ is precisely equivalent to  
the evidence integral from Eq.~(\ref{eq:bayes_theorem}) conditioned on $\ell_l$. Thus, {calculating} the posteriors for each $\ell$ demands evaluating
the evidence for each DWBA cross section generated using a distinct $\ell$ value.  

Denoting the evidence integral that corresponds to a model $\ell_l$ as $Z_l$, we can compare each value of $\ell$.
The Bayes Factor, {$B_{lk}$,} can be calculated between two angular momentum transfers which are assumed to have equal prior probabilities:
\begin{equation}
  \label{eq:bayes_factor}
  B_{lk} = \frac{Z_l}{Z_k}.
\end{equation}
Generally, if this ratio is greater than one, the data favor the transfer $\ell = l$, while values less than $1$
favor $\ell = k$. While the significance of values for $B_{lk}$ is open to interpretation, a useful
heuristic given by Jefferys \cite{Jeffreys61} is often adopted. Assuming $\ell = l$ is favored over $\ell = k$, we have the following
levels of evidence: $1 < B_{lk} < 3$ is anecdotal, $3 < B_{lk} < 10$ is substantial, $10 < B_{lk} < 30$ is strong, $30 < B_{lk} < 100$ is very strong, and $ B_{lk} > 100$ is decisive. Normalized probabilities for each transfer are given by:
\begin{equation}
  \label{eq:normalized_probabilies}
  P(\ell_l|\mathbf{D}) = \frac{Z_l}{\sum_k Z_k},
\end{equation}
where the index $k$ runs over all allowed angular momentum values. However, practically $(^{3}\textnormal{He}, d)$ reactions are highly selective, allowing us to restrict the sum to the most likely transfers with $\ell = 0 \text{-} 3$.   

By using a Bayesian model, Ref.~\cite{Marshall_2020} made it possible to incorporate optical potential uncertainties into the extraction of spectroscopic factors and assignment of $\ell$ values. However, the current data set for $^{23}$Na$(^{3}\textnormal{He}, d)$ presents challenges that require significant extensions to those previously reported methods. 

\subsection{Incorporating Relative Yields}
\label{sec:relative-yields}
Extraction of $C^2S$ for a state requires that the absolute scale of the differential cross section is known. Here we use a relative method to remove beam and target effects. Yields measured at the focal plane are normalized to the $^{23}$Na$+^{3}$He elastic scattering measured by the monitor detector positioned at $45^{\circ}$. From these normalized yields, an absolute scale is established by inferring an overall normalization  through comparison of the measured elastic scattering angular distribution collected in the focal plane to the optical model predictions. Our approach is similar in principle to those found in Refs.~\cite{hale_2001, hale_2004, vernotte_optical}.     

The present study has a set of ten elastic scattering data points, which we denote by $\frac{d Y}{d \Omega}_{\textnormal{Elastic}, j}$ for the data measured at angle $j$. From these data a posterior distribution can be found for an overall normalization parameter, $\eta$, which renormalizes the predictions of the optical model such that:
\begin{equation}
  \label{eq:optical_to_yield}
  \frac{d Y}{d \Omega}_{\textnormal{Optical}, j} = \eta \times \frac{d \sigma}{d \Omega}_{\textnormal{Optical}, j},
\end{equation}
where $\frac{d Y}{d \Omega}_{\textnormal{Optical}, j}$ is the relative yield predicted by the optical model at angle $j$. As a parameter in our model, $\eta$ needs a prior distribution. To assign equal probability on the logarithmic scale, we introduce a parameter, $g$, such that:
\begin{equation}
  \label{eq:norm_prior}
  g \sim \textnormal{Uniform}(-10, 10),
\end{equation}
where $\textnormal{Uniform}$ is the uniform distribution. $\eta$ is then defined via $\eta = 10^g$. Since $\eta$ is estimated simultaneously with $C^2S$, the uncertainty in our absolute normalization will automatically be included in the uncertainty of $C^2S$.     

\subsection{Global Potential Selection}
\label{sec:glob-potent-select}

Global optical potentials are used to construct the prior distributions for the potential parameters in our Bayesian model. Elastic scattering data were only measured for the entrance channel, since it can be gathered with the same beam energy and target as the transfer reaction of interest. As a result, our priors differ for the entrance and exit channels. For the entrance channel, mildly informative priors are selected. The depths, $V, W,$ etc., are assigned Normal distributions centered around their global values with standard deviations equal to $20 \%$ of the central value:
\begin{equation}
  \label{eq:prior_depths}
  V \sim \mathcal{N}(\mu_{global}, \{ 0.20 \, \mu_{global} \}^2).
\end{equation}
The geometric parameters, $r$ and $a$, are given priors that attempt to cover their expected physical range while still allowing the posterior to be determined by the data. Taking this range to be $r = 1.0 - 1.5$ fm and $a = 0.52 - 0.78$ fm, we can again assign Normal distributions with central values $r = 1.25$ fm and $a = 0.65$ and standard deviations of $20 \%$ the central value. Collecting the parameters for all of the potentials, the priors for the entrance channel are written compactly as:
\begin{equation}
  \label{eq:entrance_priors}
  \boldsymbol{\mathcal{U}}_{\textnormal{Entrance}} \sim \mathcal{N}(\mu_{\textnormal{central}, k}, \{0.20 \, \mu_{\textnormal{central}, k}\}^2),
\end{equation}
where the index $k$ runs over each of the potential parameters. The subscript $central$ refers to either the values taken from the selected global study for each depth or the central values $r = 1.25$ fm and $a = 0.65$ fm for the geometric parameters.

The first attempts to fit the elastic scattering data used the optical model from the lab report of Beccehetti and Greenless \cite{b_g_3he}. The imaginary depth of this potential for a beam of $^3$He on $^{23}$Na at $E_{^{3}\textnormal{He}}=21$ MeV is $36$ MeV. We note that this value is nearly twice as deep as the values reported in the more recent works of Trost \textit{et al.} \cite{TROST_1980}, Pang \textit{et al.} \cite{pang_global} and Liang \textit{et al.} \cite{Liang_2009}. Although these works use a surface potential, the work of Vernotte \textit{et al.} \cite{vernotte_optical} is parameterized by a volume depth, and also favors depths around $20$ MeV. While the starting parameters are of little consequence to standard minimization techniques, the overly deep well depth is an issue for our Bayesian analysis because it determines the prior distribution for our model. When using the deeper value {of 36 MeV} for inference, we observed that the data preferred a lower depth, thereby causing a bimodal posterior with one mode centered around the global depth and the other resulting from the influence of the data. Based on these observations, a decision was made to use the potential of Liang \textit{et al.} (Ref.~\cite{Liang_2009}) due to its applicability in the present mass and energy range {and its shallower imaginary depth of 19.87 MeV}. We have chosen to exclude the imaginary spin-orbit portion of the Liang potential because of the limited evidence presented for its inclusion in Ref.~\cite{Liang_2009}. 

The exit channel optical potential parameters must also be assigned prior distributions. Our experiment does not have data to constrain these parameters directly, but fixing these parameters in our analysis would neglect a source of uncertainty. We chose informative priors that are determined by the selected global deuteron potential. These parameters are assigned Normal priors centered around the global values and given standard deviations of $10 \%$:
\begin{equation}
  \label{eq:exit_priors}
  \boldsymbol{\mathcal{U}}_{\textnormal{Exit}} \sim \mathcal{N}(\mu_{\textnormal{global}, k}, \{0.10 \, \mu_{\textnormal{global}, k}\}^2).
\end{equation}
The selected deuteron potential is the non-relativistic \say{L} potential from Ref.~\cite{daehnick_global}. Since the region of interest is $11$-$12$ MeV, the outgoing deuterons will have an energy of $E_d \approx E_{^3\textnormal{He}} + Q_{(^3\textnormal{He}, d)} - E_x \approx 15.5$ MeV.

All of the potentials used in the following analysis are listed in Table~\ref{tab:opt_parms_na}. The bound state spin-orbit term was set to roughly satisfy $\lambda = 25$ with $\lambda \approx 180 V_{so} / V$ for values of $V$ in the above energy range. The bound state geometric parameters, all spin-orbit terms for the entrance and exit channels, and Coulomb radii were fixed in our calculations.

\begin{table*}[ht]
\centering
\begin{threeparttable}[e]
\caption{\label{tab:opt_parms_na}Optical potential parameters used in this work before inference.}
\setlength{\tabcolsep}{4pt} %adds space between columns in the table
\begin{tabular}{ccccccccccccccccc}
\toprule[1.0pt]\addlinespace[0.3mm] Interaction  & $V$ & $r_{0}$ & $a_{0}$ & $W$ & $W_{s}$ & $r_{i}$ & $a_{i}$ & $r_{c}$ & $V_{so}$ & $r_{so}$ & $a_{so}$ \\
                                                 & (MeV) & (fm) & (fm) & (MeV) & (MeV) & (fm) & (fm) & (fm) & (MeV) & (fm) & (fm)\\ \hline\hline\addlinespace[0.6mm]

\hspace{-0.1cm} $^{3}$He $+ ^{23}$Na \footnote{Global potential of Ref.~\cite{Liang_2009}.} & $117.31$ & $1.18$ & $0.67$ & & $19.87$ &$1.20$ & $0.65$ & $1.29$ & $2.08$ & $0.74$ & $0.78$ \\
$d$ $+$ $^{24}$Mg\footnote{Global potential of Ref.~\cite{daehnick_global}.} & $88.1$ & $1.17$ & $0.74$ & $0.30$  & $12.30$ & $1.32$  &  $0.73$ & $1.30$ & $6.88$ & $1.07$ & $0.66$ & \\
\hspace{0.0cm}$p$ $+$ $^{23}$Na & \footnote{Adjusted to reproduce binding energy of the final state.} & $1.25$ & $0.65$ & & & & & $1.25$ & $6.24$ & $1.25$ & $0.65$  \\[0.2ex]
\bottomrule[1.0pt]
\end{tabular}
\end{threeparttable}
\end{table*}

\subsection{Elastic Scattering}
\label{sec:elastic-scattering}

As stated in Sec.~\ref{sec:experiment-details}, elastic scattering yields were measured for $15^{\circ} \text{-} 55^{\circ}$ in $5^{\circ}$ steps and finally at $59^{\circ}$, for a total of $10$ angles. The yields at each angle were normalized to those measured by the monitor detector. A further normalization to the Rutherford cross section was applied to the elastic scattering data to ease the comparison to the optical model calculations.   

Low angle elastic scattering cross sections in normal kinematics can be collected to almost arbitrary statistical precision, with the present data having statistical uncertainty
of approximately $2 \text{-} 7 \%$. In this case, it is likely that the residuals between these data and the optical model predictions are dominated by theoretical and experimental systematic uncertainties. To account for this possibility, the Bayesian model is modified to consider an additional unobserved uncertainty in the elastic channel:
\begin{equation}
  \label{eq:elastic_unc}
  \sigma_{\textnormal{Elastic}, i}^{\prime 2} = \sigma_{\textnormal{Elastic}, i}^2 + \bigg(f_{\textnormal{Elastic}} \frac{d \sigma}{d \Omega}_{\textnormal{Optical}, i} \bigg)^2,
\end{equation}
where the experimentally measured uncertainties, $\sigma_{\textnormal{Elastic}, i}$, at angle $i$ have been added in quadrature with an additional uncertainty coming from the predicted optical model cross section. This prescription is precisely the same procedure that is used for the additional transfer cross section uncertainty from Ref.~\cite{Marshall_2020}. With only $10$ data points, an informative prior on $f_{\textnormal{Elastic}}$ is necessary to preserve the predictive power of these data. We select the form:
\begin{equation}
  \label{eq:elastic_f}
  f_{\textnormal{Elastic}} \sim \textnormal{HalfNorm}(0.10^2).
\end{equation}
This quantifies the expectation that the data will have residuals with the theoretical prediction of about $10 \%$. We found the above prior to provide the best compromise between the experimental uncertainties, which lead to unphysical optical model parameters, and less informative priors that lead to solutions above $f_{\textnormal{Elastic}} = 50 \%$ where the data become non-predictive.  

\begin{figure*}
    \centering
    \includegraphics[width=\textwidth]{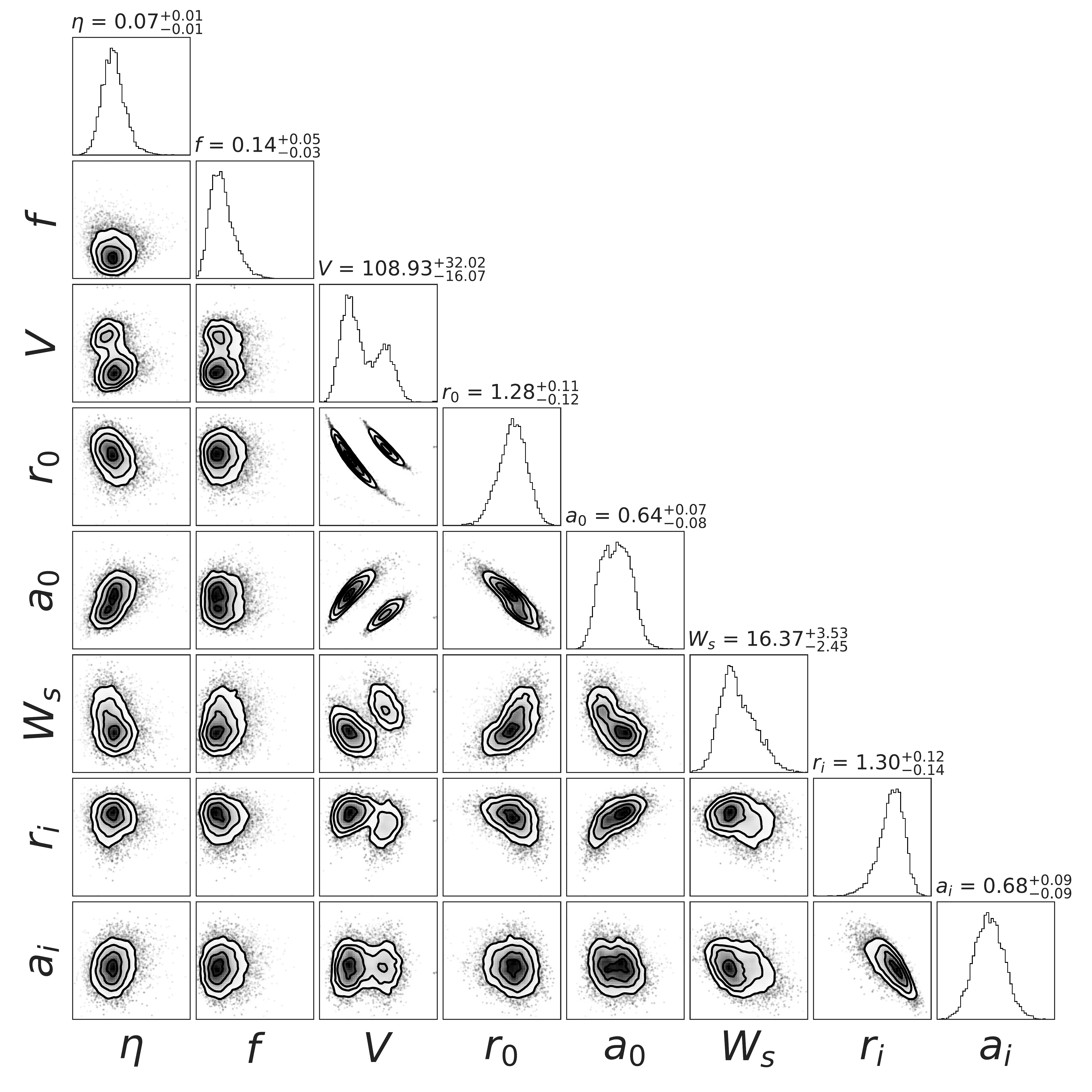}
    \caption{Pair correlation plot of the posterior samples for the nested sampling run. The discrete ambiguity is prominent in the $^{3}$He $+ ^{23}$Na data, posing a significant challenge in estimating the optical model parameter posteriors.}
    \label{fig:corner_discrete}
\end{figure*}

Once the above parameter was included, the data could be reliably fit. However, it then became clear that the discrete ambiguity posed a serious issue for the analysis. It is know (see for example Ref.~\cite{drisko_1963}) that nearly identical theoretical cross sections can be produced with drastically different potential depths due to the phase shift only differing by an additive multiple of $\pi$. Previously, Ref.~\cite{Marshall_2020} found that the biasing of the entrance channel potential priors towards their expected physical values was sufficient to remove other modes from the posterior. For the present data, the potential priors did little to alleviate the problem, as might be expected since strongly absorbed projectiles like $^{3, 4}$He suffer much worse discrete ambiguities (\cite{drisko_1963}) compared to the deuteron scattering data in Ref.~\cite{Marshall_2020}. In order to explore potential solutions, the nested sampling algorithm in \texttt{dynesty} was used to draw appropriately weighted samples from both of the modes. Nested sampling can explore multi-modal distributions with ease \cite{speagle2019dynesty}, but is not necessarily suited towards precise posterior estimation. A run was carried out with $1000$ live points, and required over $5 \times 10^6$ likelihood calls, which is nearly three times the number of samples required in other calculations. The pair correlation plot of these samples is shown in Fig.~\ref{fig:corner_discrete}, and the impacts of the discrete ambiguity can clearly be seen. 

\begin{figure}
    \centering
    \includegraphics[width=.45\textwidth]{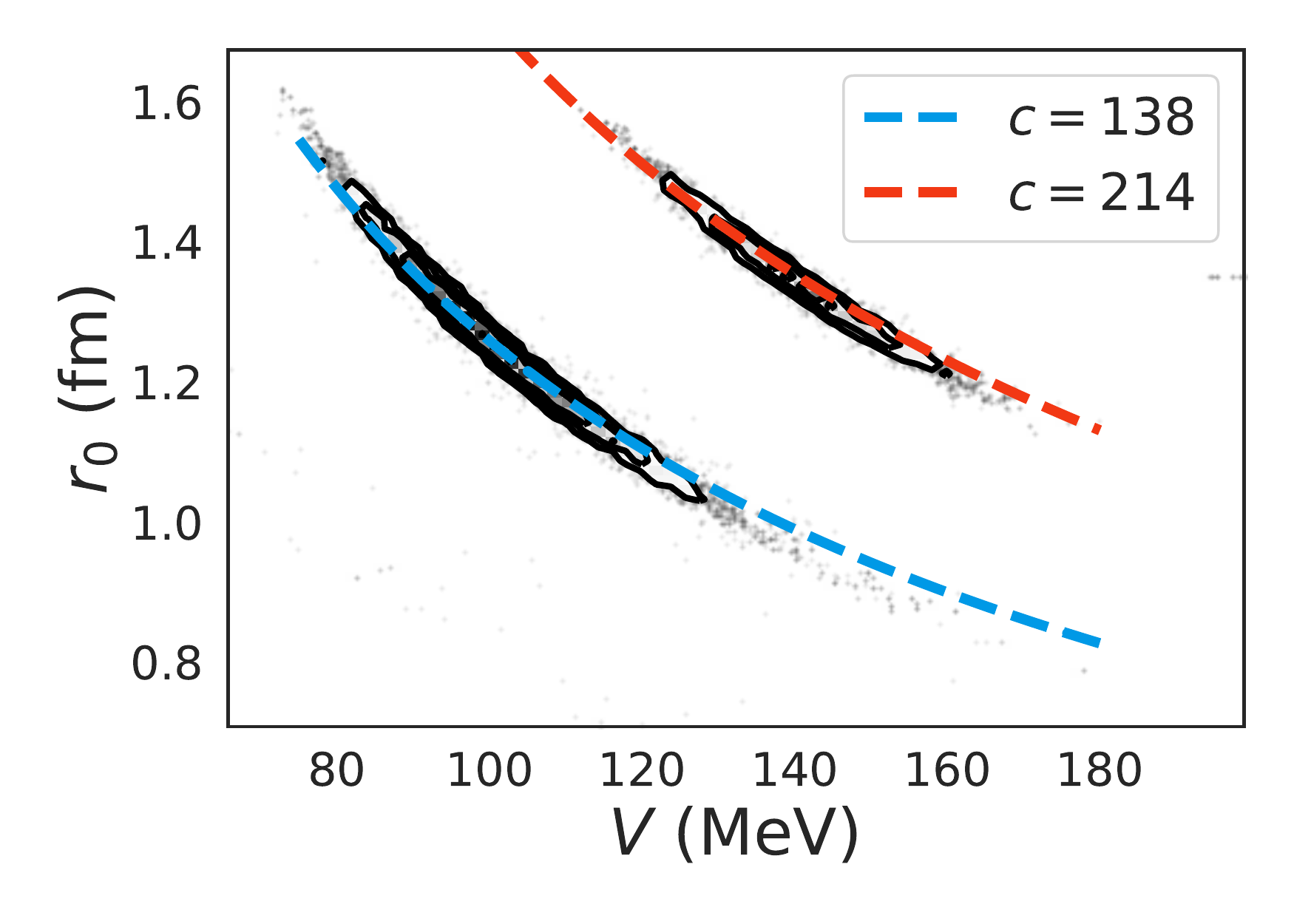}
    \caption{The discrete ambiguity as seen in the $V$ versus $r_0$ correlations between the {histogrammed} samples of the nested sampling calculation {shown in black}. The colored lines show the description of the correlation based on the analytic form $Vr_0^n = c$. The value of $c$ provides a way to distinguish these modes.  }
    \label{fig:vr_constant_samples}
\end{figure}

\begin{figure}
    \centering
    \includegraphics[width=.45\textwidth]{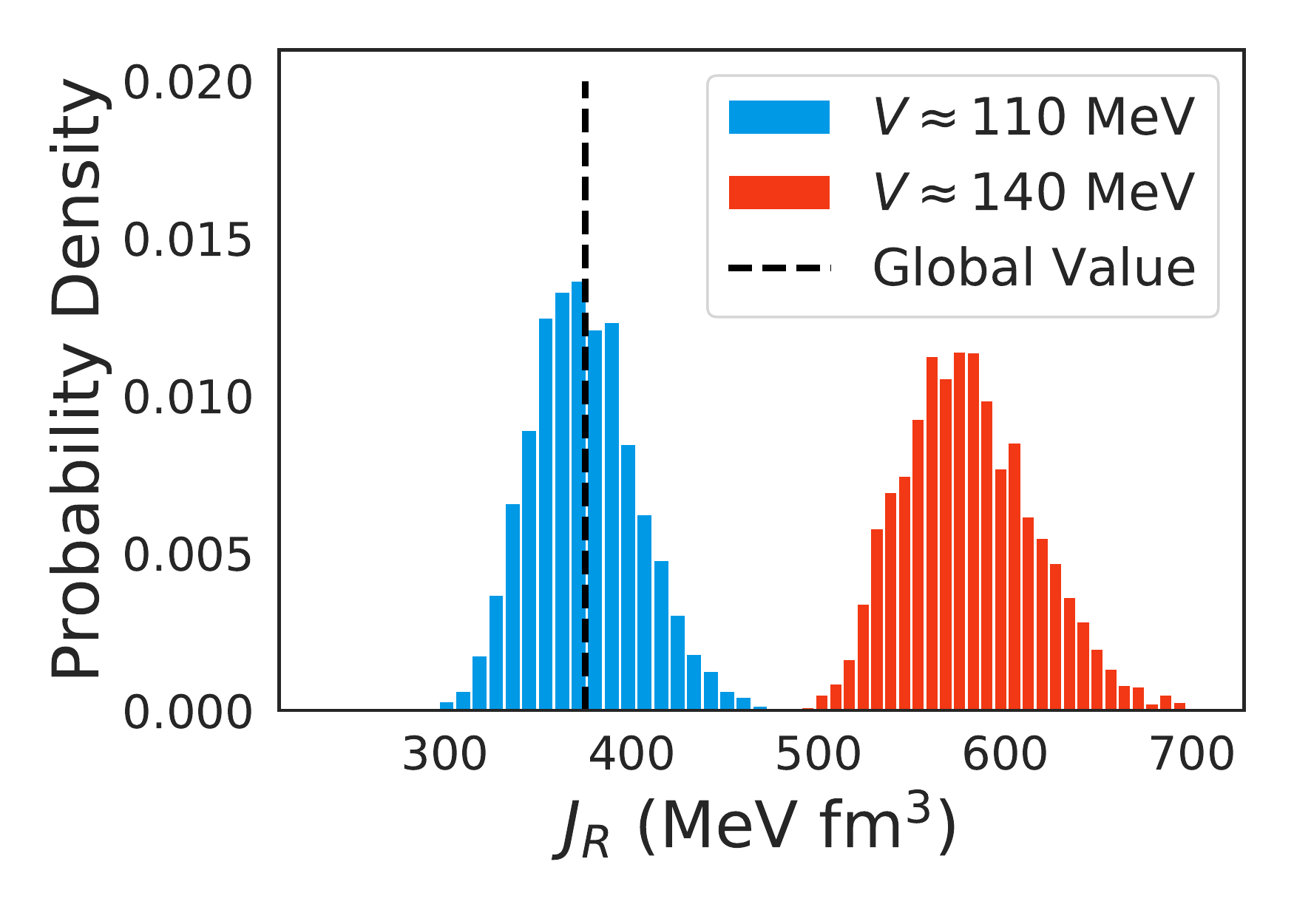}
    \caption{Values from the volume integral of the real potential as calculated using Eq.~(\ref{eq:j_analytical}) and the samples from the nested sampling calculation. The discrete ambiguity causes two well separated peaks to appear. {The black dashed line shows the value of $J_R$ for the global potential.}}
    \label{fig:j_r_hist}
\end{figure}

Two different approaches were explored to differentiate the modes. The first was a simple selection of the modes based on the continuous ambiguity,  $Vr_0^n = c$. Fig.~\ref{fig:corner_discrete} shows that the correlation between $V$ and $r_0$ can cleanly resolve the two modes, while the correlations in the other parameters have significant overlap between them. In this approach, the constant, $c$, is calculated for each mode, while the exponent is kept fixed with a value of $n=1.14$, taken from Ref.~\cite{vernotte_optical}. It was found that the correlation in the samples was well described by this relation as shown in Fig.~\ref{fig:vr_constant_samples}. Our second approach utilized the volume integral of the real potential. Ref.~\cite{varner} gives an approximate analytical form of the integral:
\begin{equation}
 \label{eq:j_analytical}
   J_R = \frac{1}{A_P A_T} \frac{4 \pi}{3} R_0^3 V \bigg[ 1 + \bigg( \frac{\pi a_0}{R_0}\bigg)^2 \bigg],
\end{equation}
where $R_0 = r_0 A_T^{1/3}$. Calculating $J_R$ for the samples in each mode resulted in two clearly resolved peaks, as shown in Fig.~\ref{fig:j_r_hist}. 

After comparing these two methods, it was decided to use the first one, and exclude the other modes via a uniform distribution based on the relationship between $V$ and $r_0$. This calculation had the advantages of being relatively simple and only involving two parameters. The method based on $J_R$ has the advantage that the global values well predict the location of the peak, but its dependence on $a_0$ makes its possible effect on the posterior less clear. Integrating the $Vr_0^n$ relation into the Bayesian method requires a probability distribution be specified. A uniform distribution that covered $\pm 30 \%$ around $c$ of the physical mode was chosen. We have intentionally avoided the word prior because this condition clearly does not represent a belief about the parameter $c$ before inference. Rather, this is a \textit{constraint} enforced on the posterior to limit the inference to the physical mode \cite{Wu_2019}. It should be emphasized that the posterior distributions of all the parameters will be conditioned on $c$, i.e., $P(\theta|D, c)$. The constraint is written:
\begin{equation}
    \label{eq:c_constraint}
    c \sim \textnormal{Uniform}(c_0(1-0.30), c_0(1+0.30)),
\end{equation}
where $c_0$ is the value that is roughly centered around the lower mode. In this case $c_0 = 132.9$. As long as the distribution in Eq.~(\ref{eq:c_constraint}) covers all of the physical mode and excludes the unphysical ones, the value of $c_0$ and the width of the distribution should be understood to be arbitrary.

\subsection{Transfer Considerations}

Transfer cross sections are calculated using the zero-range approximation with the code \texttt{FRESCO}. The zero-range approximation is necessary in the current context because of the number of function evaluations that are needed to compute the posterior
distributions (for this work $2 \times 10^6$). For the volume integral of the proton-deuteron interaction, $D_0$, we use a value of $D_0 = -172.8$ MeV fm$^{3/2}$ \cite{bassel}. $D_0$ is calculated theoretically and has a dependence on the selected nucleon-nucleon interaction. We added a $15 \%$ uncertainty using a parameter $\delta D_0$ to account for the spread observed between different {theoretical} models in Refs.~\cite{all_norms_3He}. {A similar estimate for the $D_0$ uncertainty was made in Ref.~\cite{bertone}.}

The residuals between the transfer cross section and DWBA calculations will be impacted not only by the experimental and optical model uncertainties, but by any deficiency in the reaction theory. If we do not acknowledge that the DWBA residuals could be greater than the uncertainties coming from counting statistics, then we would be assuming that the transfer data are a meaningful constraint on the optical model parameters. If this were the case, each state would have its own set of optical model parameters that have been incorrectly adjusted to best reproduce the observed angular distribution. To avoid this issue, we add an additional theoretical uncertainty in quadrature with the experimental uncertainties, similar to our procedure for the elastic scattering. Using the same functional form as Eq.~(\ref{eq:elastic_unc}), we define a fraction of the DWBA cross section, with the weakly informative prior:
\begin{equation}
  \label{eq:f_transfer}
  f \sim \textnormal{HalfNorm}(1.0^2),
\end{equation}
meaning that our expectation for the fractional uncertainty on the DWBA cross section at each angle is $f < 1$.  

A majority of the states of astrophysical interest lie above the proton threshold, and are therefore unbound. For bound states, calculation of the overlap functions, which determine $C^2S$, is done by using a single particle potential with its Woods-Saxon depth adjusted to reproduce the binding energy of the state. For unbound states, an analogous procedure would be to adjust the well depth to produce a resonance centered around $E_r$. \texttt{FRESCO} does not currently support a search routine to vary $V$ to create a resonance condition, meaning that $V$ would have to be varied by hand until a phase shift of $\pi/2$ is observed. Such a calculation is obviously time consuming and computationally infeasible in the current work. An alternative is the weak binding approximation. This approach assumes that the wave function of resonance scattering resembles the wave function of a loosely bound particle, typically with a binding energy on the order of $E_{bind} = 1$ keV. Studies have shown that this approximation performs well for states within $\approx 500$ keV of the particle threshold, and reproduce the unbound calculations to within $1 \%$ \cite{Kankainen_2016, KAHL_2019}. There are indications that the validity of this approximation depends on the $\ell$ value. The reasoning is that states with higher $\ell$ values more closely resemble bound states, due to the influence of the centrifugal barrier, and therefore are better described by the approximation \cite{Poxon_Pearson_2020}. For this work, DWBA calculations for states above the proton threshold were carried out with the weak binding approximation. The error arising from use of the approximation is considered negligible in the current context.            

Further complications arise from the non-zero ground state of $^{23}$Na ($J^{\pi} = 3/2^+$). In this case, angular distributions can be characterized by a mixture of $\ell$ transitions. Although in principle every allowed $\ell$ transition can contribute, practically speaking, it is difficult to unambiguously determine all but the lowest two $\ell$ contributions because of the rapidly decreasing cross section with increasing $\ell$ \cite{hodgson1971}.
Ignoring the light particle spectroscopic factor, the relationship between the experimentally measured differential cross section and the DWBA prediction can be expressed as: 
\begin{equation}
  \label{eq:mixed_l}
  \frac{d \sigma}{d \Omega}_{exp} = C^2S \bigg[ \alpha \frac{d \sigma}{d \Omega}_{\textnormal{DWBA}, \ell_1} + (1-\alpha) \frac{d \sigma}{d \Omega}_{\textnormal{DWBA}, \ell_2}   \bigg],
\end{equation}
where $\alpha$ is defined such that $C^2S_{\ell_1} = C^2S \alpha$ 
and $C^2S_{\ell_2} = C^2S (1 - \alpha)$ \cite{vernotte_cs}. Note that the values for $\ell$ must still obey parity conservation, meaning the most probable combinations for $(^3$He$, d)$ are $\ell = 0 \oplus 2$ and $\ell = 1 \oplus 3$. Incorporating multiple $\ell$ transfers into the Bayesian framework requires assigning a prior to $\alpha$. The above definitions make it clear that $\alpha = [0, 1]$; therefore, an obvious choice is:
\begin{equation}
    \label{eq:alpha_prior}
    \alpha \sim \textnormal{Uniform}(0, 1).
\end{equation}

\subsection{Bayesian Model for $^{23}$N\lowercase{a}$(^{3}\textnormal{He}, d)^{24}$M\lowercase{g}}
\label{sec:spec_factors}

Before explicitly defining the Bayesian model for the DWBA analysis, the points made above are reiterated for clarity. 

\begin{enumerate}
    \item The measured elastic scattering uncertainties have been added in quadrature with an inferred theoretical uncertainty.
    \item The $^3$He optical model has a severe discrete ambiguity. A constraint based on the continuous ambiguity has been added to the model to select the physical mode.
    \item Due to the non-zero spin of the ground state of $^{23}$Na, the transfer cross section can have contributions from multiple $\ell$ values.
    \item Only the two lowest $\ell$ values are considered for a mixed transition, with the relative contributions weighted according to a parameter $\alpha$ that is uniformly distributed from $0$ to $1$.
\end{enumerate}

Folding these additional parameters and considerations into the Bayesian model of Ref.~\cite{Marshall_2020} gives:
\begin{align}
  \label{eq:dwba_model_na}
 & \textnormal{Parameters:} \nonumber \\
 & n = 1.14 \nonumber \\
 & c_0 = 132.9 \nonumber \\
 & \textnormal{Priors:} \nonumber \\
 & \boldsymbol{\mathcal{U}}_{\textnormal{Entrance}} \sim \mathcal{N}(\mu_{\textnormal{central}, k}, \{0.20 \, \mu_{\textnormal{central}, k}\}^2) \nonumber \\
 & \boldsymbol{\mathcal{U}}_{\textnormal{Exit}} \sim \mathcal{N}(\mu_{\textnormal{global}, k}, \{0.10 \, \mu_{\textnormal{global}, k}\}^2) \nonumber \\
 & f \sim \textnormal{HalfNorm}(1) \nonumber \\
 & f_{\textnormal{Elastic}} \sim \textnormal{HalfNorm}(0.10^2) \nonumber \\
 & \delta D_0^2 \sim \mathcal{N}(1.0, 0.15^2) \nonumber \\
 & C^2S \sim \textnormal{HalfNorm}(1.0^2) \nonumber \\
 & g \sim \textnormal{Uniform}(-10, 10) \nonumber \\
 & \textnormal{Functions:} \\
 & \eta = 10^{g}  \nonumber \\
 & c = \mathcal{U}_{\textnormal{Entrance}, (k=0)} \big(\mathcal{U}_{\textnormal{Entrance}, (k=1)}\big)^n \nonumber \\
 & \frac{d Y}{d \Omega}^{\prime}_{\textnormal{Optical}, j} = \eta \times \frac{d \sigma}{d \Omega}_{\textnormal{Optical}, j} \nonumber \\
 & \frac{d Y}{d \Omega}^{\prime}_{\textnormal{DWBA}, i} = \eta \times \delta D_0^2 \times C^2S \times \frac{d \sigma}{d \Omega}_{\textnormal{DWBA}, i} \nonumber \\
 & \sigma_i^{\prime 2} = \sigma_{\textnormal{Transfer}, i}^2 +  \bigg(f\frac{d Y}{d \Omega}^{\prime}_{\textnormal{DWBA}, i}\bigg)^2 \nonumber \\
 & \sigma_{\textnormal{Elastic}, i}^{\prime 2} = \sigma_{\textnormal{Elastic}, i}^2 + \bigg(f_{\textnormal{Elastic}} \frac{d Y}{d \Omega}_{\textnormal{Optical}, i} \bigg)^2 \nonumber \\
 & \textnormal{Likelihoods:} \nonumber \\
 & \frac{d Y}{d \Omega}_{\textnormal{Transfer}, i} \sim \mathcal{N}\bigg(\frac{d Y}{d \Omega}^{\prime}_{\textnormal{DWBA}, i}, \sigma_i^{\prime \, 2} \bigg) ,  \nonumber \\
 & \frac{d Y}{d \Omega}_{\textnormal{Elastic}, j} \sim \mathcal{N}\bigg(\frac{d Y}{d \Omega}^{\prime}_{\textnormal{Optical}, j}, \sigma_{\textnormal{Elastic}, i}^{\prime 2} \bigg) ,  \nonumber \\
 & \textnormal{Constraint:} \nonumber \\
 & c \sim \textnormal{Uniform}(c_0(1-0.30), c_0(1+0.30)), \nonumber
\end{align}
where the index $k$ runs over the optical model potential parameters, $i$ and $j$ denote the elastic scattering and transfer cross section angles, respectively, and $\mathcal{U}_{\textnormal{Entrance}, (k=0,1)}$ are the real potential depth and radius for the entrance channel.  In the case of a mixed $\ell$ transfer, the model has the additional terms:
\begin{align}
  \label{eq:model_mixed_l}
  &\textnormal{Prior:}  \nonumber \\
  &\alpha \sim \textnormal{Uniform}(0, 1)  \nonumber \\
  &\textnormal{Function:} \\
  &\frac{d Y}{d \Omega}^{\prime}_{\textnormal{DWBA}, i} =  \eta \times \delta D_0^2 \times C^2S  \nonumber \\ 
  &\times \bigg[ \alpha \frac{d \sigma}{d \Omega}_{\textnormal{DWBA}, \ell_1} + (1-\alpha) \frac{d \sigma}{d \Omega}_{\textnormal{DWBA}, \ell_2}   \bigg] \nonumber,
\end{align}
where the definition for $\frac{d Y}{d \Omega}^{\prime}_{\textnormal{DWBA}, i}$ is understood to replace all other occurrences of that variable in Eq.~(\ref{eq:dwba_model_na}). Note that the individual cross sections, $\frac{d \sigma}{d \Omega}_{\textnormal{DWBA}, \ell_1}$ and $\frac{d \sigma}{d \Omega}_{\textnormal{DWBA}, \ell_2}$, are calculated simultaneously using same sampled values for the optical potential.  

\subsection{Results}
\label{sec:results}

\begin{figure}
    \centering
    \includegraphics[width=.45\textwidth]{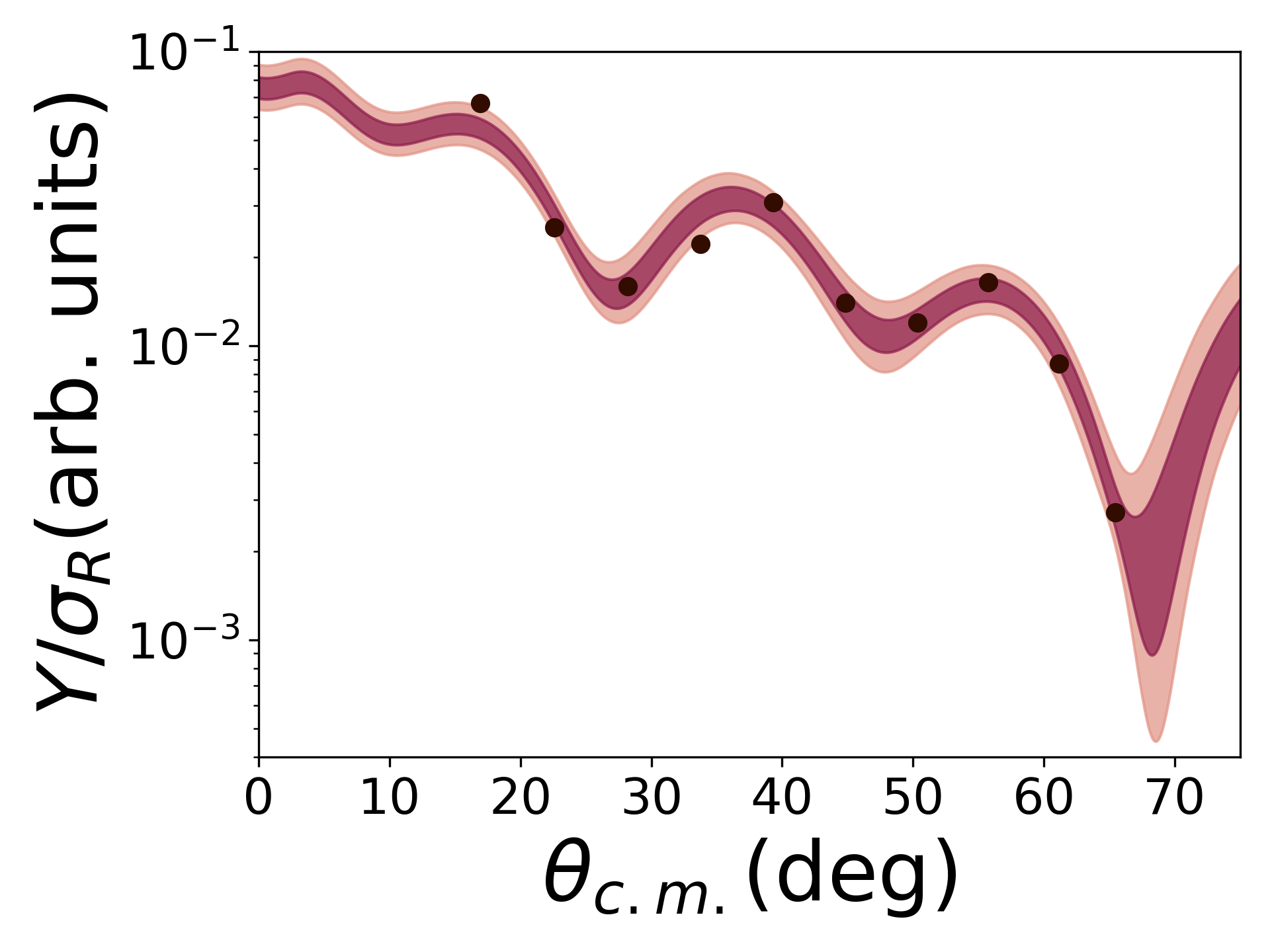}
    \caption{The credibility intervals obtained for the elastic scattering fit compared to the measured yields relative to Rutherford scattering ($Y_R/\sigma_R$). The dark and light purple bands show the $68 \%$ and $95 \%$ credibility intervals, respectively. The measured error bars are smaller than the points, while the adjusted uncertainty of Eq.~(\ref{eq:elastic_unc}) that is inferred from the data is not shown.}
    \label{fig:elastic_fit_na}
 \end{figure}

\begin{figure*}
   \captionsetup[subfigure]{labelformat=empty}
  \centering
    \vspace{-1\baselineskip}
    \begin{subfigure}[t]{0.30\textwidth}
        \includegraphics[width=\textwidth]{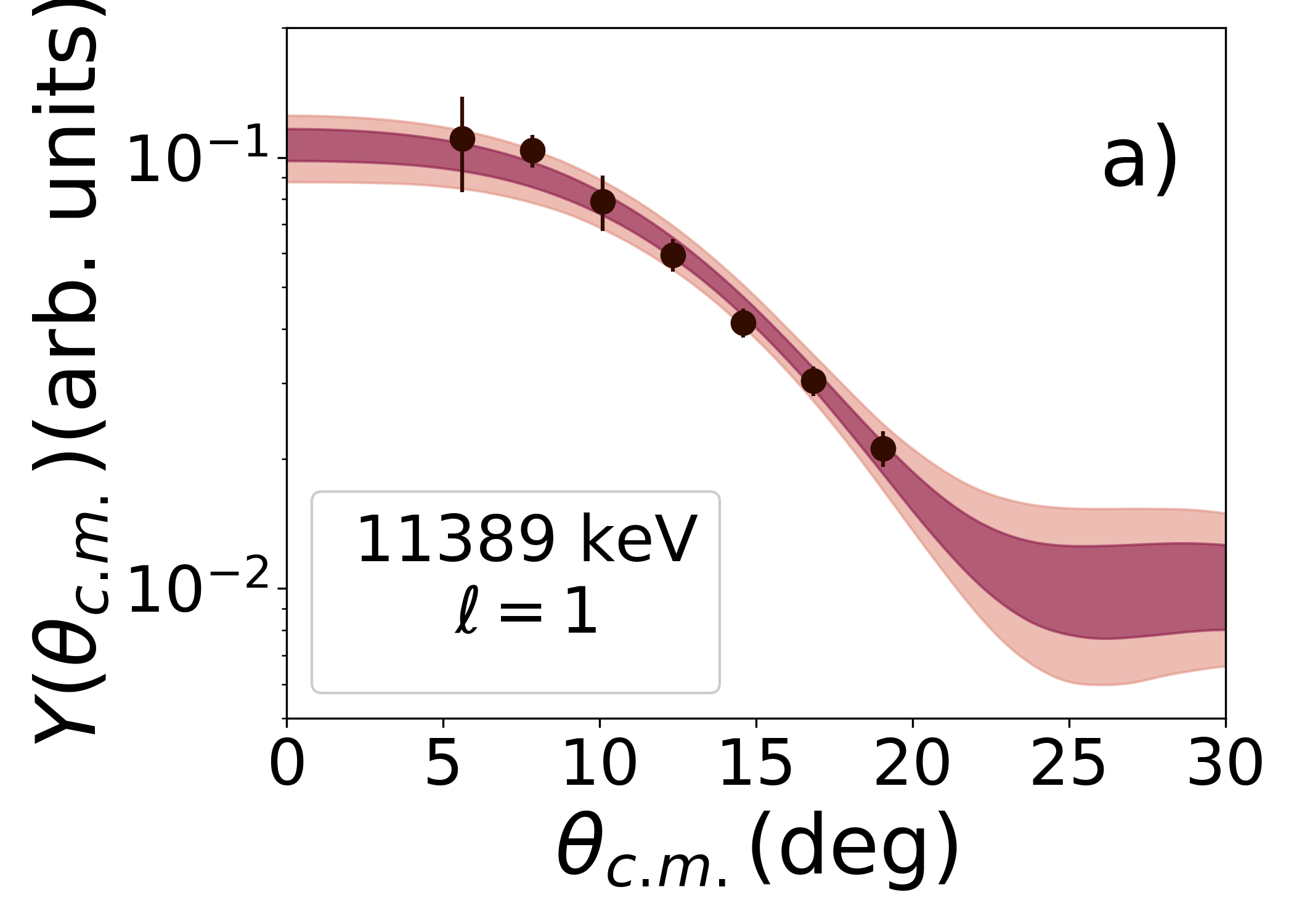}
        \caption{\label{fig:11390_fit}}
      \end{subfigure}
          \vspace{-1\baselineskip}
    \begin{subfigure}[t]{0.30\textwidth}
      \includegraphics[width=\textwidth]{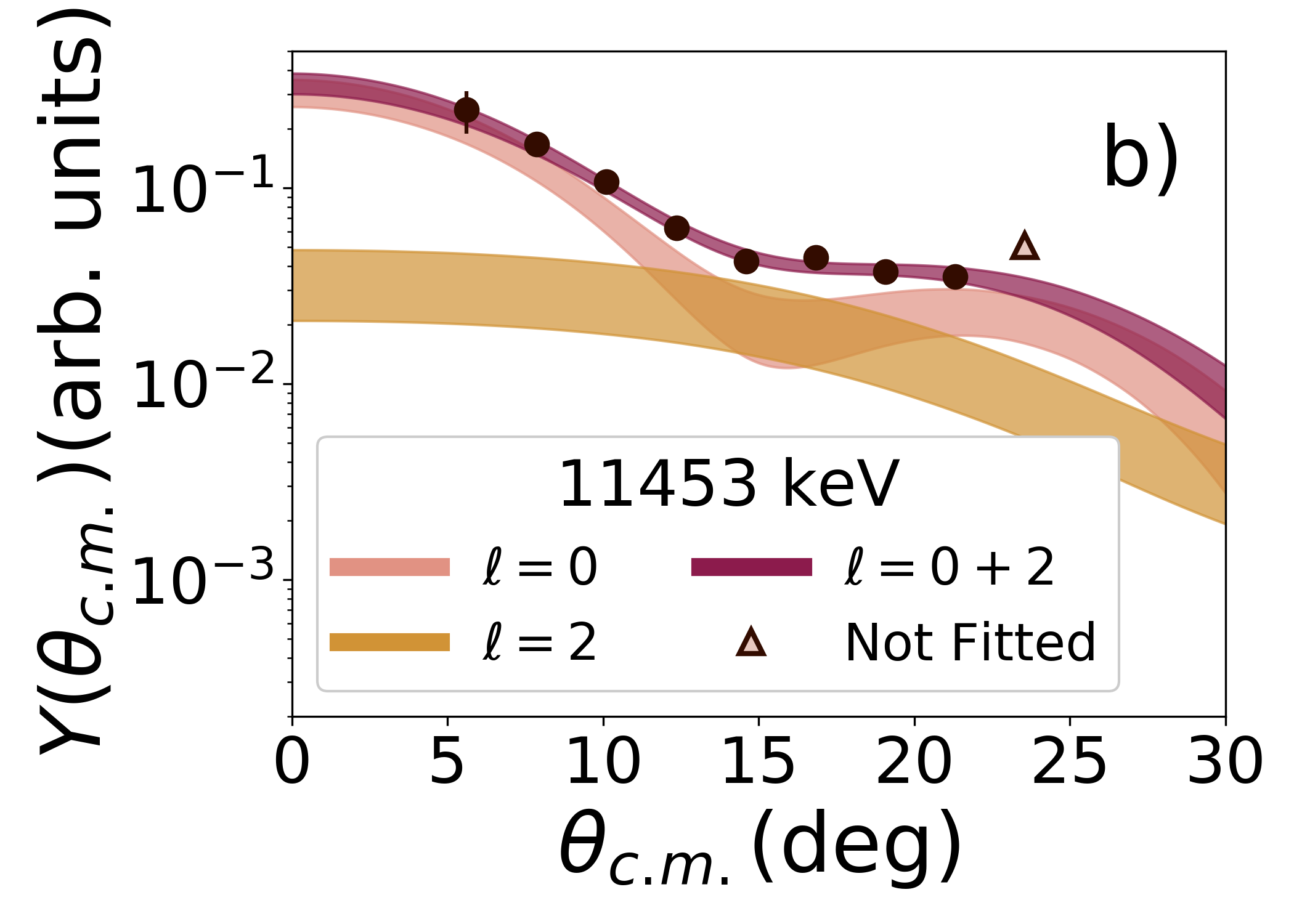}
          \vspace{-1\baselineskip}
      \caption{\label{fig:11453_fit}}
    \end{subfigure}
    \vspace{-1\baselineskip}
    \begin{subfigure}[t]{0.30\textwidth}
      \includegraphics[width=\textwidth]{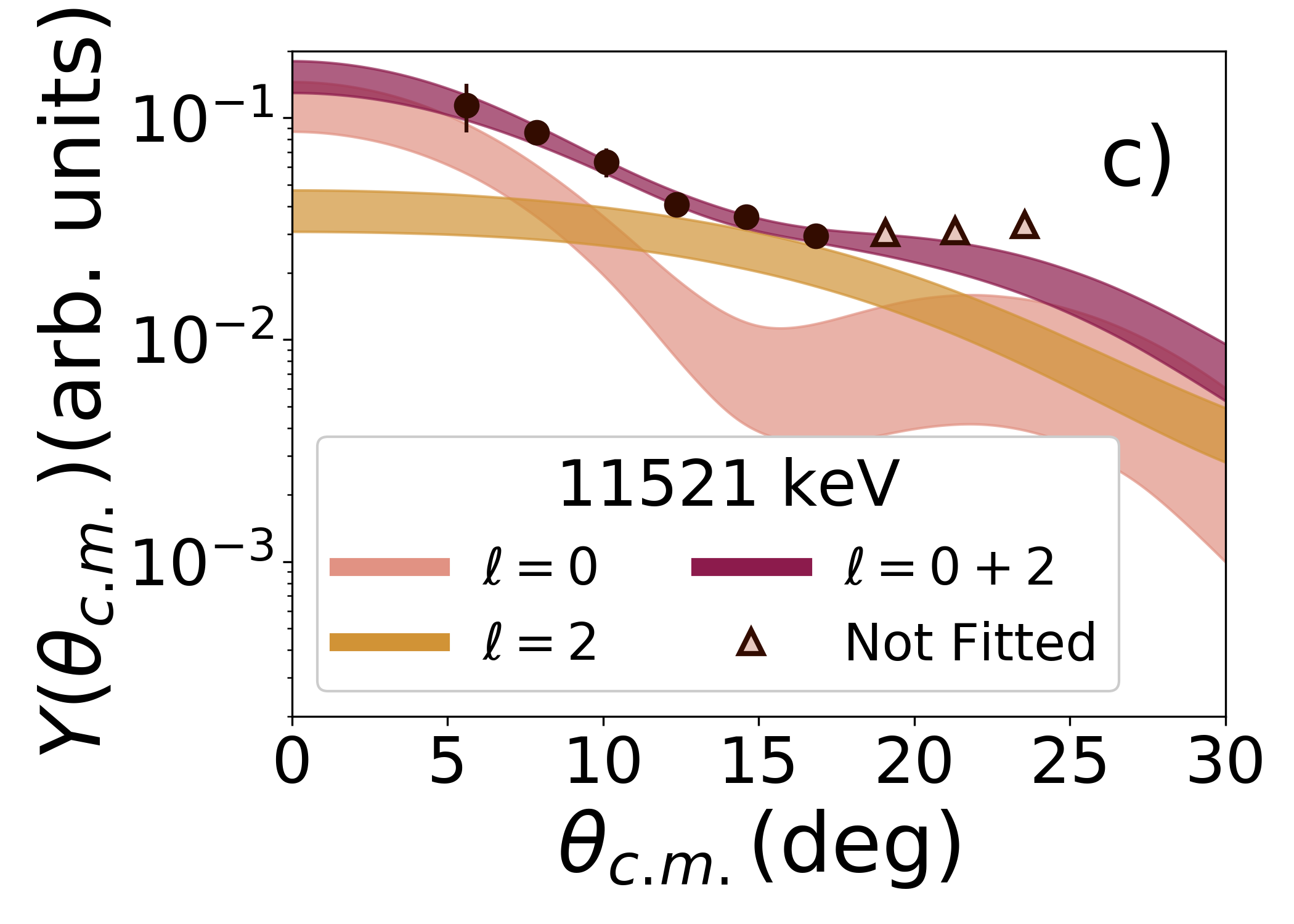}
          \vspace{-1\baselineskip}
      \caption{\label{fig:11521_fit}}
    \end{subfigure}
    \begin{subfigure}[t]{0.30\textwidth}
      \includegraphics[width=\textwidth]{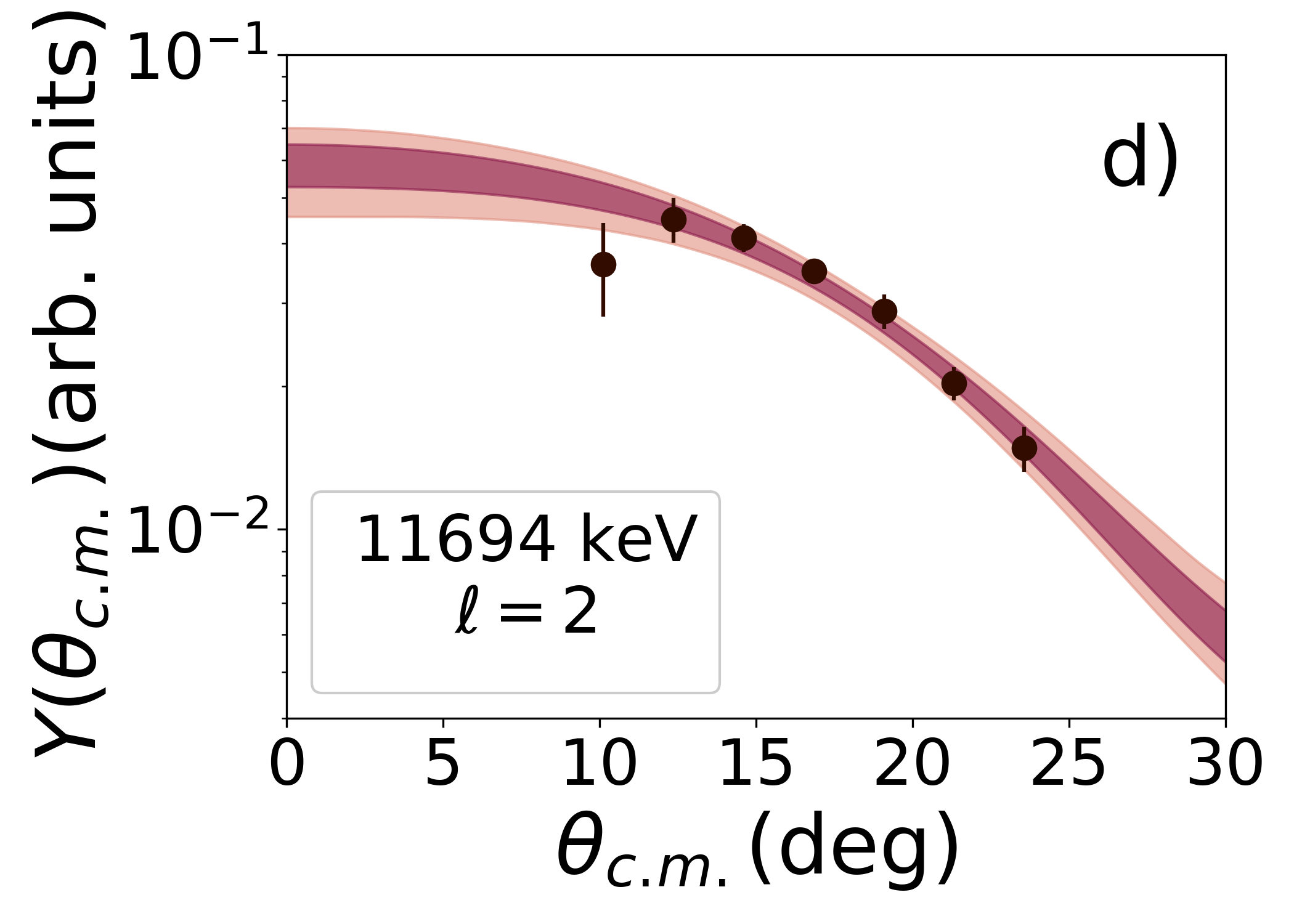}
          \vspace{-1\baselineskip}
      \caption{\label{fig:11695_fit}}
    \end{subfigure}
    \vspace{-1\baselineskip}
    \begin{subfigure}[t]{0.30\textwidth}
      \includegraphics[width=\textwidth]{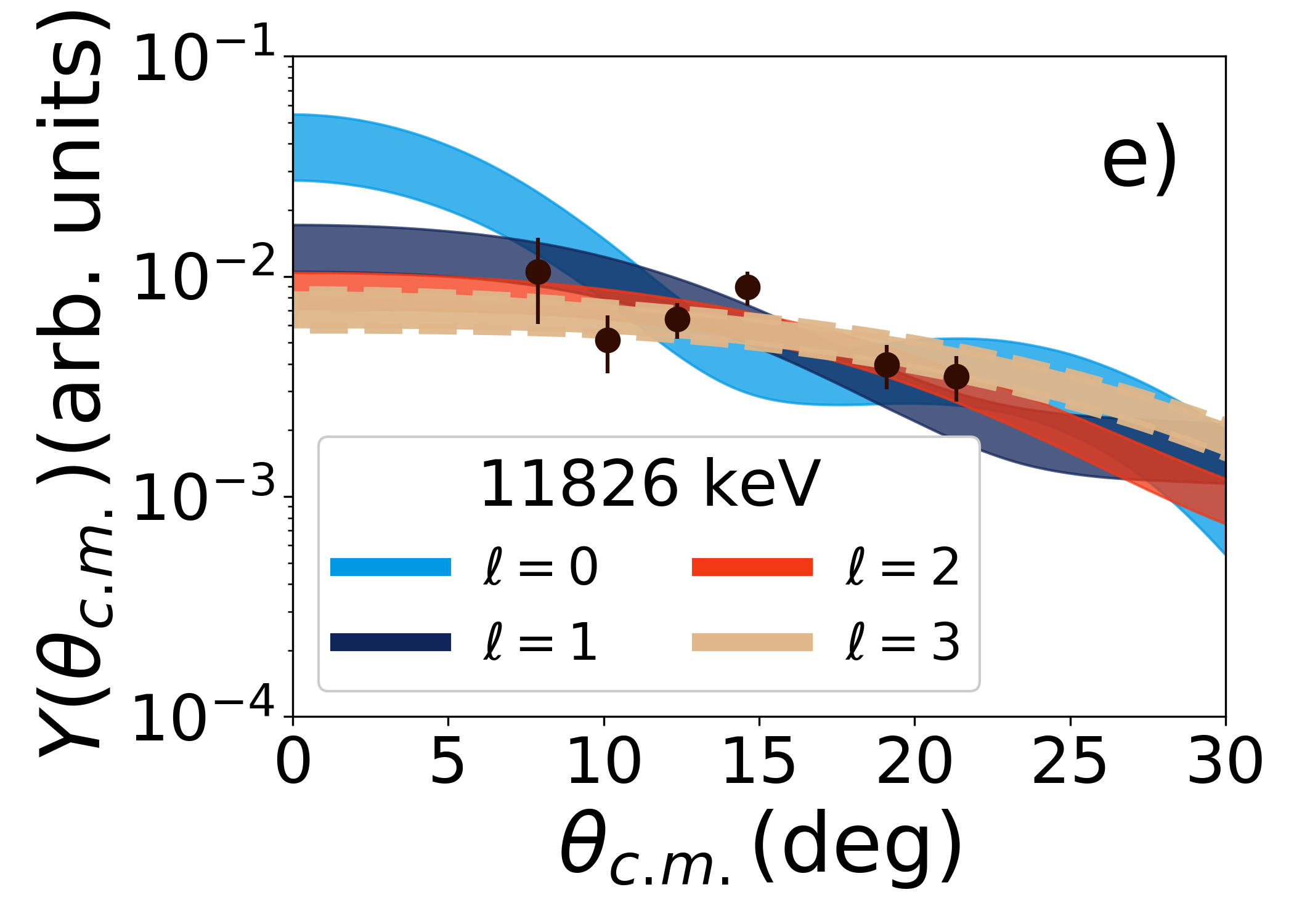}
          \vspace{-1\baselineskip}
      \caption{\label{fig:11824_fit}}
    \end{subfigure}
    \begin{subfigure}[t]{0.30\textwidth}
      \includegraphics[width=\textwidth]{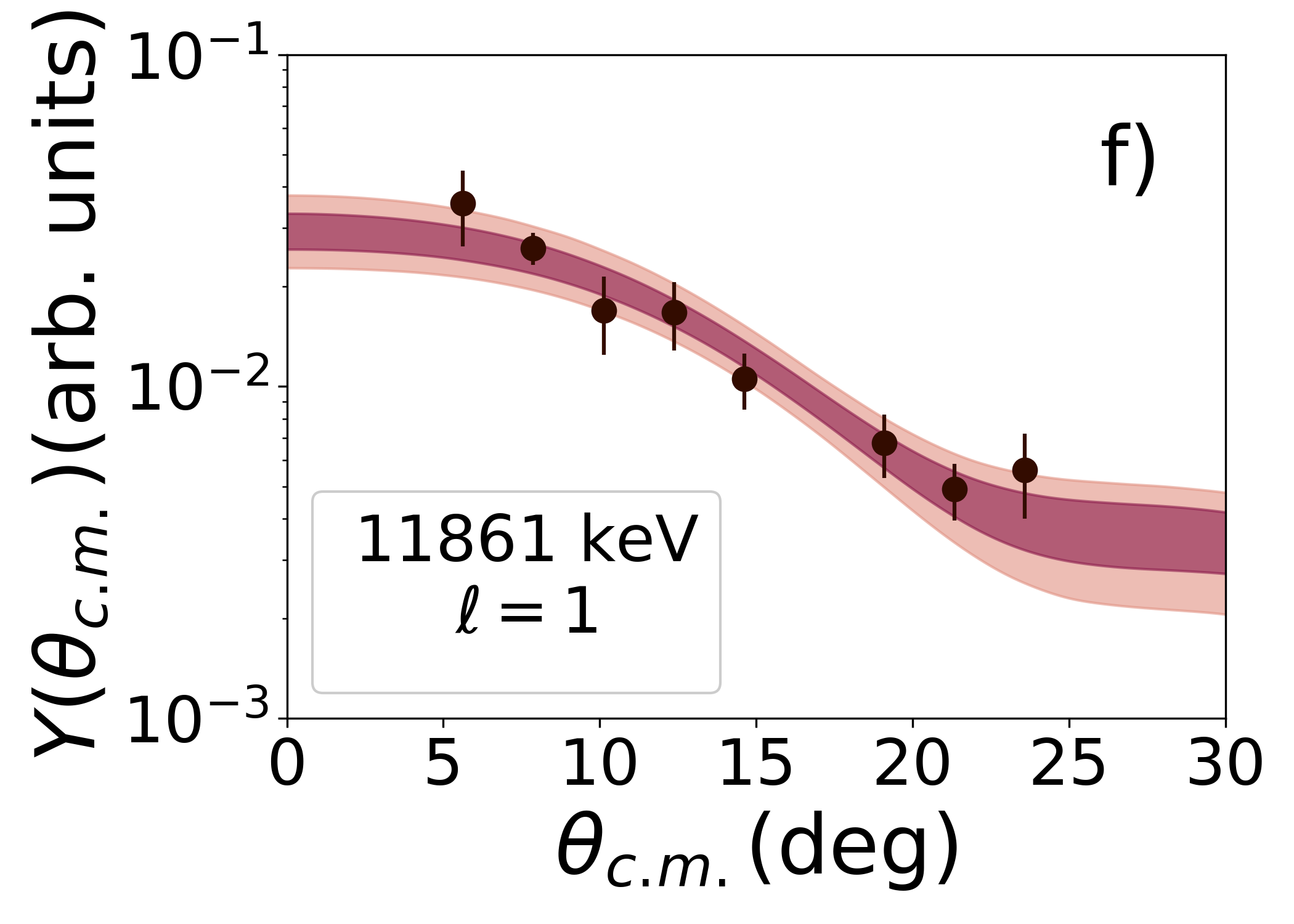}
          \vspace{-1\baselineskip}
      \caption{\label{fig:11860_fit}}
    \end{subfigure}
  \centering
    \begin{subfigure}[t]{0.30\textwidth}
      \includegraphics[width=\textwidth]{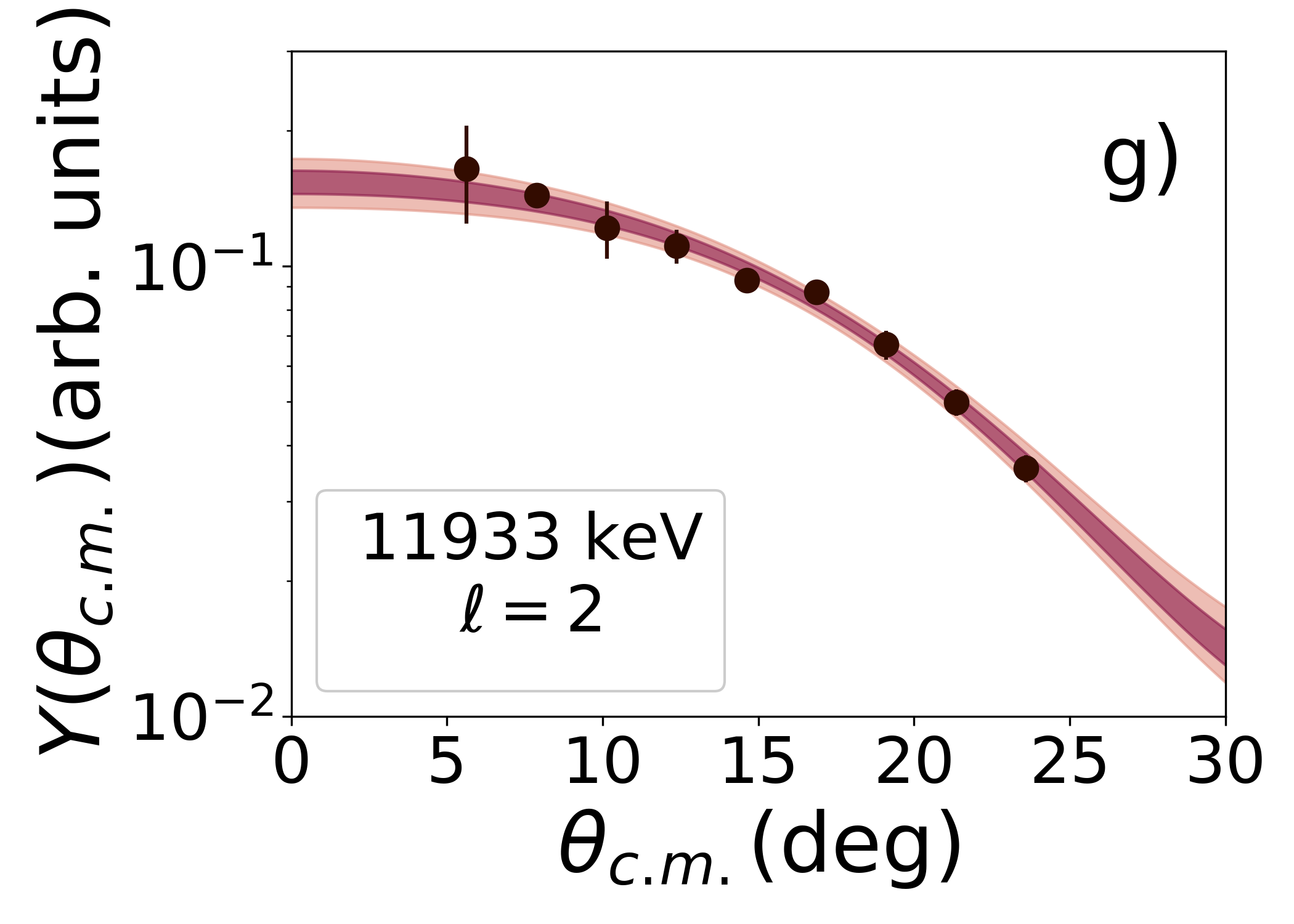}
          \vspace{-1\baselineskip}
      \caption{\label{fig:11933_fit}}
    \end{subfigure}
    \begin{subfigure}[t]{0.30\textwidth}
      \includegraphics[width=\textwidth]{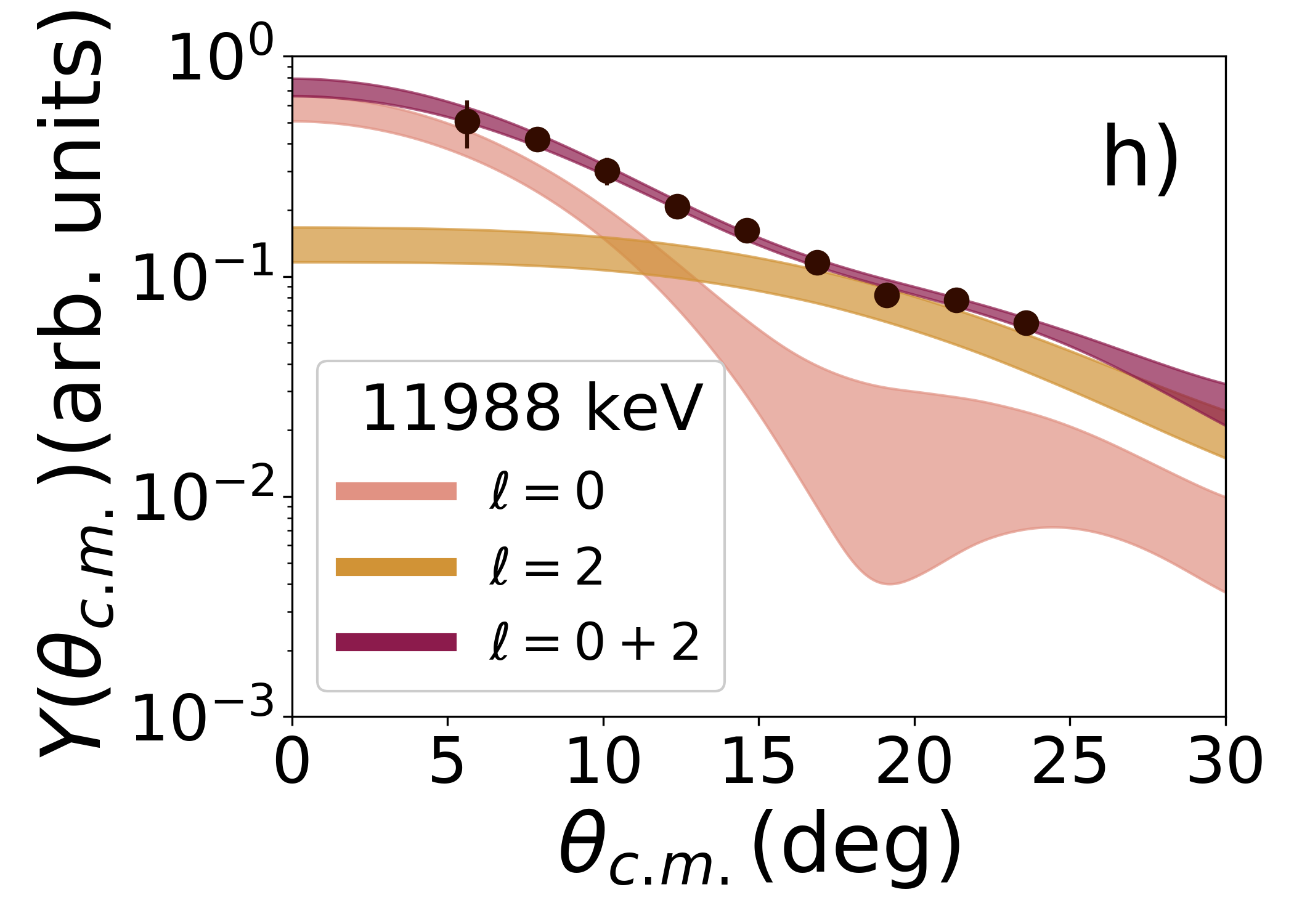}
          \vspace{-1\baselineskip}
          \caption{\label{fig:11988_fit}}
    \end{subfigure}
    \begin{subfigure}[t]{0.30\textwidth}
      \includegraphics[width=\textwidth]{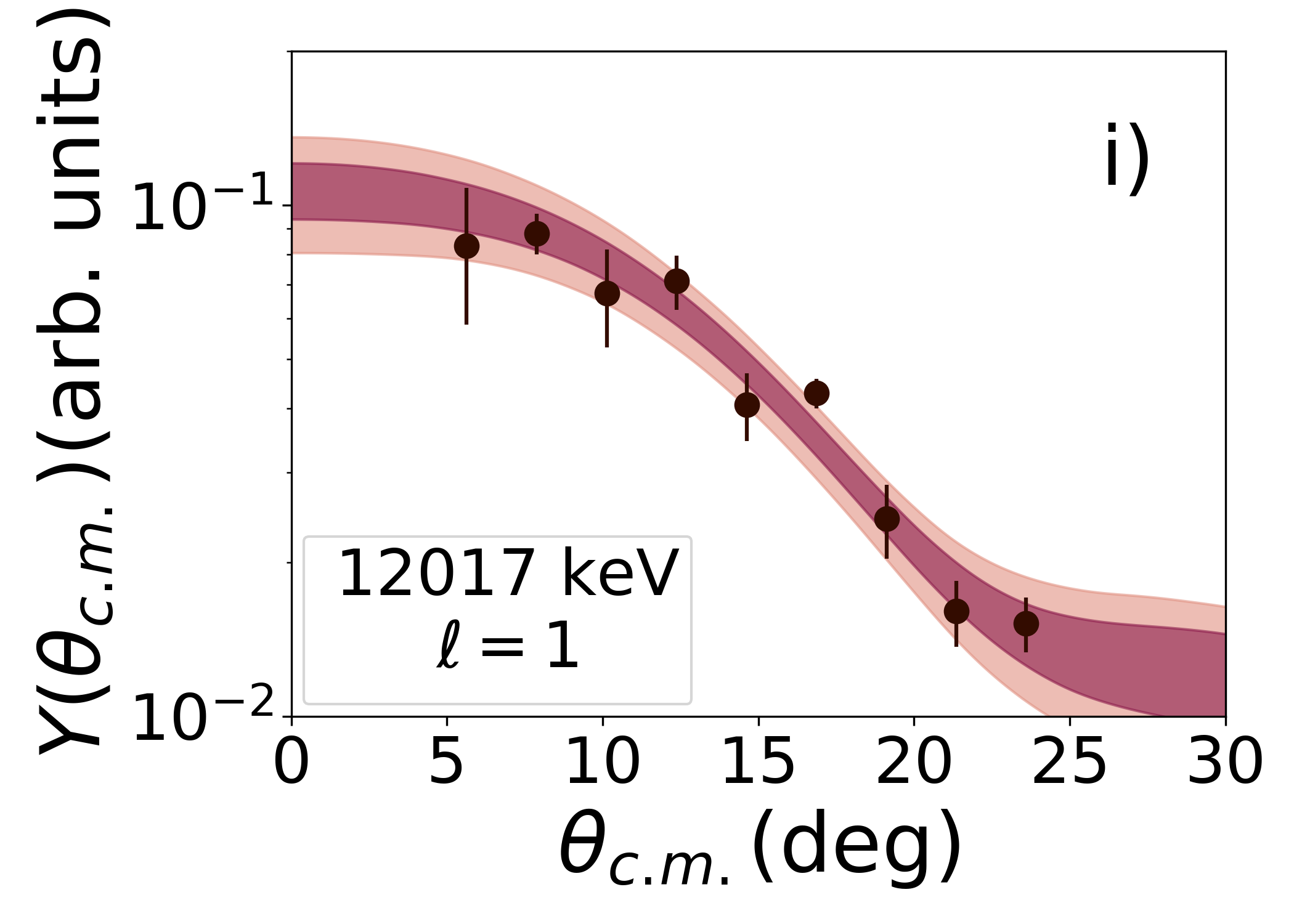}
          \vspace{-1\baselineskip}
      \caption{\label{fig:12017_fit}}
    \end{subfigure}
            \begin{subfigure}[t]{0.30\textwidth}
      \includegraphics[width=\textwidth]{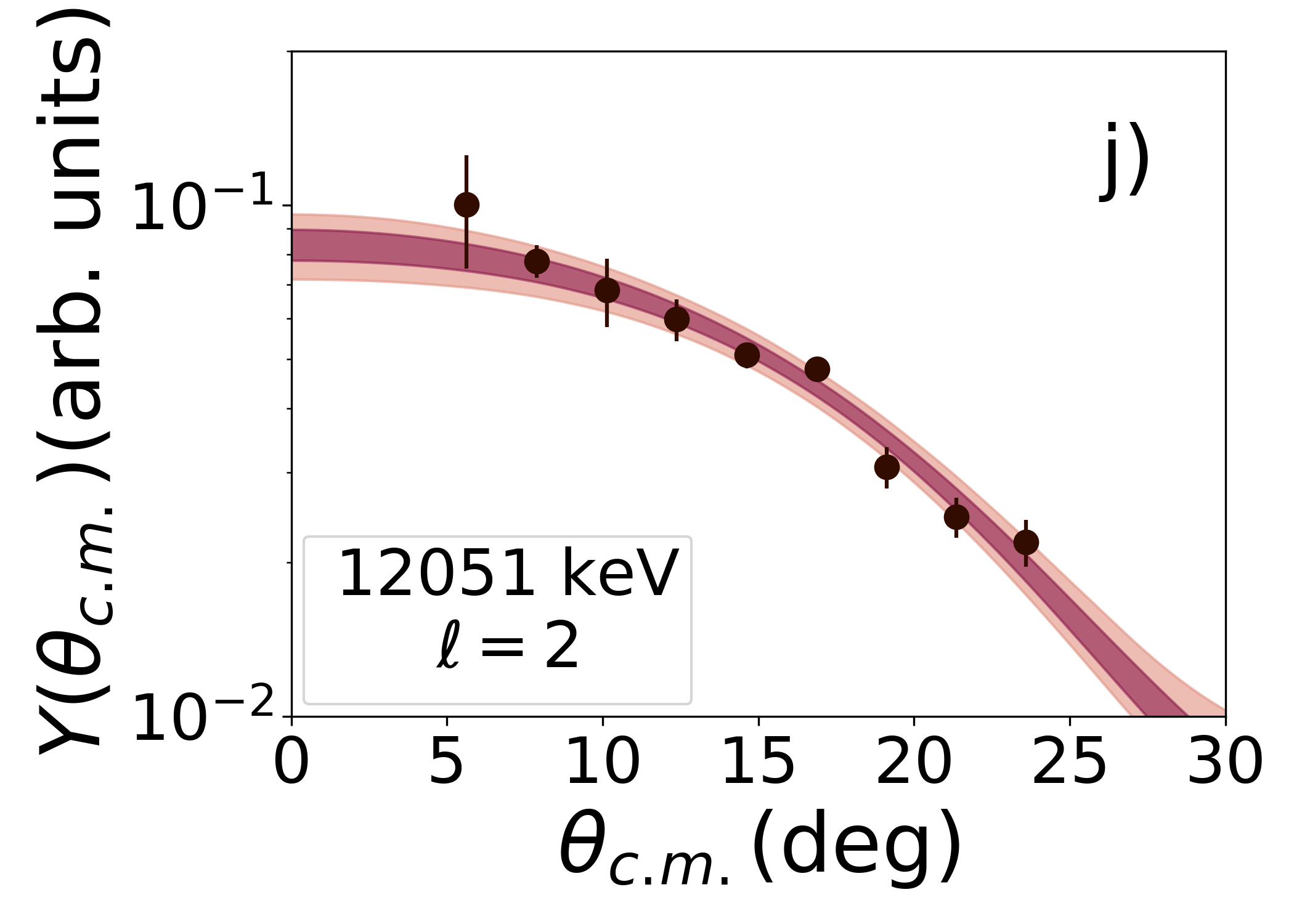}
          \vspace{-1\baselineskip}
      \caption{\label{fig:12051_fit}}
    \end{subfigure}
            \begin{subfigure}[t]{0.30\textwidth}
      \includegraphics[width=\textwidth]{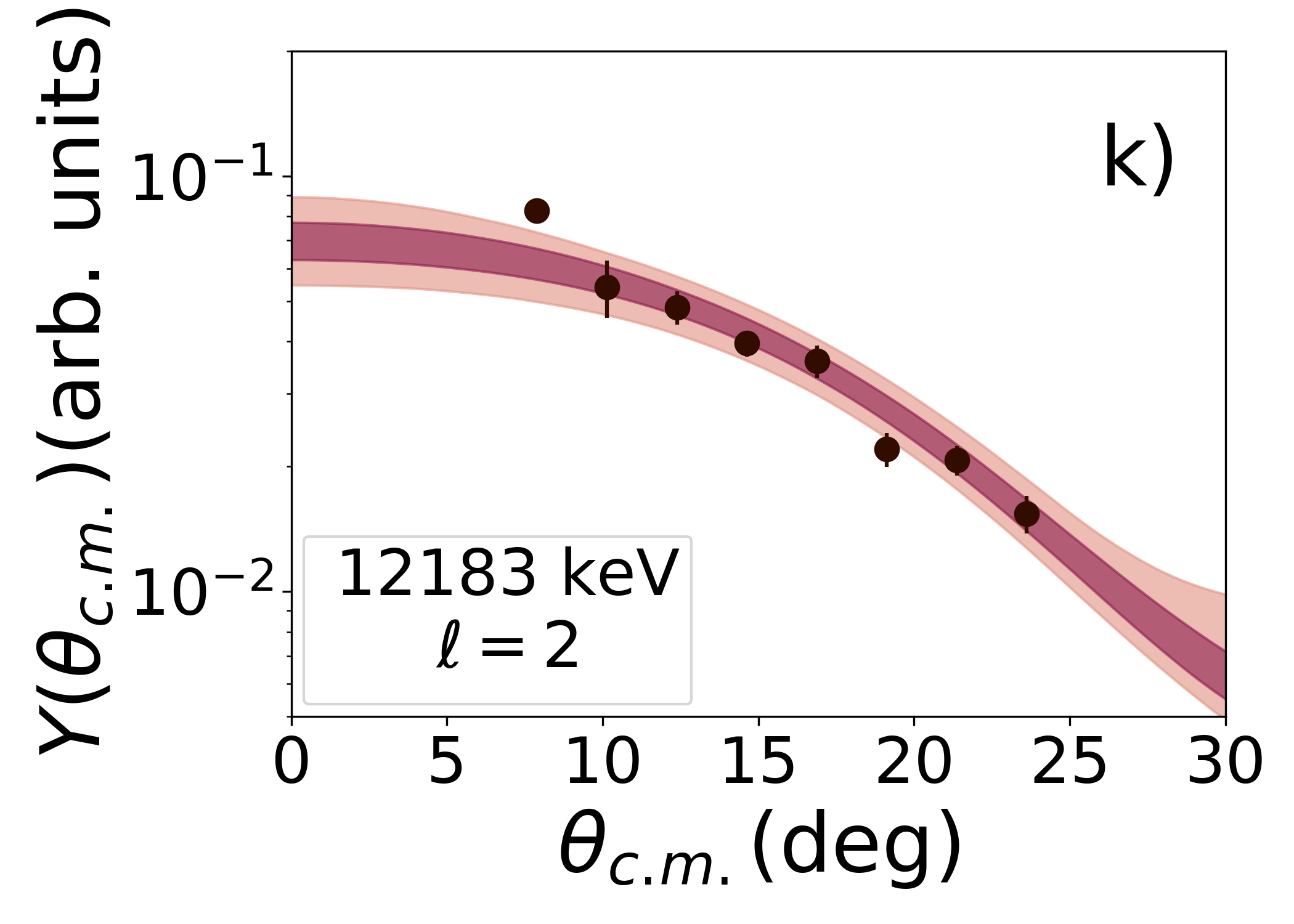}
          \vspace{-1\baselineskip}
      \caption{\label{fig:12183_fit}}
    \end{subfigure}
    \vspace{-1.8\baselineskip}
    \caption{\label{fig:states} DWBA calculations for the states of $^{24}$Mg . The $68 \%$ and $95 \%$ credibility intervals are shown in purple and light purple, respectively. Only the data points shown in black were considered in each calculation, with the triangles being excluded based on the cross section increasing after the minimum value was reached. For the {$11826$} keV state, the $68 \%$ bands are shown for all of the $\ell$ transfers between {$0\text{-}3$}.}    
\end{figure*}

The above Bayesian model was applied to the eleven states observed in the astrophysical region of interest. For each state, affine invariant MCMC \cite{ensemble_mcmc}, as implemented in the \texttt{python} package \texttt{emcee} \cite{emcee} was run with $400$ walkers taking $8000$ steps, giving a total of $3.2 \times 10^6$ samples. Of these samples, the first $6000$ steps were discarded as burn in, and the last $2000$ steps were thinned by $50$ for $16000$ final samples. The effective sample size was estimated to be greater than $2000$ based on the calculated autocorrelation of $\approx 400$ steps. These $16000$ samples were used to estimate the posterior distributions for $C^2S$, and to construct the differential cross sections shown in Fig.~\ref{fig:states}. An example of the simultaneous fit obtained for the elastic scattering data is shown in Fig.~\ref{fig:elastic_fit_na}. All of the data have been plotted as a function of their relative value (Sec.~\ref{sec:relative-yields}). Data points were only fit up to the first minimum in the cross section, the region where DWBA is expected to be most applicable \cite{thompson_nunes_2009}.  The normalization $\eta$ was found to be $\eta = 0.075^{+0.007}_{-0.006}$, which shows that the absolute scale of the data, despite the influence of the optical model parameters, can be established with a $9 \%$ uncertainty.

Values obtained for $(2J_f+1)C^2S$ in this work are listed in Table \ref{tab: na_c2s_table}, were the term $(2J_f+1)C^2S$ is constant for all possible values of $J_f$ for the final state if it is populated by the same $j = \ell \oplus s$ transfer. There is general agreement between our values and those of Ref.~\cite{hale_2004}, which provides further evidence that the absolute scale of the data is well established. However, for the three $2^+$ states that show a mixture of $\ell = 0 \oplus 2$, the current values are consistently lower. In these cases, the Bayesian method demonstrates that considerable uncertainty is introduced when a mixed $\ell$ transfer is present. The origin of this effect merits a deeper discussion, which we will now present. 

First, consider that the posterior distributions for $(2J_f+1)C^2S$ from states with unique $\ell$ transfers were found to be well described by log-normal distributions. Estimations of these distributions can be made by deriving the log-normal parameters $\mu$ and $\sigma$ from the MCMC samples. These parameters are in turn related to the median value of the log-normal distribution by $med.=\exp{(\mu)}$ and its factor uncertainty, $f.u.=\exp{(\sigma)}$. The $med.$ and $f.u.$ quantities are listed in Table \ref{tab: na_c2s_table}. It can be seen that states that have a unique $\ell$ transfer show factor uncertainties of $f.u. \approx 1.30$, or, rather, a $30 \%$ uncertainty. On the other hand, states that show a mixed $\ell$ transition vary from $f.u. = 1.4 \text{-} 2.0$. It was found that the individual $\ell$ components, which are the quantities relevant to the reaction rate, have a large factor uncertainty and deviate strongly from a log-normal distribution. However, their sum shares the same properties as the states with a single $\ell$ transfer. In other words, the total spectroscopic factor still has a $30 \%$ uncertainty. Since the total spectroscopic factor is the quantity that determines the relationship between the theoretical calculations and the data, its uncertainty is similar to a single $\ell$ spectroscopic factor, $30 \%$. For the mixed $\ell$ case, the individual components are terms in a sum that produces the theoretical prediction. The mean value of this sum grows linearly with each term, while the uncertainty grows roughly as the square root of the sum of the squares. It is this fact that requires, without appealing to the current Bayesian methods, the individual $\ell$ components to have a greater percentage uncertainty than their sum. Since previous studies, like those of Ref.~\cite{hale_2004}, assume a constant uncertainty with the extraction of spectroscopic factors, each $\ell$ component is assumed to have the same percentage uncertainty. The above discussion highlights that this assumption cannot be true, regardless of the statistical method. The influence of optical model parameters limits the precision of the total normalization of the cross section; thereby, giving an upper limit on the precision that can be expected from the components. These results indicate that applying a standard $\chi^2$ fit to a mixed $\ell$ transfer might not accurately extract the individual spectroscopic factors if optical model uncertainties are ignored.            

We will now discuss our results, and summarize the previously reported information for each of these states.

% \begin{figure}
%     \centering
%     \includegraphics[width=.8\textwidth]{11390_log_norm.pdf}
%     \caption{The posterior samples of $(2J_f+1)C^2S$ for the $11390$-keV state. The dark blue dashed line shows the corresponding log-normal distribution.}
%     \label{fig:11390_log_normal}
% \end{figure}

% \begin{figure}
%     \centering
%     \includegraphics[width=\textwidth]{11453_sf_log_norm_big_fig.pdf}
%     \caption{Plots of the three posterier distributions associated with the $11453$-keV spectroscopic factor. The top two plots show the individual $\ell$ $(2J_f+1)C^2S$ samples and their corresponding log-normal distributions. The bottom shows the sum of the two components and its log-normal distribution. It can be seen that the sum has the same factor uncertainty as transfers described by a single $\ell$ value.}
%     \label{fig:11453_log_normal}
% \end{figure}

\begin{table*}[t]
  \centering
  \setlength{\tabcolsep}{5pt}
  \caption{ \label{tab: na_c2s_table} The values of $(2J_f+1)C^2S$ that were derived in this work compared to those of Ref.~\cite{hale_2004}. All values for this work give the $68 \%$ credibility interval from the posterior estimation. Additionally, the parameters of the corresponding log-normal distribution are listed. All spin parity information, except that of the $11825$-keV state, is taken from Ref.~\cite{firestone_2007}, and are updated based on the current observations.}
\begin{threeparttable}
  \begin{tabular}{lllllll}
    \toprule
    \toprule
    $E_x$ (keV) & $J^{\pi}$ & $\ell$ \footnote{$+$ in the context of mixed $\ell$ transfers is simply a delineation between each $\ell$ component.} & $(2J_f+1)C^2S$ & $med.$   & $f.u.$& Ref.~\cite{hale_2004}  \\ \hline \vspace{-2mm}
  \\\vspace{2mm}
$11389$     & $1^-$    & $1$  &$0.066^{+0.021}_{-0.015}$ & $0.067$ & $1.30$ & $0.06$ \\ \vspace{2mm}
$11453$     & $2^+$    & $0+2$ & $0.14^{+0.05}_{-0.04}$ + $0.05^{+0.03}_{-0.02}$ &   $0.14$ + $0.048$      & $1.39$ + $2.00$ & $0.24$ + $0.16$ \footnote{Ref.~\cite{hale_2004} assumed a doublet. The $(2J_f+1)C^2S$ values were taken from these two states.}  \\ \vspace{2mm}
$11521$     & $2^+$    & $0+2$ & $0.05^{+0.03}_{-0.02}$ + $0.057^{+0.024}_{-0.018}$ &  $0.055$ + $0.056$ & $1.61$ + $1.51$ & $0.10$ \footnote{Ref.~\cite{hale_2004} assumed a doublet, with a portion of the strength assigned to a negative parity state.} \\ \vspace{2mm}
$11694$     & $4^+$    & $2$ & $0.085^{+0.025}_{-0.018}$ & $0.086$  & $1.29$ & $0.11$ \\ \vspace{2mm} 
{$11826$}         &    & $0$ & $0.023^{+0.012}_{-0.007}$ & $0.024$  & $1.52$ & $0.039$ \\ \vspace{2mm}
        &    & $1$ & $0.010^{+0.004}_{-0.003}$ & $0.010$  & $1.40$ & $0.009$ \\ \vspace{2mm}
        &    & $2$ & $0.014^{+0.005}_{-0.003}$ & $0.014$  & $1.36$ & $0.015$ \\ \vspace{2mm}
        &    & $3$ & $0.025^{+0.009}_{-0.006}$ & $0.025$  & $1.36$ & $0.024$ \\ \vspace{2mm}
$11861$     & $1^-$    & $1$ & $0.022^{+0.007}_{-0.005}$ & $0.022$  & $1.32$ & $0.026$ \\ \vspace{2mm}
$11933$     & $(2 \text{-} 4)^+$ & $2$ & $0.23^{+0.07}_{-0.05}$ & $0.24$ & $1.30$ & $0.25$      \\ \vspace{2mm}
$11988$     & $2^+$ & $0+2$ & $0.26^{+0.10}_{-0.07}$ + $0.24^{+0.10}_{-0.07}$  & $0.26$ + $0.24$ & $1.40$ + $1.45$ & $0.42$ + $0.33$      \\ \vspace{2mm}
$12017$     & $3^-$ & $1$ & $0.20^{+0.06}_{-0.04}$ & $0.20$ & $1.30$ & $0.13$      \\ \vspace{2mm}
$12051$     & $4^+$ & $2$ & $0.13^{+0.04}_{-0.03}$ & $0.14$ & $1.30$ & $0.13$      \\ \vspace{2mm}
$12183$     & $(1,2^+)$ & $2$ & $0.12^{+0.04}_{-0.03}$ & $0.12$ & $1.34$ & $0.13$      \\
    \bottomrule
    \bottomrule
\end{tabular}
\end{threeparttable}
\end{table*}

\subsubsection{The $11389$-keV State; $-303$-keV Resonance}

This state has been reported in several studies, and is known to have a spin parity of $J^{\pi}=1^-$ {\cite{zwieglinski_1978}}. Our measurements
confirm an $\ell=1$ nature to the angular distribution, making it a candidate for a subthreshold $p$-wave resonance. A higher lying state with unknown spin-parity has been reported in Ref.~\cite{vermeer_1988} at $E_x = 11394(4)$ keV. The current evaluation states that the
$^{25}\textnormal{Mg}(^{3}\textnormal{He}, ^{4}\!\textnormal{He})^{24}\textnormal{Mg}$ measurement of Ref.~\cite{El_Bedewi_1975} also observes this
higher state at $11397(10)$ keV, but their angular distribution gives an $\ell=1$ character, indicating it would be compatible with the lower $J^{\pi}=1^-$ state. Ref.~\cite{hale_2004} finds a similar peak in their spectrum, but considered it a doublet because of the ambiguous shape of the angular distribution, which was caused primarily by the behavior of the data above $20^{\circ}$. Due to our angular distribution not having these higher angles, and considering the excellent agreement between our data and an $\ell=1$ transfer, only the state at $11389.2(12)$ keV with $J^{\pi}=1^-$ was considered to be populated. The present calculation assumes a $2p_{3/2}$ transfer and is shown in Fig.~\ref{fig:11390_fit}.

\subsubsection{The $11453$-keV State; $-240$-keV Resonance}

Two states lie in the region around $11.45$ MeV, with the lower assigned $J^{\pi}=2^+$ and the upper $J^{\pi}=0^+$. The only study that reports a definitive observation of the $0^+$, $11460(5)$ keV state is the $(\alpha,\alpha_0)$ of Ref.~\cite{goldberg_54}. The current study and that of Ref.~\cite{hale_2004} indicate that there is a state around $E_x = 11452$ keV that shows a mixed $\ell = 0 + 2$ angular distribution. Since the ground state of $^{23}$Na is non-zero, this angular distribution can be the result of a single $2^+$ state, and the $\ell=2$ component cannot be unambiguously identified with the higher lying $0^+$ state. The $(p,p^{\prime})$ measurement of Ref.~\cite{zwieglinski_1978} also notes a state at $11452(7)$ keV with $\ell=2$. The excellent agreement between our excitation energy and the gamma ray measurement of Ref.~\cite{endt_1990} leads us to assume the full strength of the observed peak comes from the $2^+$ state. The calculation shown in Fig.~\ref{fig:11453_fit} assumes transfers with quantum numbers $2s_{1/2}$ and $1d_{5/2}$.

\subsubsection{The $11521$-keV State; $-172$-keV Resonance}

Another sub-threshold $2^+$ state lies at $11521.1(13)$ keV. It should be noted that another state with unknown spin-parity was observed at $11528(4)$ keV in Ref.~\cite{vermeer_1988}, but has not been seen on other studies. Ref.~\cite{vermeer_1988} reports a measured $\Gamma_{\gamma}/\Gamma \approx 1$ for this new state, making it a candidate for an unnatural parity $^{24}$Mg state. The present angular distribution, Fig.~\ref{fig:11521_fit}, is indicative of a mixed $\ell = 0 + 2$ assignment. Thus, the observation is associated with the $2^+$ state at $11521.1(13)$ keV, and transfers were calculated using $2s_{1/2}$ and $1d_{5/2}$.      

\subsubsection{The $11694$-keV State; $1$-keV Resonance}

For our measurement this state was partially obscured by a contaminant peak from the ground state of $^{17}$F coming from $^{16}$O$(^{3} \textnormal{He}, d)^{17}$F for $\theta_{Lab} < 9^{\circ}$. Previous measurements have established a firm $4^+$ assignment, and our angular distribution is consistent with an $\ell =2$ transfer. The fit for a $1d_{5/2}$ transfer is shown in Fig.~\ref{fig:11695_fit}.  

\subsubsection{The $11826$-keV State; $133$-keV Resonance}
\begin{figure}
  \centering
  \includegraphics[width=.45\textwidth]{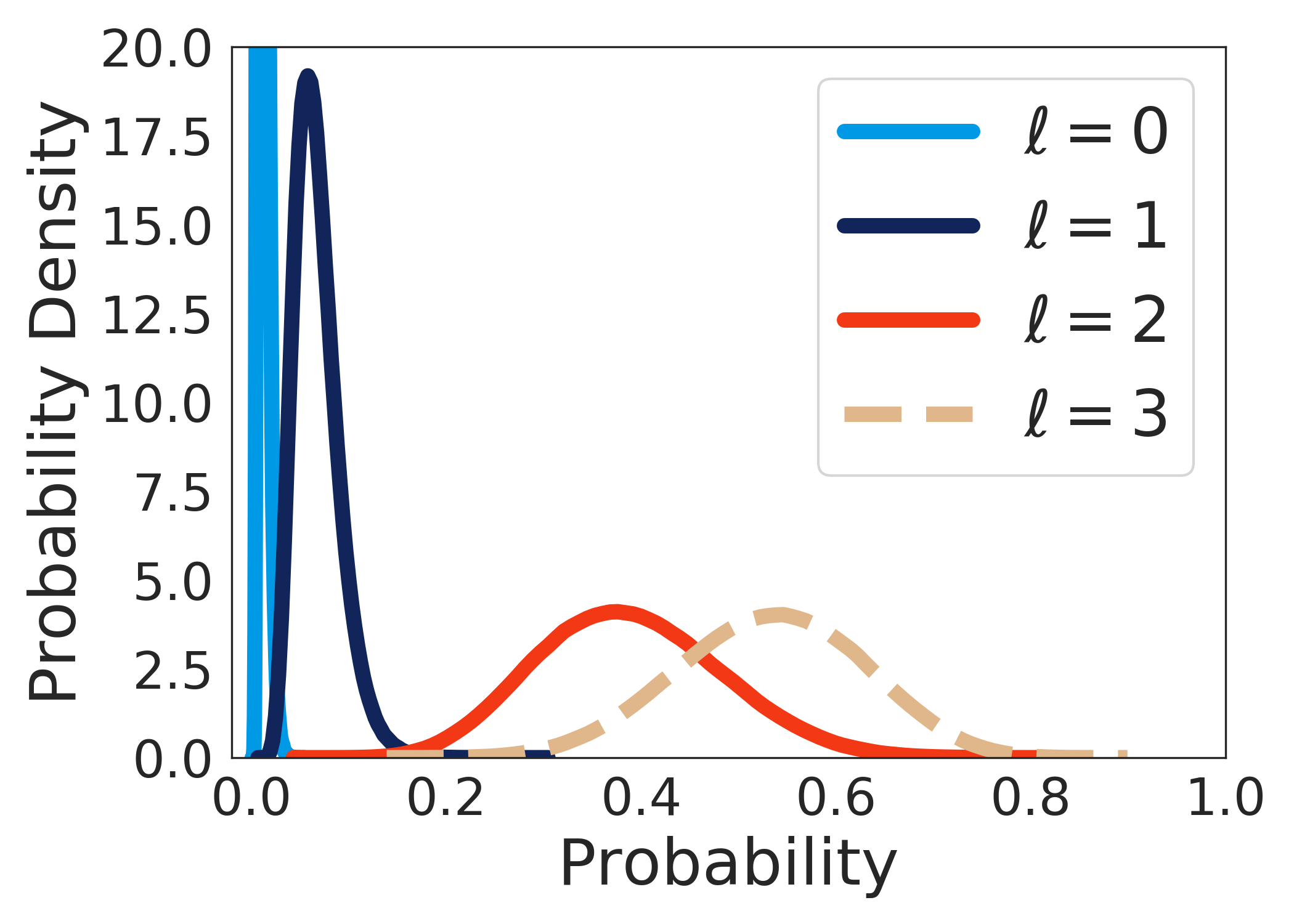}
  \caption{The distributions from the nested sampling algorithm for the most likely $\ell$ values for the $11826$-keV state.}
  \label{fig:l_comp_probs_na}
\end{figure}

This state is also obscured at several angles by the fifth excited state of $^{15}$O. The previous constraints on its spin parity come from the comparison of the extracted spectroscopic factors for each $\ell$ value in Ref.~\cite{hale_2004} and the upper limits established in Ref.~\cite{Rowland_2004} and subsequently Ref.~\cite{Cesaratto_2013}.
This DWBA analysis finds an angular distribution consistent with Ref.~\cite{hale_2004}, which it should be noted experienced similar problems with the nitrogen contamination, but with the Bayesian model comparison methods presented in Sec.~\ref{sec:overv-bayes-dwba}, constraints can be set based purely on the angular distribution. All of the considered $\ell$ transfer are shown in Fig.~\ref{fig:11824_fit}, and were calculated assuming $2s_{1/2}$, $2p_{3/2}$, $1d_{5/2}$, and $1f_{7/2}$ transfers, respectively. The results of the nested sampling calculations, which give the relative probabilities of each transfer, are presented in Table~\ref{tab:probs} and shown in Fig.~\ref{fig:l_comp_probs_na}. The adopted values were taken to be the mean of these distributions instead of the median as in Ref.~\cite{Marshall_2020}. Since the statistical errors of the nested sampling are normally distributed in $\ln Z$, the resulting probabilities are distributed log-normally. The choice of the mean instead of the median then amounts to selecting the arithmetic mean instead of the geometric mean, which ensures $\sum_{\ell} P(\ell) = 1$.

\begin{table*}[]
  \centering
  \setlength{\tabcolsep}{12pt}
    \caption{\label{tab:probs} Results of the model comparison calculations for the $11826$ keV state. For each $\ell$ value, We list the $\log{Z}$ value calculated with nested sampling, the median Bayes factor when compared to the most likely transfer $\ell=3$, and the mean probability of each transfer.}

  \begin{tabular}{llll}
    \toprule
    \toprule
    $\ell$     & $\ln{Z}_{\ell}$              & $B_{3 \ell}$         & $P(\ell)$         \\ \hline \vspace{-2mm}
    \\
           $0$    &      $44.226(294)$           & $47.79$            & $1 \%$           \\ 
           $1$    &      $45.990(289)$           & $8.20$             & $7 \%$ \\ 
           $2$    &      $47.762(323)$           & $1.39$             & $39 \%$ \\
           $3$    &      $48.093(293)$           & $1.00$             & $53 \%$ \\ 
    \bottomrule
    \bottomrule
\end{tabular}
\end{table*}

\subsubsection{The $11861$-keV State; $168$-keV Resonance}

There are two states within a few keV of one another reported to be in this region. One is known to have $ J^{\pi}=1^-$ {\cite{zwieglinski_1978}}, and has been populated in nearly all of the experiments listed in Table~\ref{tab:energy_comp}. The other state is reported to decay to the $6^+$, $8114$-keV state, with a $\gamma$-ray angular distribution that favors an assignment of $8^+$ \cite{branford_1972}. The later polarization measurements of Ref.~\cite{wender_1978} support the assignment of $8^+$. For our experiment, the tentative $8^+$ state is likely to have a negligible contribution to the observed peak, and the angular distribution in Fig.~\ref{fig:11860_fit} is consistent with a pure $\ell=1$ transfer. The calculation assumed $2p_{5/2}$.

\subsubsection{The $11933$-keV State; $240$-keV Resonance} 

The $11933$-keV State does not have a suggested spin assignment in the current ENSDF evaluation \cite{firestone_2007}. However, the earlier compilation of Ref.~\cite{endt_eval_1990} lists a tentative $(2-4)^+$. The compilation assignment is justified from two pieces of evidence. First, the 
$\ell=2$ angular distribution observed in the $(^4 \textnormal{He}, ^3\textnormal{He})$ measurement of Ref.~\cite{El_Bedewi_1975} suggests $(0\text{-}4)^+$. Second, the $0^+$ and $1^+$ assignments are ruled out from the observed $\gamma$-decays to the $J^{\pi} = 2^+$, $1368$-keV and $J^{\pi} = 4^+$, $4122$-keV states observed in Ref.~\cite{Berkes_1964}. Our measurement indicates an $\ell=2$ transfer. Based on these observations, and the satisfactory ability to describe the angular distribution with $\ell=2$, a $1d_{5/2}$ transfer was calculated, and is shown in Fig.~\ref{fig:11933_fit}. It should also be noted that Schmalbrock \textit{et. al} suggested that this state could be the analogue to a $T=1$ state with spin $3^+$ in $^{24}$Na \cite{schmalbrock_1983}.    

\subsubsection{The $11988$-keV State; $295$-keV Resonance}

As can be seen in Table \ref{tab:energy_comp}, the $11988$-keV State has been observed in multiple experiments, including the high precision $\gamma$-ray measurement of Ref.~\cite{endt_1990}. A spin parity of $2^{+}$ has been assigned based on the inelastic measurement of Ref.~\cite{zwieglinski_1978}. The current fit is shown in Fig.~\ref{fig:11988_fit} and assumes a mixed $\ell = 0+2$ transition with $2s_{1/2}$ and $1d_{5/2}$.

\subsubsection{The $12017$-keV State; $324$-keV Resonance}

The $12017$-keV state is known to have $J^{\pi}=3^-$, which was established from the angular distributions of Ref.~\cite{Kuperus_1963, Fisher_1963} and confirmed by the inelastic scattering of Ref.~\cite{zwieglinski_1978}. Our angular distribution is consistent with an $\ell=1$ transfer, and was calculated assuming $2p_{3/2}$. The fit is shown in Fig.~\ref{fig:12017_fit}.

\subsubsection{The $12051$-keV State; $359$-keV Resonance}

The angular distribution of $\alpha$-particles from $^{23}$Na$(p, \alpha)$ measured in Ref.~\cite{Fisher_1963} established $J^{\pi}=4^+$ for the $12051$-keV state, which was later confirmed by the inelastic scattering of Ref.~\cite{zwieglinski_1978}. The angular distribution of the present work is well described by a transfer of $1d_{5/2}$, which is shown in Fig.~\ref{fig:12051_fit}.   

\subsubsection{The $12183$-keV State; $491$-keV Resonance}

Ref.~\cite{MEYER_1972} observed that the $12183$-keV state $\gamma$-decays to $0^+$, $2^+$, and $1^{+}$ states, which permits values of $(1,2^{+})$. The angular distribution of Ref.~\cite{hale_2004} permits either $\ell = 2$ or $\ell = 0+2$ transfers, which requires the parity of this state be positive. The current work finds an angular distribution consistent with a pure $\ell=2$ transfer. The calculation of the $1d_{5/2}$ transfer is shown in Fig.~\ref{fig:12183_fit}.

%\textit{11729 State}-This state was not observed in the present study. It has a firm spin parity of $0^+$, established in the $(\alpha,\alpha_0)$ measurement of Ref.~\cite{goldberg_54}.

\section{Proton Partial Widths}
\label{sec:prot-part-widths}
The spectroscopic factors extracted in Sec.~\ref{sec:spec_factors} are only an intermediate step in the calculation of the $^{23}$Na$(p, \gamma)$ reaction rate. From the proton spectroscopic factors of this work, proton partial widths can be calculated using
\begin{equation}
  \label{eq:partial-widths}
  \Gamma_p = C^2S \Gamma_{\textnormal{sp}},
\end{equation}
where $\Gamma_{\textnormal{sp}}$ is the single-particle partial width. If there is a mixed $\ell$ transfer, then the total proton width is calculated using:
\begin{equation}
    \label{eq:mixed_l_proton_width}
    \Gamma_p = \sum_{\ell} \Gamma_{p, \ell}.
\end{equation}
However, for our case the $\ell=2$ single particle widths, $\Gamma_{sp}$, are typically two orders of magnitude lower than the $\ell=0$ ones, making them negligible in the calculations presented below.

\subsection{Bound State Uncertainties}

There are additional sources of uncertainty impacting the determination of $\Gamma_p$. One of the largest is the bound state parameters used to define the overlap function. Since the overlap function is extremely sensitive to the choice of Woods-Saxon radius and diffuseness parameters, the extracted spectroscopic factor can vary considerably. This dependence has been discussed extensively in the literature, for a review, see Ref.~\cite{2014_Tribble}. Ref.~\cite{Marshall_2020} confirmed this strong dependence in a Bayesian framework. If the uncertainties of $C^2S$ are independent from those of $\Gamma_{sp}$, then single-particle transfer reaction experiments that determine spectroscopic factors will be unable to determine $\Gamma_p$ with the precision needed for many astrophysics applications.

Ref.~\cite{bertone} noted an important consideration for the calculation of $\Gamma_p$ from $C^2S$ and $\Gamma_{sp}$. If these quantities are calculated using the \textit{same} bound state potential parameters, the variation in $C^2S$ is anticorrelated with that of $\Gamma_{sp}$. Thus, the product of these two quantities, i.e., $\Gamma_{p}$, has a reduced dependence on the chosen bound state potentials. Using the same bound state parameters for both quantities, Refs.~\cite{hale_2001, hale_2004} found variations in $\Gamma_p$ of $\approx 5 \%$. With the Bayesian methods of this study, we investigate whether this anticorrelation still holds in the presence of optical model uncertainties.

\begin{figure}
    \centering
    \includegraphics[width=.45\textwidth]{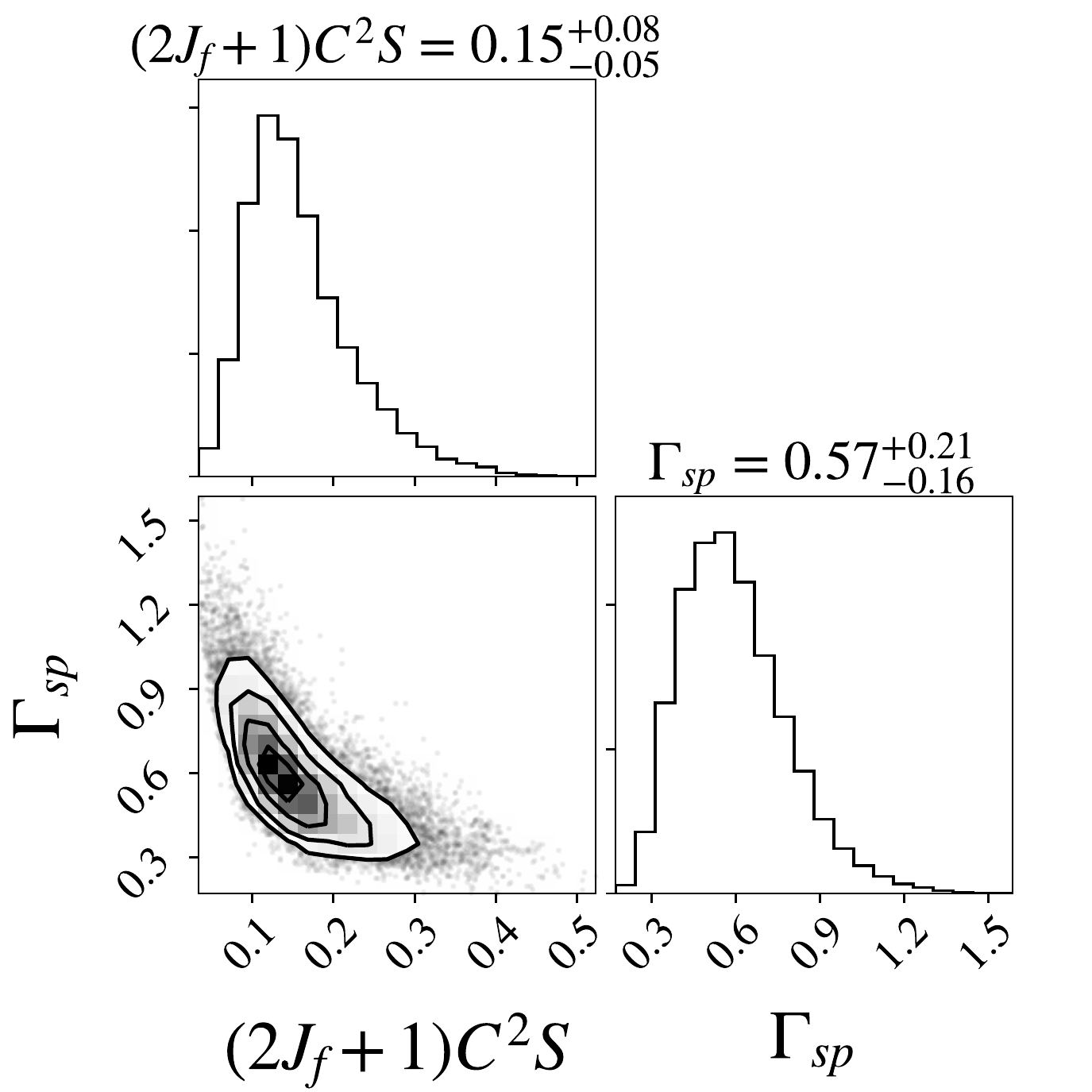}
    \caption{Pair correlation plot for the MCMC posterior samples of $\Gamma_{sp}$ and $(2J_f+1)C^2S$ for the $12051$-keV state. A strong anticorrelation exists when the same bound state parameters are used to calculate both quantities, resulting in $\Gamma_p$ having less sensitivity to these parameters.}
    \label{fig:corner_g_sp_sf}
\end{figure}

The code \texttt{BIND} calculates $\Gamma_{\textnormal{sp}}$ for a resonance at energy $E_r$ with a Woods-Saxon potential. For additional details on this code see Ref.~\cite{ILIADIS_1997}. Modifications were made to the code so that it could be run on a set of tens of thousands of bound state samples to produce a set of $\Gamma_{sp}$ samples. Due to the numerical instability of the integration for low energy resonances, the potential impact of the weak binding approximation, and the difficulties for mixed $\ell$ transitions, the state selected for this calculation needs to have a $ 500 \gtrapprox E_r \gtrapprox 100$ keV, $\ell \geq 2$, and a known spin parity. The only such state is at $E_x = 12051$ keV ($E_r = 359$ keV). A new MCMC calculation was carried out using the same model as Eq.~(\ref{eq:dwba_model_na}) with the additional parameters for the bound state $r_0$ and $a_0$. These were given priors:
\begin{align}
    \label{eq:bound_state_priors}
    & r_0 \sim \mathcal{N}(1.25, 0.125^2) \\
    & a_0 \sim \mathcal{N}(0.65, 0.065^2). \nonumber 
\end{align}
The sampler was again run with $400$ walkers taking $8000$ steps. The final $2000$ steps were thinned by $50$ giving $16000$ posterior samples. These samples were then plugged into \texttt{BIND} to produce the $16000$ samples of $\Gamma_{sp}$. Since these samples all come directly from the MCMC calculation they naturally account for the variations in the optical model parameters as well as $C^2S$. First it is worth establishing the bound state parameters influence on the uncertainty of $C^2S$. The log-normal distribution well described these samples and had a factor uncertainty of $f.u.=1.50$ increased from $f.u.=1.30$ in the case of fixed bound state parameters. The pair correlation plot for $(2J_f+1)C^2S$ versus $\Gamma_{sp}$ is shown in Fig.~\ref{fig:corner_g_sp_sf}. The resulting distribution gives $(2J_f+1)\Gamma_{p} = 0.083^{+0.025}_{-0.018}$ eV, while the value calculated using fixed bound state parameters gives $(2J_f+1)\Gamma_{p} = 0.082^{+0.025}_{-0.018}$ eV.

The cancellation between the variation in $\Gamma_{sp}$ and $C^2S$ is nearly exact in this case, with the resulting uncertainty being $30 \%$ in both calculations. The quantum numbers of the bound state, $n$ and $j$, can also a have a dramatic effect on the extracted spectroscopic factor. Repeating the above calculation assuming a $2d_{5/2}$ state instead of a $1d_{5/2}$ causes $C^2S$ to drop to a value $50 \%$ lower. Once again, taking the MCMC samples of the bound state geometric parameters and running \texttt{BIND} with these parameters as well as $n = 2$ gives $(2J_f+1)\Gamma_{p} = 0.090^{+0.029}_{-0.020}$ eV.  
This relation still requires further study using Bayesian methods, particularly the influence of the bound state quantum numbers $n$ and $j$, which cannot be determined from the transfer data, but for the present work the potential influence of the bound state parameters on $\Gamma_p$ is considered negligible compared to the those of the optical model.     

\subsection{Subthreshold Resonances}

Three of the observed states lie close enough to the proton threshold to be astrophysically relevant. The penetrability, $P_{\ell}$, is undefined for $E_r < 0$, and therefore $\Gamma_{sp}$ cannot be calculated for subthreshold states. Instead these resonances will be integrated using $\theta^2 = C^2S \theta_{sp}^2$. $\theta_{sp}^2$ can be calculated using the fits provided in either Ref.~\cite{ILIADIS_1997} or Ref.~\cite{BARKER_1998}. We have adopted the fit of Ref.~\cite{ILIADIS_1997}. It should be noted that the fit of Ref.~\cite{ILIADIS_1997} was derived using the bound state parameters $r_0 = 1.26$ fm and $a_0 = 0.69$ fm which differ from those used in this work. The impact of this difference was investigated by using higher lying states where values of $\theta_{sp}^2$ could also be calculated using \texttt{BIND}. The maximum observed deviation was $10 \%$, which is in decent agreement with the expected accuracy of the fit as mentioned in Ref.~\cite{ILIADIS_1997}. The values of $\theta^2$ for this work are shown in Table \ref{tab:subthresh_resonance_table}.

\begin{table*}[t]
\centering
  \setlength{\tabcolsep}{10pt}
  \caption{\label{tab:subthresh_resonance_table} Reduced width calculations for the observed subthreshold resonances. All $\theta_{sp}^2$ values were calculated using the fit of Ref.~\cite{ILIADIS_1997} and should be considered to have a $10 \%$ systematic uncertainty. The $68 \%$ credibility intervals of the samples are presented in the fifth column.}
  \begin{tabular}{lllll}
    \toprule
    \toprule
    $E_x$(keV) & $E_r$(keV)  & $J^{\pi}$   & $\theta_{sp}^2$ & $(2J_f + 1) \theta^2 $   \\ \hline
    \\ [-1.5ex]
    
$11389.2(12)  $      &    $-303.5(12)$     &  $1^-$      &  $0.738$      &     $0.049^{+0.016}_{-0.011}$   \\ [0.8ex] 
$11452.9(4)   $      &   $-239.8(4)$       &  $2^+$      &  $0.654$      &     $0.09^{+0.03}_{-0.03}$            \\ [0.8ex]
$11521.1(13)  $      &   $-171.6(13)$      &  $2^+$ &  $0.639$  &    $0.035^{+0.018}_{-0.013}$              \\ [0.8ex]
    \bottomrule
    \bottomrule
  \end{tabular}
\end{table*}

\subsection{Resonances Above Threshold}

Eight resonances were observed above the proton threshold and below $500$ keV. Except for $E_r = 2$, all of the $\Gamma_{sp}$ values were calculated using \texttt{BIND}. \texttt{BIND} calculations were carried out with the Woods-Saxon potential parameters $r_0 = 1.25$ fm, $a_0 = 0.65$ fm, $r_c = 1.25$ fm, $V_{so} = 6.24$, and channel radius of $1.25$ fm. The low resonance energy of $E_r = 2$ presented numerical challenges for \texttt{BIND}, so it was calculated using the fit of Ref.~\cite{ILIADIS_1997}. Our results are shown in Tablet~\ref{tab:gamma_p_table}.  

\begin{table*}
\centering
  \setlength{\tabcolsep}{8pt}
  \caption{ \label{tab:gamma_p_table} Proton partial widths derived from this work. The values of $\Gamma_{sp}$ from \texttt{BIND} are listed for reference. $(2J_f+1)\Gamma_p$ values are given in terms of their $68 \%$ credibility intervals.}
\begin{threeparttable}
  \begin{tabular}{llllll}
    \toprule
    \toprule
    $E_x$(keV)      & $E_r$(keV)   & $J^{\pi}$          & $\Gamma_{sp}$(eV)                     & $(2J_f + 1) \Gamma_p $(eV) This Work & { $(2J_f + 1) \Gamma_p $(eV) Previous Work}      \\ \hline
                                                                                                                                                                                                         \\ [-1.5ex]
    $11694(4)     $ & $1(4)$       & $4^+$              & $2.589 \times 10^{-140}$ \footnote{Calculated using $\theta_{sp}^2$ from the fit of Ref.~\cite{ILIADIS_1997} to avoid the numerical instability of \texttt{BIND} at $2$ keV. An additional $10 \%$ systematic uncertainty should be considered.} & $2.2^{+0.6}_{-0.5} \times 10^{-141}$ &                                                                 \\ [0.8ex]
    $11826(3)     $ & $133(3)$     & $\ell=0$           & $1.092 \times 10^{-03}$               & $2.6^{+1.3}_{-0.8} \times 10^{-5}$   &                                                                 \\ [0.8ex]
                    &              & $\ell=1$           & $2.314 \times 10^{-04}$               & $2.2^{+0.9}_{-0.6} \times 10^{-6}$   &                                                                 \\ [0.8ex]
                    &              & $\ell=2$           & $4.949 \times 10^{-06}$               & $6.7^{+2.4}_{-1.7} \times 10^{-8}$   & {$1.23^{+0.49}_{-0.45} \times 10^{-8}$ \footnote{Derived from resonance strengths reported in Ref.~\cite{BOELTZIG_2019} and $\Gamma_{\gamma}/\Gamma$ values from Ref.~\cite{vermeer_1988}}} \\ [0.8ex]
                    &              & $\ell=3$           & $6.157 \times 10^{-08}$               & $1.5^{+0.5}_{-0.4} \times 10^{-9}$   &                                                                 \\ [0.8ex]
    $11860.8(14)  $ & $168.1(14)$  & $1^-$              & $5.894 \times 10^{-3}$                & $1.3^{+0.4}_{-0.3} \times 10^{-4}$   & {$1.8(4) \times 10^{-4}$ \footnote{{Derived in Ref.~\cite{hale_2004}, which should be consulted for details}.}}              \\ [0.8ex]
    $11933.06(19) $ & $240.37(19)$ & $(2 \text{-} 4)^+$ & $1.034 \times 10^{-2}$                & $2.4^{+0.7}_{-0.5} \times 10^{-3}$   & {$1.2(2)$ \footnotemark[2]}                              \\ [0.8ex]
    $11988.45(6)  $ & $295.76(6)$  & $2^+$              & $15.39$                               & $4.0^{+1.5}_{-1.1}$                  &                                                                 \\ [0.8ex]
    $12016.8(5)   $ & $324.1(5)$   & $3^-$              & $8.550$                               & $1.7^{+0.5}_{-0.4}$                  &                                                                 \\ [0.8ex]
    $12051.3(4)   $ & $358.6(4)$   & $4^+$              & $6.141 \times 10^{-1}$                & $8.2^{+2.5}_{-1.8} \times 10^{-2}$   &                                                                 \\ [0.8ex]
    $12183.3(1)   $ & $490.6(1)$   & $(1,2)^+$          & $9.318$                               & $1.1^{+0.4}_{-0.3}$                  &                                                                 \\ [0.8ex]

    \bottomrule
    \bottomrule
  \end{tabular}
  
\end{threeparttable}
\end{table*}

\subsection{Discussion}

The literature for $\omega \gamma$ values is extensive. Ref.~\cite{hale_2004} compiled and corrected previous measurements for stopping powers and target stoichiometry. Using those compiled values as well as the recent measurement of Ref.~\cite{BOELTZIG_2019}, comparisons can be made between the results of the current work and previous measurements. We choose to compare $(2J_f + 1) \Gamma_p$ values deduced from $\omega \gamma$ measurements instead of transforming our $(2J_f + 1) \Gamma_p$ values into their associated $\omega \gamma$. Knowledge of $\Gamma_{\gamma}/\Gamma$ is required in order to carry out a comparison, which limits us to a select few of the many measured resonances.  

\subsubsection{$133$-keV Resonance}

The $133$-keV resonance was measured directly at a significance greater than $2 \sigma$ for the first time at LUNA and is reported in Ref.~\cite{BOELTZIG_2019}. The value from that work is $\omega \gamma =  1.46^{+0.58}_{-0.53} \times 10^{-9}$ eV. Using $\Gamma_{\gamma}/\Gamma = 0.95(4)$ from Ref.~\cite{vermeer_1988} implies $(2J_f+1)\Gamma_p = 1.23^{+0.49}_{-0.45} \times 10^{-8}$ eV. The upper limit reported in Ref.~\cite{Cesaratto_2013} can also be used for comparison and yields $(2J_f+1)\Gamma_p \leq 4.35 \times 10^{-8}$ eV. The closest value from this work is the $\ell = 2$ transfer which gives $(2J_f+1)\Gamma_p = 6.0^{+2.1}_{-1.5} \times 10^{-8}$ eV. The disagreement between our value and that of LUNA is stark, and a significant amount of tension exists with the upper limit of Ref.~\cite{Cesaratto_2013}. 

\subsubsection{$168$-keV Resonance}

Ref.~\cite{hale_2004} derived a proton width of $(2J_f+1)\Gamma_p = 1.8(4) \times 10^{-4}$ eV for the $168$-keV Resonance using $\omega \gamma_{(\alpha, \gamma)}$, $\omega \gamma_{(p, \alpha)}$, and $\Gamma$. This value is in good agreement with the current work $(2J_f+1)\Gamma_p = 1.3^{+0.4}_{-0.3} \times 10^{-4}$ eV.

\subsubsection{$240$-keV Resonance}

Using the resonance strength measured in Ref.~\cite{BOELTZIG_2019} of $\omega \gamma = 4.8(8) \times 10^{-4}$ eV and $\Gamma_{\gamma}/\Gamma > 0.7$ from Ref.~\cite{vermeer_1988}, $(2J_f+1)\Gamma_p$ has a lower limit of $3.8(6) \times 10^{-3}$ eV, which is in mild tension with the transfer value of $2.5(7) \times 10^{-3}$ eV.

\subsubsection{$295$-keV Resonance}

Ref.~\cite{BOELTZIG_2019} measured {$\omega \gamma = 1.08(19) \times 10^{-1}$} eV, while Ref.~\cite{vermeer_1988} gives $\Gamma_{\gamma}/\Gamma = 0.70(9)$. In this case, $(2J_f+1)\Gamma_p = 1.2(2) $ eV. The current value is in significant disagreement with $(2J_f+1)\Gamma_p = 4.0^{+1.5}_{-1.1}$ eV.

\subsubsection{$491$-keV Resonance}

The $490$-keV Resonance is considered a standard resonance for the $^{23}$Na$(p, \gamma)$ reaction, and has a value of $9.1(12) \times 10^{-2}$ eV \cite{PAINE_1979}. Unfortunately, $\Gamma_{\gamma}/\Gamma$ is not known. However, an upper limit for $\omega \gamma_{(p, \alpha)}$ has been set at $\leq 0.011$ eV \cite{hale_2004}. The ratio of the two resonances strengths can set an upper limit for $\Gamma_{\alpha}/\Gamma_{\gamma}$:
\begin{equation}
    \label{eq:resonance_strength_ratio}
    \frac{\omega \gamma_{(p, \alpha)}}{\omega \gamma_{(p, \gamma)}} = \frac{\Gamma_{\alpha}}{\Gamma_{\gamma}}.
\end{equation}
Plugging in the values gives $\Gamma_{\alpha}/\Gamma_{\gamma} \leq 0.12 $. Assuming $\Gamma_p \ll \Gamma_{\gamma}$, $\Gamma_{\gamma}/\Gamma \geq 0.89$. The current value for $(2J_f+1)\Gamma_p = 1.1^{+0.4}_{-0.3}$ eV which can be compared to the upper limit of the standard resonance of $(2J_f+1)\Gamma_p = 0.82(11)$ eV. If we assume the $\alpha$ channel is completely negligible, $(2J_f+1)\Gamma_p = 0.73(10)$ eV. The standard resonance value appears to be consistent with the current work.

\subsection{Final Remarks on Proton Partial Widths}

The above comparisons make it clear that the agreement between the current experiment and previous measurements is inconsistent. Of particular concern are the $133$-keV and $295$-keV resonances, in which the disagreement is at a high level of significance. However, the measurement of Ref.~\cite{BOELTZIG_2019} at LUNA used the $295$-keV resonance as a reference during the data collection on the $133$-keV resonance, which could explain some correlation between those resonance strengths when compared to this work. On the other hand, LUNA's value of $\omega \gamma = 1.08(19) \times 10^{-1}$ eV is in excellent agreement with the value given in the compilation of Endt \cite{endt_eval_1990}, $\omega \gamma = 1.05(19) \times 10^{-1}$ eV, which normalized the value of Ref.~\cite{Switkowski_1975} to the standard resonance at $491$-keV. These comments are not meant to brush aside the serious issues that come with extracting proton partial widths from DWBA calculations, but to highlight that any comparison between direct and indirect measurements involves data from several sources, each of which have their own systematic uncertainties complicating the conclusions that can be drawn. There is a need for detailed, systematic studies to determine the reliability of $\Gamma_p$ values extracted from transfer reactions at energies relevant to astrophysics. 

It is also worth reiterating the comment first made in Ref.~\cite{Marshall_2021}, the updated resonance energy of $133$ keV compared to the previously assumed $138$ keV could impact the assumption of a thick target yield curve made in Ref.~\cite{Cesaratto_2013, BOELTZIG_2019}. The significantly lower energy has the potential to move the beam off of the plateau of the yield curve, further affecting the extracted resonance strength, but the magnitude of this effect is difficult to estimate. However, the measurement of Ref.~\cite{Cesaratto_2013} made at LENA has an upper limit that is consistent with the LUNA value and is in tension with the current work. Importantly, their upper limit also assumed the $138$-keV resonance energy, but used a much thicker target ($\approx 30$ keV) than the LUNA measurement ($\approx 15$ keV) making it less sensitive to the resonance energy shift. Again it should be mentioned that all of this discussion presupposes that the proton state has $\ell = 2$ and that our observed angular distribution arises completely from a direct reaction mechanism. If the spin is one of the other possible values, the current results will differ by over an order of magnitude, which could indicate the observed yields have significant contributions from a compound reaction mechanism.

\section{The $^{23}$N\lowercase{a}$(p, \gamma)$ and $^{23}$N\lowercase{a}$(p, \alpha)$  Reaction Rates}
\label{sec:23nap-gamma-23nap}
There exists a formidable amount of data relevant to the $^{23}$Na$(p, \gamma)$ and $^{23}$Na$(p, \alpha)$ reaction rates. The values compiled in Ref.~\cite{hale_2004} make up the majority of the current STARLIB rates \cite{sallaska_2013}. A detailed reanalysis of these rates is likely needed, but is well beyond the scope of the current work. As such, we focus our efforts on showing the astrophysical implications of the results presented above. To do this we construct two updated versions of the rates in Ref.~\cite{BOELTZIG_2019}, which are themselves updates of STARLIB Version v6.5 \cite{sallaska_2013}. The first update (called New1) uses {all of our} recommended resonance energies presented in Sec.~\ref{sec:recommended-energies} {and scales the STARLIB proton partial widths for consistency}. The second update (called New2) is a more exploratory study that, {in addition to the updated energies,} replaces the resonance strength {for the $133$-keV resonance} reported in Ref.~\cite{BOELTZIG_2019} {with the proton partial widths measured in this work (Table \ref{tab:gamma_p_table}) using the probabilities for $\ell=0\text{-}3$ transfers from Table \ref{tab:probs}}. {New2} also makes corrections to subthreshold resonances involved in the $(p, \alpha)$ rate. All rates and their uncertainties were calculated using the Monte-Carlo reaction rate code \texttt{RatesMC} \cite{LONGLAND_2010_1}.    

%\footnote{Current version of STARLIB is available at https://github.com/Starlib/Rate-Library}

\subsection{Energy Update}
\label{sec:energy-update}

\begin{figure*}
    \centering
    \includegraphics[width=.45\textwidth]{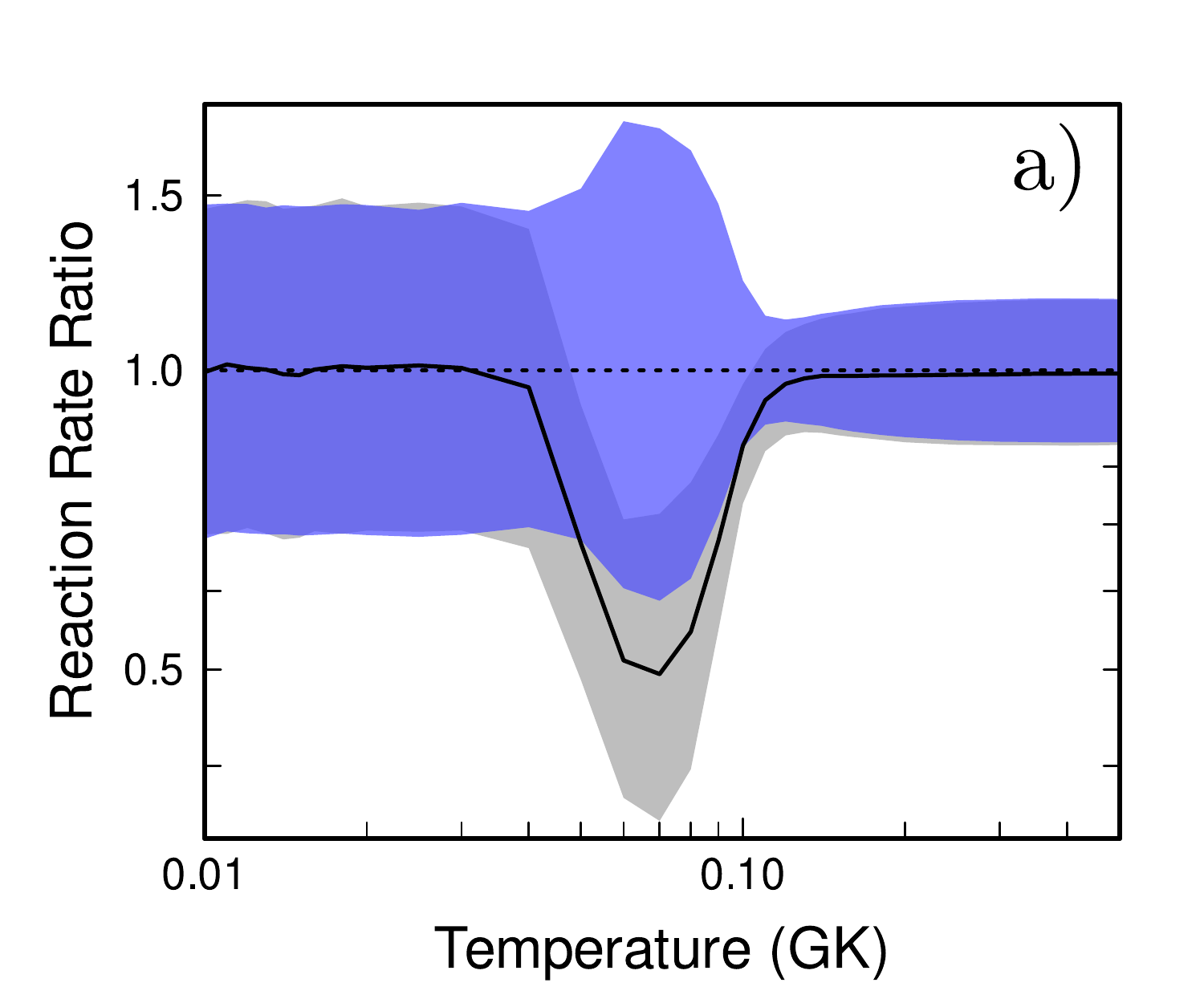}\includegraphics[width=.45\textwidth]{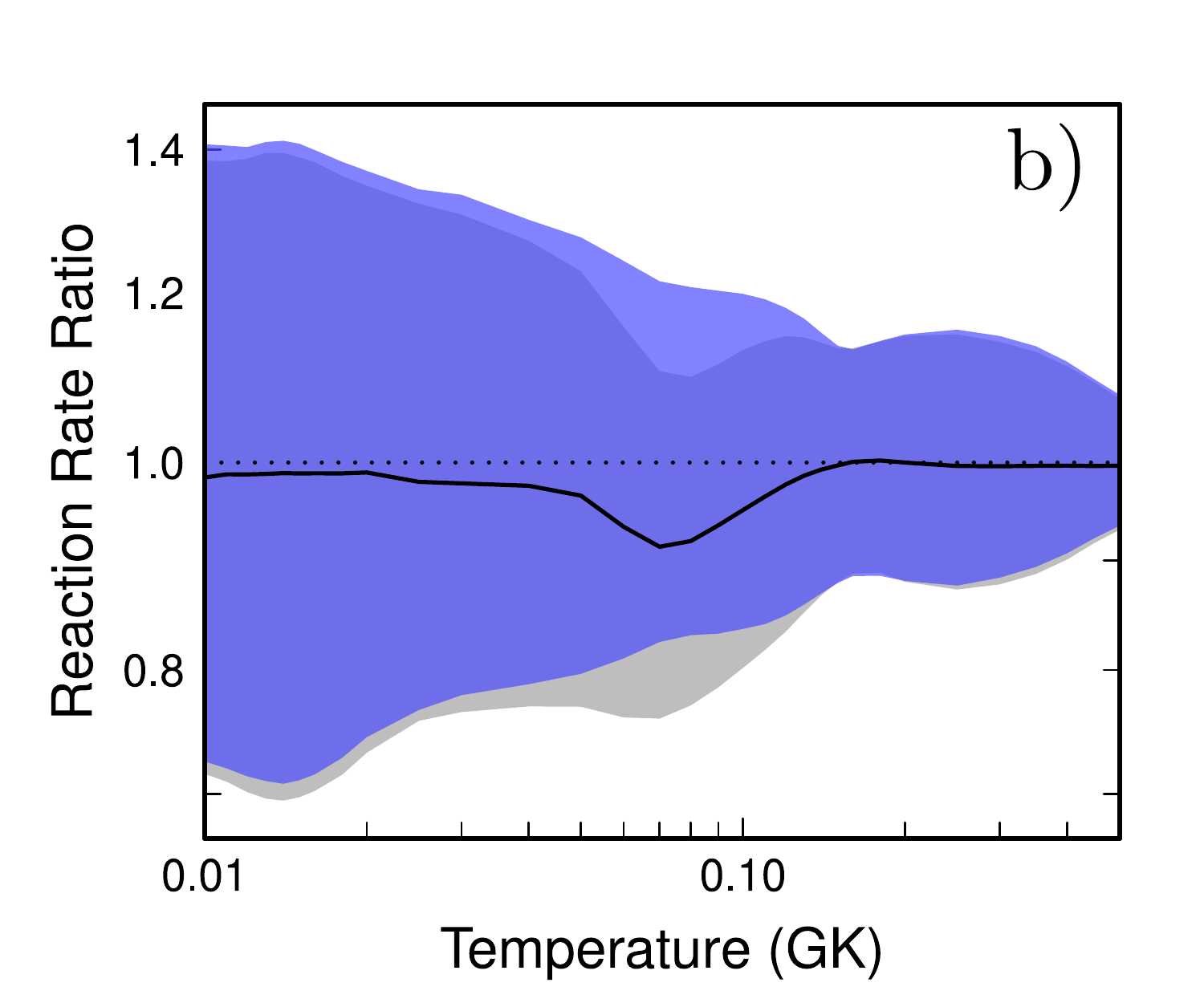}
    \caption{New1 reaction rates {normalized to their median}. (left) The $(p, \gamma)$ reaction rate {taken from Ref.\cite{Marshall_2021}}. The blue contours show the relative uncertainty as a function of temperature. The gray contour is the recommend rate of Ref.~\cite{BOELTZIG_2019} {normalized to the updated rate's median}. Both contours show $68 \%$ coverage. (right) The reaction rate ratio plot for the $(p, \alpha)$ rate.}
    \label{fig:compare_p_g_energy_update}
\end{figure*}

The resonance energies presented in Table \ref{tab:comp_and_recommended} were substituted into the \texttt{RatesMC} input files provided by STARLIB for the $(p, \gamma)$ and $(p, \alpha)$ rates. Particle partial widths were scaled as needed to reflect the new energies. {Normalizing the rates to their own median produces the reaction rate ratios shown in Fig.~\ref{fig:compare_p_g_energy_update}}. The blue contours centered around one show the $68 \%$ coverage of the rates of this work, while the gray contour is the ratio of the rate as determined by LUNA \cite{BOELTZIG_2019} to the updated rate. The influence of the new energy for the $133$-keV resonance on the $(p, \gamma)$ rate can be clearly seen. Recall that the resonance energy enters the rate exponentially, and in this case the $5$-keV shift in energy is responsible for the rate increasing by a factor of $2$ for temperatures of $70 \text{-} 80$ MK. The impact of the new energies on the $(p, \alpha)$ rate are more modest. A factor of $1.25$ increase is observed as a result of the lower energy for the $168$-keV resonance resulting from the exclusion of Hale's measurement from the weighted average. The updated rate is still well within the uncertainty of the current STARLIB rate.    

\subsection{Partial Widths Update}
\label{sec:part-widths-update}

The partial widths extracted in this study are consistent with those reported in Ref.~\cite{hale_2004}. However, it was found that $\theta^2$ value for the $-304$-keV resonance was erroneously translated using the value of $(2J+1)C^2S$ instead of $C^2S$ in Ref.~\cite{ILIADIS_2010_3}, making this subthreshold $p$-wave resonance appear $\times 3$ stronger. When corrected, the sub-threshold region is dominated primarily by the two $s$-wave resonances at $-240$ keV and $-172$ keV, and the rate at lower temperatures is increased.

\begin{figure*}
    \centering
    \includegraphics[width=.45\textwidth]{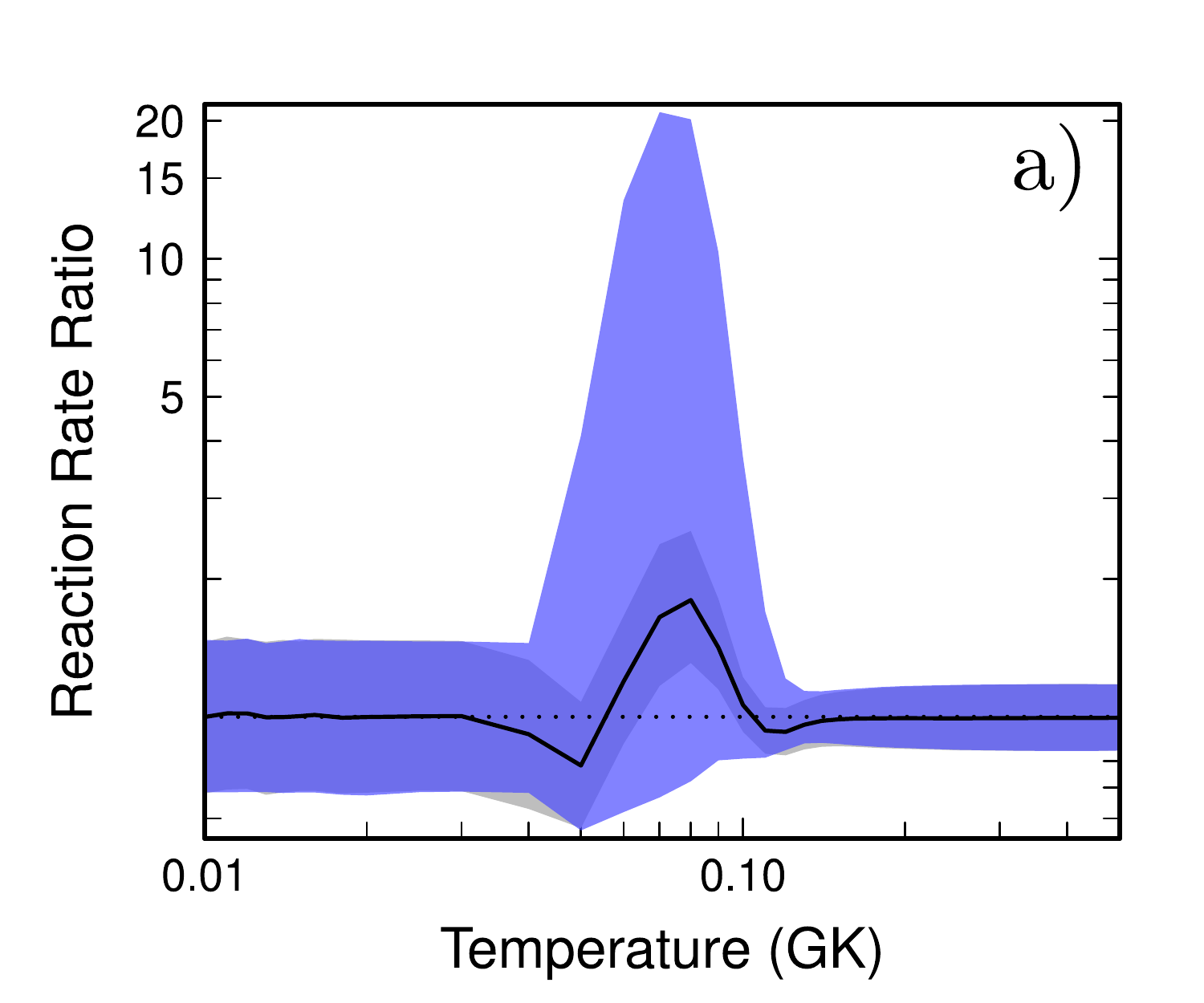}\includegraphics[width=.45\textwidth]{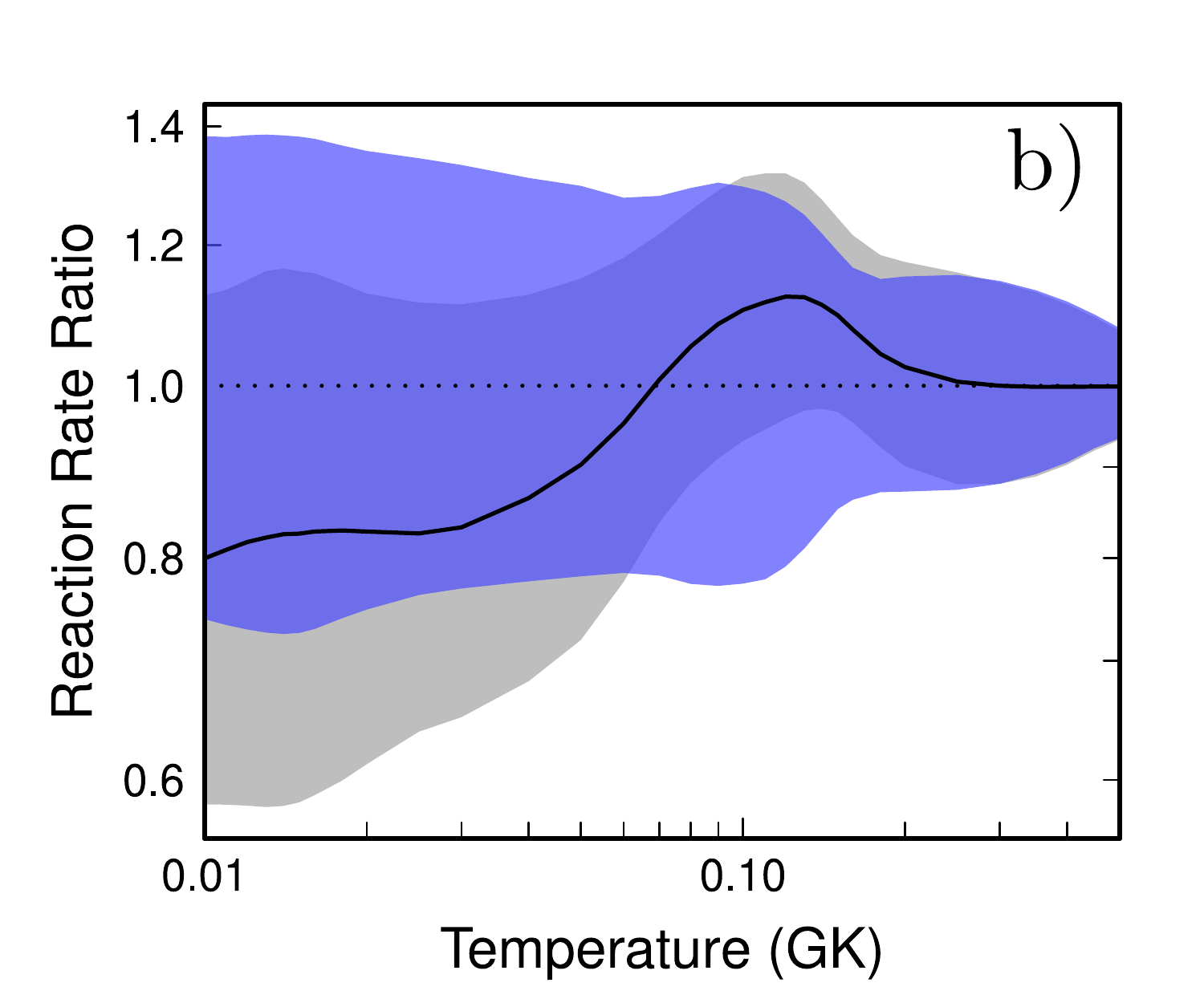}
    \caption{New2 reaction rates {normalized to their median}. The gray contour is the recommend rate of Ref.~\cite{BOELTZIG_2019} {normalized to the New2 rate's median} (left) The $(p, \gamma)$ reaction rate ratio. Large variations are seen around 80 MK due to the uncertain spin parity assignment for the $133$-keV resonance. (right) The reaction rate ratio plot for the $(p, \alpha)$ rate.}
    \label{fig:compare_p_g_energy_update_New2}
\end{figure*}

In the case of the $133$-keV resonance, we substitute our proton partial widths weighted by the probabilities given in Table.~\ref{tab:probs}. Folding different $\ell$ probabilities estimated directly from transfer reactions is only possible due to the Bayesian methods developed in Ref.~\cite{Marshall_2020} and the Monte-Carlo reaction rate developed in Ref.~\cite{LONGLAND_2010_1, 2014_Mohr}. The net effect is a dramatically more uncertain rate in the temperature ranges relevant to globular cluster nucleosynthesis, as can be seen in Fig.~\ref{fig:compare_p_g_energy_update_New2}. {New2 uses the same energy value updates as New1.}   

\subsection{AGB Models}

% ---------------------intro-----------------------------------------------------------------------------------------

The impact of our updated sodium destruction rates was examined in the context of intermediate mass (M $\gtrsim 4M_{\odot}$, depending on metallicity) AGB stellar environments. AGB models that are sufficiently massive to enter the thermally pulsing AGB (also dependent on initial metallicity, see Ref.~\cite{Karakas22paper1}) can activate the NeNa cycle within the intermittent hydrogen burning shell for temperatures greater than $~15$ MK. Hydrogen burning can also occur at the base of the convective envelope if temperatures exceed $50$ MK, with the NeNa cycle operating for $T > 80-100$ MK. This process is known as hot bottom burning (HBB) and can lead to significant enhancement of hydrogen burning products in the envelope \cite{Siess10, Ventura13, Cristallo15, Karakas16, Pignatari16, Karakas18, Cinquegrana22paper2}. AGB stars that undergo HBB provide a possible explanation for the Na-O abundance anomaly of globular clusters, though they and other models cannot account for all observations \cite{problems_with_hbb}. The competition between $^{23}$Na production via the NeNa cycle and $^{23}$Na destruction via the proton capture channels in question is not limited exclusively to the thermally pulsing AGB. $^{23}$Na produced during the main sequence can be mixed to the stellar surface during the first and second (for intermediate mass stars) dredge up events. For further details of AGB evolution and nucleosynthesis, we refer the reader to Refs.~\cite{Busso99, Herwig05, Nomoto13, Karakas14, Karakas22paper1, Cinquegrana22paper2, Ventura22nuc}. Specifically for $^{23}$Na production in AGB stars, see Refs.~\cite{Forestini97, Mowlavi99sodium, Mowlavi1999third, Karakas03neon, Ventura06, Cristallo06, Siess07chemical, Izzard07, Doherty13, Slemer16, DAntona16nuc}. 

The models discussed in this section all achieve temperatures sufficient for HBB, but of varying efficiencies. We choose a range of initial masses and metallicities (where $Z$ denotes the initial mass fraction of all elements heavier than helium): 4 and 7\(M_\odot\) at $Z=0.001$, 4\(M_\odot\) at $Z=0.0028$, 6 and 8\(M_\odot\) at $Z=0.014$ (solar metallicity; \citealt{Asplund09}). The evolutionary properties of the $Z=0.014$ models were previously published in Refs.~\cite{Karakas14He, Karakas16}, the $Z=0.0028$ model in Ref.~\cite{Karakas18} and the 7\(M_\odot\), $Z=0.001$ model in Ref.~\cite{Fishlock14}. We note that in general, lower metallicities and higher initial masses will produce higher temperatures at the base of the envelope.
The lowest metallicity at 7\(M_\odot\) contains the highest temperatures, the lower mass model, 4\(M_\odot\), at the same metallicity contains the coolest. 

% ---------------------the models----------------------------------------------------------------------------

The evolutionary sequences were run with the Monash stellar evolution code (\cite{Lattanzio86, Frost96, Karakas07}; described most recently in Ref.~\cite{Karakas22paper1, Karakas14He} and Ref.~\cite{2022arXiv220801859C}) and post processing nucleosynthesis code (see Refs.~\cite{Cannon93, Lattanzio96, Lugaro12, Cinquegrana22paper2}). Briefly, the evolution of the models is run from the zero-age main sequence to the tip of the thermally pulsing AGB. Mass loss on the red giant branch (RGB) is only included in the 4M, Z=0.001 model. The quantity of mass lost from intermediate mass stars is typically insignificant on the RGB owing to their short lifetimes in this phase. For the 4\(M_\odot\), $Z=0.001$ model, the approximation of Ref.~\cite{Reimers75} is used, with parameter $\eta_R$=0.477 based on Ref.~\cite{Mcdonald15mass}. For AGB mass-loss, we use the semi-empirical mass-loss rate in all the models \cite{Vassiliadis93}, except for the 4$M_{\odot}$, $Z=0.001$ model in which we use the method of Ref.~\cite{Blocker95a} with $\eta = 0.02$ (see treatment of mass-loss in Ref.~\cite{Karakas18} for intermediate-mass AGB stars). We treat convection using the Mixing-Length Theory (MLT) of convection, with the MLT parameter, $\alpha _{\rm MLT}$, set to 1.86. We assume instantaneous mixing in convective regions and use the method of \textit{relaxation} \cite{Lattanzio86} to determine the borders of convective regions. We use the \AE SOPUS low temperature opacity tables of Ref.~\cite{Marigo09} and the OPAL opacities \cite{Iglesias96} for high temperature regions. The evolution code follows six isotopes, $^{1}$H, $^{3}$He, $^{4}$He, $^{12}$C, $^{14}$N and $^{16}$O, adopting the rates of Refs.~\cite{Harris83thermonuclear, Fowler75thermonuclear, Caughlan88}. 

These sequences are then fed into a post processing nucleosynthesis code, which follows 77 isotope species from hydrogen to sulfur, with a few iron-peak nuclei \cite{Karakas10Up}. We use solar scaled initial compositions based on the solar abundances of Ref.~\cite{Asplund09}. Excluding the $^{23}$Na proton capture rates investigated in this paper, the $Z=0.001$ and $Z=0.014$ models use the nuclear reaction rates from the 2016 default JINA REACLIB database \cite{Cyburt10}. For the $Z=0.0028$ model, the rates were updated to the 2021 default set from the same database.  

For all models, we ran three calculations with different sets of median rates: LUNA’s latest experimental results (LUNA), the energy updates from this paper (New1, Sec.~\ref{sec:energy-update}), and the partial width updates (New2, Sec.~\ref{sec:part-widths-update}). LUNA and New1 differ only in the temperature range of 60-80 million K, as seen in Fig.~\ref{fig:compare_p_g_energy_update}. In this range, the New1 sodium destruction rates are faster due to the influence of the $133$-keV resonance in the $^{23}$Na$(p, \gamma)$ reaction. 
% ---------------------results------------------------------------------------------------------------------------------

%In Fig.~\ref{fig:m6z014_detail}, we show the surface abundances of isotopes $^{23}$Na, $^{20}$Ne and $^{24}$Mg as a function of model number (which we use as a proxy for evolutionary time). Each row corresponds to a different stellar mass.

For each model, we also calculate the stellar yields. The yields are the integrated mass expelled from the model over its lifetime, where a positive yield for an isotope indicates net production of that species, a negative indicates net destruction of that species. These are shown in Table~\ref{tab:yields}. The largest impact of the New1 rates is found in the 8\(M_\odot\), solar metallicity model, which holds a 5$\%$ variation in yields between New1 and LUNA. This is most likely due to this particular model experiencing the largest duration of HBB within the temperature range for which the difference between New1/New2 and LUNA is at a maximum. The next largest variation is seen by the 6\(M_\odot\), solar metallicity model which shows a 2$\%$ difference between the yields. The variation for both 4\(M_\odot\) models and the 7\(M_\odot\), $Z=0.001$ model are less than 1$\%$. The $^{20}$Ne abundances mirror those of $^{23}$Na, where higher quantities of $^{20}$Ne are found with the New1 rates. There is almost no difference in $^{24}$Mg between the rates for any of the models. It would therefore appear that most of the variation in the $^{23}$Na yields between the rates is coming from the small variations in the median $^{23}$Na$(p, \alpha)$ rate.

\begin{table*}[t]
\centering
  \setlength{\tabcolsep}{0pt}
  \caption{\label{tab:yields} Stellar yields for stable isotopes of interest {for various masses and metallicities}.}
  \begin{tabular}{l|ccc|ccc|ccc}
    \toprule
    \toprule
        &                     & $^{23}$Na           &                     &                     & $^{20}$Ne           &                     &                     & $^{24}$Mg           &                     \\
        & LUNA                & New1                & New2                & LUNA                & New1                & New2                & LUNA                & New1                & New2                \\ \hline
                                                                                                                                                                                                              \\ [-1.5ex]
m4z001  & $1.65\times 10^{-5}$  & $1.64\times 10^{-5}$  & $1.61\times 10^{-5}$  & $-1.29\times 10^{-6}$ & $-1.29\times 10^{-6}$ & $-9.48\times 10^{-7}$ & $-1.69\times 10^{-6}$ & $-1.69\times 10^{-6}$ & $-1.68\times 10^{-6}$ \\ [1.2ex]
m7z001  & $-1.05\times 10^{-5}$ & $-1.05\times 10^{-5}$ & $-1.02\times 10^{-5}$ & $2.19\times 10^{-5}$  & $2.20\times 10^{-5}$  & $2.12\times 10^{-5}$  & $-2.31\times 10^{-4}$ & $-2.31\times 10^{-4}$ & $-2.31\times 10^{-4}$ \\ [1.2ex]
m4z0028 & $2.48\times 10^{-5}$  & $2.47\times 10^{-5}$  & $2.47\times 10^{-5}$  & $2.01\times 10^{-5}$  & $2.00\times 10^{-5}$  & $2.02\times 10^{-5}$  & $-7.94\times 10^{-6}$ & $-7.94\times 10^{-6}$ & $-7.93\times 10^{-6}$ \\ [1.2ex]
m6z014  & $9.08\times 10^{-5}$  & $8.89\times 10^{-5}$  & $9.19\times 10^{-5}$  & $3.37\times 10^{-6}$  & $4.79\times 10^{-6}$  & $2.57\times 10^{-6}$  & $-2.68\times 10^{-4}$ & $-2.68\times 10^{-4}$ & $-2.69\times 10^{-4}$ \\ [1.2ex]
m8z014  & $1.05\times 10^{-4}$  & $1.00\times 10^{-4}$  & $9.28\times 10^{-5}$  & $2.37\times 10^{-5}$  & $2.76\times 10^{-5}$  & $2.73\times 10^{-5}$  & $-1.79\times 10^{-3}$ & $-1.79\times 10^{-3}$ & $-1.80\times 10^{-3}$ \\ 
    \bottomrule
    \bottomrule
  \end{tabular}
\end{table*}

\begin{figure}
  \centering
  \includegraphics[width=0.5\textwidth]{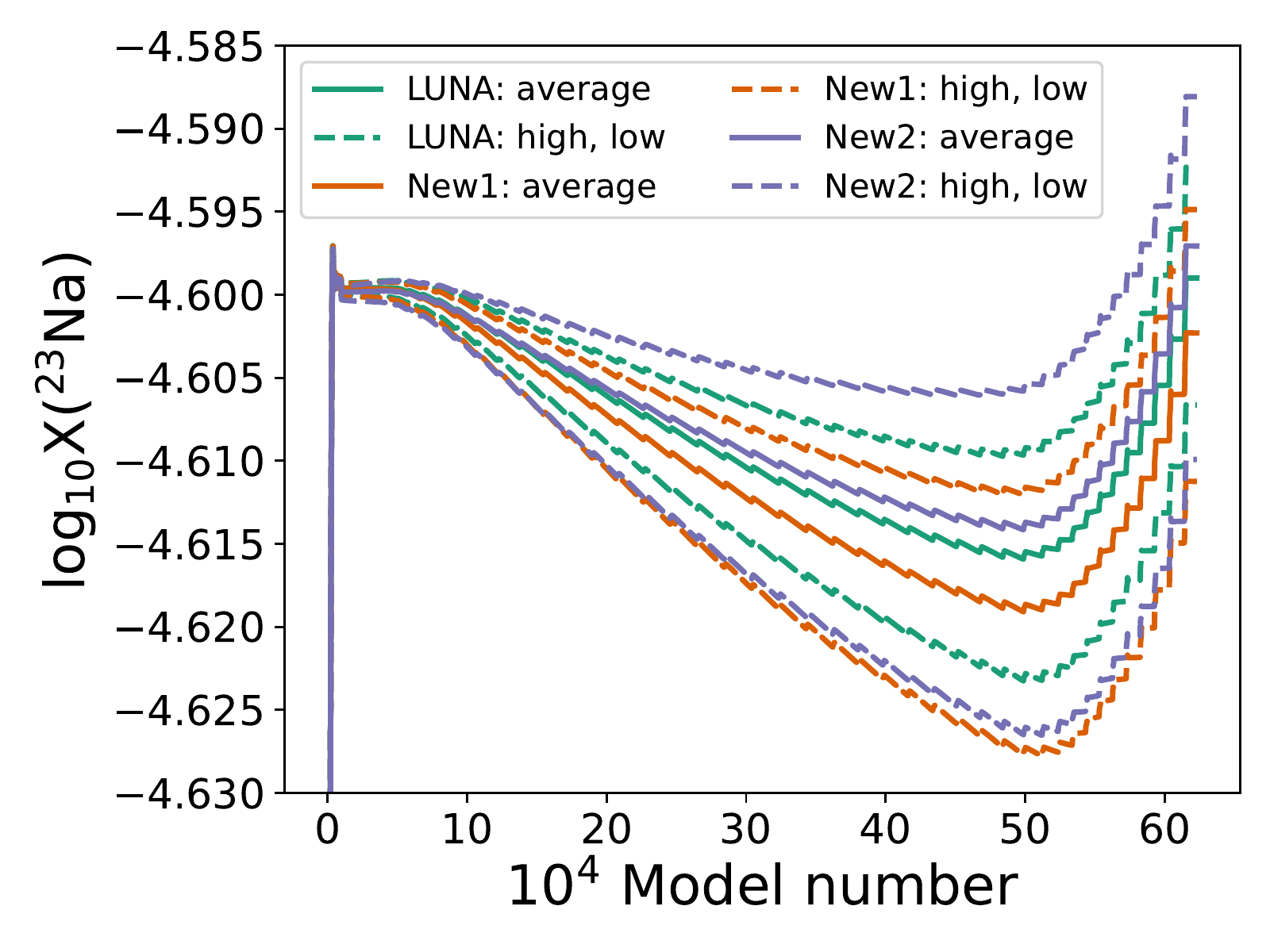}
  \caption{$^{23}$Na surface abundance for median {(solid line)} and low/high ({dashed lines}) reaction rates for LUNA, New1, and New2 (see text for details). {The high rates for the $^{23}$Na$(p, \gamma)$ and $^{23}$Na$(p, \alpha)$ lead to lower $^{23}$Na surface abundances and vice versa for the low rates}. Model number is a proxy for time. The impact of rate uncertainties shown here is small.}
  \label{fig:m6z014_detail}
\end{figure}

For one chosen model, 6\(M_\odot\) and $Z=0.014$, we ran a further six calculations (two combinations for each set of rates), to estimate the potential impact of the rate uncertainties. For each of the three rates, a high and low rate were run. These correspond to the $16^{\text{th}}$ and $84^{\text{th}}$ percentile of both the $(p, \gamma)$ and $(p, \alpha)$ rates. Thus, the high rate has an increase in both destructive rates and the low rate has a corresponding decrease in both. We show these results in Figure.~\ref{fig:m6z014_detail}. Even with the conservative uncertainties of New2, there appears to be very little impact on $^{23}$Na production.    

% ---------------------discussion-------------------------------------------------------------------------------------

\subsection{Discussion}

The $^{23}$Na abundances and the initial mass and metallicity thresholds for which the updated rates show maximum variation should be considered qualitatively. There are various uncertainties in stellar modeling that directly impact the temperature at the base of the convective envelope.  These uncertainties can skew the exact amount of $^{23}$Na that is destroyed. For example, we use the MLT to treat convective regions in the Monash code. Other methods, such as the Full Spectrum of Turbulence \cite{Canuto91, Canuto96} used in the ATON code \cite{Ventura98, Ventura20}, are known to produce higher temperatures at the base of the convective envelope. Consequently, HBB occurs at a lower initial stellar mass \cite{Ventura18}. The choice of mass loss rate on the AGB will also impact HBB. The mass loss rate of Ref.~\cite{Vassiliadis93} is slower than that of Ref.~\cite{Blocker95a} when used in intermediate-mass stellar models (e.g., see discussion in Ref.~\cite{Karakas16}). The mass loss rate of Ref.~\cite{Vassiliadis93} results in more thermal pulses, a longer AGB lifetime, and consequently the base of the envelope will spend longer at higher temperatures. Hence the 4$M_{\odot}$ model of $Z=0.0028$ from Ref.~\cite{Karakas18} achieves much higher temperatures at the base of the envelope compared to the model of the same mass and metallicity evolved with the mass loss rate of Ref.~\cite{Blocker95a}.

There is a general pattern of faster $^{23}$Na destruction with the New1 rates as opposed to LUNA in AGB models that spend significant time during HBB in the key temperature range of $60$-$80$ million K. However, the exact initial mass and metallicity thresholds for which HBB at this temperature range occurs is heavily dependent on the stellar evolution mode utilized alongside the chosen input physics.
Simple single zone calculations still indicate the importance of the $^{20}$Ne$(p, \gamma)$, $^{23}$Na$(p, \alpha$), and $^{23}$Na$(p, \gamma)$ reactions rates, but the above calculations emphasize that stellar modeling uncertainties dominate once a polluter candidate is chosen.     

\section{Conclusions and Outlook}
\label{sec:conclusions-outlook}

Utilizing the high resolution capabilities of the TUNL SPS, astrophysically important excited states in $^{24}$Mg were populated via the $(^3 \text{He}, d)$ transfer reaction. Careful calibration and compilation of previous results give a significantly lower resonance energy for the $133$-keV resonance. This resonance has the single largest contribution to the $(p, \gamma)$ reaction rate at temperatures important for globular cluster nucleosynthesis. Angular distributions were analyzed using the Bayesian DWBA methods of Ref.~\cite{Marshall_2020}, and spectroscopic factors were extracted. Methods were developed to deal with the additional challenges presented by $^{23}$Na$(^3$He$, d)^{24}$Mg: mixed $\ell$ transfers, a severe discrete ambiguity, and data that needed absolute scaling established during the fitting. These advances mean that our analysis is the first of its kind, where Bayesian methods were used to accurately determine uncertainties at every step of the analysis of a transfer reaction. As a result of the above effort, astrophysical reaction rates derived from such experiments will naturally reflect the underlying nuclear physics uncertainties. 

The astrophysical impact of these uncertainties was briefly investigated. Our work indicates that the unknown astrophysically conditions still dominate the total uncertainty. However, in a given environment significant variation still exists due to uncertainties in the NeNa cycle. 
The results of this experiment indicate that uncertainties are still present in both the $^{23}$Na$(p, \gamma)$ and $^{23}$Na$(p, \alpha)$ reaction rates. 

{The direct capture component of the rate, which dominates at temperatures lower than 60 MK and is significant up to 70 MK, has recently been updated by Boeltzig \textit{et. al} \cite{Boeltzig_2022}. Larger uncertainties on the direct capture component were found due to previously neglected interference effects and an assumption of larger uncertainties on $C^2S$. Updated spectroscopic factors for bound states could significantly alter the behavior of the low temperature portion of the reaction rate.} We suggest that future work focus on the direct capture component of the $^{23}$Na$(p, \gamma)$ rate, the precise energy determination of the $133$-keV resonance, and the sub-threshold region of $^{23}$Na$(p, \alpha)$. At this time our knowledge of the Na-O anticorrelation in globular clusters is still limited by the nuclear physics.

\section*{Acknowledgments}

The authors would like to thank Christian Iliadis for his valuable input and the TUNL technical staff for their assistance during the experiment. This material is based upon work supported by the U.S. Department of Energy, Office of Science,
Office of Nuclear Physics, under Award No. DE-SC0017799 and Contract No. DE-FG02-97ER41041. G.C.C and A.I.K are supported by the Australian Research Council Centre of Excellence for All Sky Astrophysics in 3 Dimensions (ASTRO 3D), through project number CE170100013.

\section*{Appendix A: The Energies Reported by Hale \textit{\lowercase{et al}}.}
\label{sec:hale_discussion}

As first reported in Ref.~\cite{Marshall_2021}, a significant disagreement exists between the results of our measurement and those of Ref.~\cite{hale_2004}. Of particular concern is the state corresponding to the $138$-keV resonance, whose mean values falls $\approx 9$ keV below what is reported in Ref.~\cite{hale_2004}. A disagreement of this magnitude is of particular concern since the previous measurement was also performed at TUNL using the SPS. 

Studying the information reported in Ref.~\cite{hale_2004}, measured energies are reported for a single region of the focal plane that covers $\approx 400$ keV. These energies are extracted from a $3^{\textnormal{rd}}$ order polynomial calibration based on twelve states surrounding the mentioned region. Of these twelve states, the most interior, i.e, the states that begin and end the interpolated region, were states identified as $11330$ keV and $12184$ keV, respectively. Comparing the spectrum from this work and that shown in Fig.~3 of Ref.~\cite{hale_2004}, the state labeled $11330$ keV in their spectrum corresponds to the state identified as $11317(3)$ keV in this work. Ref.~\cite{firestone_2007} lists two states around this energy range, one with $E_x = 11314.4(15) $ keV and the other $E_x = 11330.2(10)$ keV. Neither of these states has an unambiguous spin parity assignment in the current evaluation, but the preceding compilation of Ref.~\cite{endt_eval_1990} identified the lower energy state ($11314$ keV) as $(3,4)^+$  and the higher ($11330$ keV) as $(2^+ \text{-} 4^+)$. These assignments seem to be in tension with the $(p, p^{\prime})$ angular distribution of Ref.~\cite{zwieglinski_1978}, which assigns the lower lying state $\ell=3$ giving $J^{\pi} = 3^{-}$. However, Ref.~\cite{Warburton_1981} reports $\log ft = 5.19(14)$ for $^{24}$Al$(\beta^+)$ ( ground state $J^{\pi} = 4^+$), which based on the empirical rules derived in Ref.~\cite{Raman_1973} requires an allowed decay giving $(3, 4, 5)^+$ for this state. In light of these discrepancies, it is hard to reach a firm conclusion about the identity of the state populated in this work and Ref.~\cite{hale_2004}.

One method to investigate the disagreement is to recalibrate our data using the calibration states of the previous study. This cannot be considered a one-to-one comparison because of the Bayesian method used to calibrate the focal plane and the different focal plane detectors used in each study, but it should show the impact of misidentifying the states around $11320$ keV. To be specific, we consider two sets of energies:
\begin{enumerate}
    \item The adopted results of this work from Sec.~\ref{sec:energy_cal_na} (Set $\#1$).
    \item The peak centroids of this work energy calibrated using the calibration states of Hale \textit{et al.} (Set $\#2$).
\end{enumerate}
The results shown in Table \ref{tab:hale_comp_energies} report these two sets of energies and compares them to Ref.~\cite{hale_2004}. Using the same calibration for our data (Set $\#2$) produces consistent results with the Ref.~\cite{hale_2004}.

The above discussion presents the evidence that led to the decision to exclude excitation energies of Ref.~\cite{hale_2004} from the recommended energies of the current work. There is a reasonable cause to do this at the current time, but further experiments are needed to firmly resolve this issue. 

\begin{table}
\centering
\setlength{\tabcolsep}{8pt}
\caption{ \label{tab:hale_comp_energies} Comparison of the $^{24}$Mg excitation energies measured in this work (Set $\# 1$), the excitation energies derived from our data if the calibration of Hale \textit{et al.} is used (Set $\# 2$), and finally the energies Hale \textit{et al.} reported in Ref.~\cite{hale_2004}. All energies are in units of keV. These results indicate that the state close to $11320$ keV was previously misidentified, and, as a result, led to systematically higher excitation energies.}
\begin{threeparttable}
\begin{tabular}{lllllllll}
\toprule
\toprule
Set No.~$1$ (keV) & Set No.~$2$ (keV) &  Hale \textit{et al.} \cite{hale_2004} (keV) \\ \hline
$11688.7(14$) & $11695(3)$     & $11698.6(13)$ \\
$11823(3)$    & $11828(3)$     & $11831.7(18)$ \\
$11857(3)$    & $11860.1(19)$  & $11862.7(12)$ \\
$11935(3)$    & $11937.5(17)$  & $11936.5(12)$ \\
              &                & $11965.3(12)$ \footnote{Ref.~\cite{hale_2004} reports this state, which appears as an unresolved peak in their spectrum. The current study does not find a corresponding peak in the same region.} \\
$11989.3(14)$ & $11991.2(17)$  & $11992.9(12)$ \\
$12014(3)$    & $12016.2(16)$  & $12019.0(12)$ \\
$12050(3)$    & $12051.4(17)$  & $12051.8(12)$ \\
\bottomrule
\bottomrule
\end{tabular}
\end{threeparttable}
\end{table}

% \begin{tablenotes}
% \item[$\dagger$] 
% \end{tablenotes}

\raggedbottom
%\pagebreak

%merlin.mbs apsrev4-1.bst 2010-07-25 4.21a (PWD, AO, DPC) hacked
%Control: key (0)
%Control: author (8) initials jnrlst
%Control: editor formatted (1) identically to author
%Control: production of article title (-1) disabled
%Control: page (0) single
%Control: year (1) truncated
%Control: production of eprint (0) enabled
%

\end{document}